\documentclass[structabstract]{aa}  
\usepackage{graphicx}
\usepackage{amsmath,amssymb}
\usepackage[applemac]{inputenc}
\usepackage[modulo, switch]{lineno}
\usepackage{xcolor}
\usepackage{txfonts}
\usepackage{natbib}
\usepackage{hyperref}
\bibpunct{(}{)}{;}{a}{}{,} 
\newcommand{\Kepler}{\textit{Kepler}}
\newcommand{\kepler}{\textit{Kepler} }
\newcommand{\p}{\textit{p} }
\newcommand{\Dnu}{\Delta \nu}
\newcommand{\numax}{\nu_\mathrm{max}}
\newcommand{\dnua}{\delta\nu_\mathrm{01}}
\newcommand{\dnub}{\delta\nu_\mathrm{02}}

\newcommand{\dtheta}{\boldsymbol{\theta}}
\newcommand{\likel}{\mathcal{L} \left( \dtheta \right)}
\newcommand{\lnlikel}{\Lambda \left( \dtheta \right)}
\newcommand{\constr}{\mathcal{L}^*}
\newcommand{\normal}{\mathcal{N}_j \,( h_\mu, h_\sigma )}
\newcommand{\uniform}{\mathcal{U}_j \,(h_l, h_u )}
\newcommand{\supergauss}{\mathcal{S}_j \,(h_c, h_w, h_\sigma)}
\newcommand{\prior}{\pi \left( \dtheta \mid \mathcal{M} \right)}
\newcommand{\evid}{\mathcal{E}}
\newcommand{\model}{\mathcal{M}}
\newcommand{\nnest}{N_\mathrm{nest}}
\newcommand{\nlive}{N_\mathrm{live}}
\newcommand{\msame}{M_\mathrm{same}}
\newcommand{\minit}{M_\mathrm{init}}
\newcommand{\matt}{M_\mathrm{attempts}}
\newcommand{\nclust}{N_\mathrm{clust}}
\newcommand{\nmin}{N_\mathrm{min}}
\newcommand{\ndata}{N_\mathrm{data}}

\newcommand{\diamonds}{\textsc{Diamonds}}

\begin{document}
   \title{DIAMONDS: A new Bayesian nested sampling tool\thanks {Software package available at the \diamonds\,\,code website: \href{https://fys.kuleuven.be/ster/Software/Diamonds/}{https://fys.kuleuven.be/ster/Software/Diamonds/}.} \\}\subtitle{Application to Peak Bagging of solar-like oscillations}

   \author{E. Corsaro\inst{1}\fnmsep\thanks{e-mail: enrico.corsaro@ster.kuleuven.be}
          \and
          J. De Ridder\inst{1}
          }

   \institute{\inst{1}Instituut voor Sterrenkunde, KU Leuven, Celestijnenlaan 200D, B-3001 Leuven, Belgium
             }

   \date{Received ; accepted }

 
  \abstract
{Thanks to the advent of the space-based missions \textit{CoRoT} and NASA's \Kepler, the asteroseismology of solar-like oscillations is now at the base of our understanding about stellar physics. The \kepler spacecraft, especially, is releasing excellent photometric observations of more than three years length in high duty cycle, which contain a large amount of information that has not yet been investigated.}
{To exploit the full potential of \kepler light curves, sophisticated and robust analysis tools are now required more than ever. Characterizing single stars with an unprecedented level of accuracy and subsequently analyzing stellar populations in detail are fundamental to further constrain stellar structure and evolutionary models.}
{We developed a new code, termed \diamonds, for Bayesian parameter estimation and model comparison by means of the nested sampling Monte Carlo (NSMC) algorithm, an efficient and powerful method very suitable for high-dimensional and multi-modal problems. A detailed description of the features implemented in the code is given with a focus on the novelties and differences with respect to other existing methods based on NSMC. \diamonds\,\,is then tested on the bright F8 V star KIC~9139163, a challenging target for peak-bagging analysis due to its large number of oscillation peaks observed, which are coupled to the blending that occurs between $\ell=2,0$ peaks, and the strong stellar background signal. We further strain the performance of the approach by adopting a 1147.5 days-long \kepler light curve, accounting for more than 840,000 data bins in the power spectrum of the star.}
   {The \diamonds\,\,code is able to provide robust results for the peak-bagging analysis of KIC~9139163, while preserving a considerable computational efficiency for identifying the solution at the same time. We test the detection of different astrophysical backgrounds in the star and provide a criterion based on the Bayesian evidence for assessing the peak significance of the detected oscillations in detail. We present results for 59 individual oscillation frequencies, amplitudes and linewidths and provide a detailed comparison to the existing values in the literature, from which significant deviations are found when a different background is used. Lastly, we successfully demonstrate an innovative approach to peak bagging that exploits the capability of \diamonds\,\,to sample multi-modal distributions, which is of great potential for possible future automatization of the analysis technique.}
   {}

   
   \authorrunning{E. Corsaro and J. De Ridder}
   \titlerunning{DIAMONDS: a new Bayesian Nested Sampling tool}
   \maketitle
%

\section{Introduction}
The advent of the space-based photometric missions, \textit{CoRoT} \citep{Baglin06,Michel08} and NASA's \kepler \citep{Borucki10,Koch10}, has revolutionized the asteroseismology of stars exhibiting solar-like oscillations, a type of stellar oscillation that is stochastically excited and intrinsically damped, which was observed for the first time in the Sun \citep[e.g. see][for a summary on solar-like oscillations]{BK03,BK06}. 

Since May 2009, the \kepler spacecraft in particular has been providing an incredible amount of high-quality data, which allows us to study the asteroseismic properties of a large sample of low-mass stars \citep[e.g. see][for a review]{Garcia11review}. As a result, asteroseismology now confirms its key role in improving our understanding of stellar structure and evolution. 

In addition to the ensemble studies of stellar populations, which became possible for the first time already from the beginning of the \kepler mission \citep{Chaplin11Sci}, the need for investigating detailed asteroseismic properties, such as frequencies, lifetimes and amplitudes of individual oscillations, is becoming even more important at present times. The increasing observing length of the \kepler light curves is opening up possibilities for extremely detailed data analysis and modeling of stars similar to our Sun. These studies are able to yield constraints on fundamental stellar properties, such as mass, radius, and age, the internal structure, and composition \citep[e.g.][]{CD04}. Detailed modeling of some bright targets observed by \kepler have already been published \citep[e.g. see][]{Metcalfe10modeling,Metcalfe12modeling}.

However, measuring a complete set of characteristic asteroseismic parameters from informative power spectra that contain tens of resolved oscillations, while being able to retrieve results in a reasonable amount of time in parallel, is still a very challenging task to be accomplished. This analysis, often referred to as \textit{peak bagging} \citep[e.g. see][]{App03PB}, involves fitting models in high-dimensional spaces and typically deals with pronounced degeneracies that convey a number of possible solutions. In addition, as peak bagging also implies the asteroseismic identification of the oscillation peak, the use of some model selection criterion is often required. For this purpose, Bayesian statistics offers a valuable choice \cite[e.g. see][]{Sivia06,Trotta08,Benomar09,Gruberbauer09,Kallinger10CoRoT, Handberg11,Corsaro13}.

Nevertheless, Bayesian methods are typically more computationally demanding than standard classical approaches, such as maximum likelihood estimators (MLE) \citep[e.g. see][]{Anderson90,Toutain94}. Since the amount of asteroseismic data to be investigated is very large, adopting more efficient Bayesian techniques can significantly reduce the time required for performing an entire peak-bagging analysis. 

In this paper, we present a new code termed \diamonds, based on the nested sampling Monte Carlo \citep[NSMC,][hereafter SK04]{Skilling04}, a powerful and efficient method for inference analyses of high-dimensional and multi-modal problems that incorporates a robust Bayesian approach \citep[see also][hereafter M06, S07, FH08, F09, FS13, respectively]{Mukherjee06,Shaw07,Feroz08,Feroz09,Feroz13}. We show how the code can be used as a tool for the peak bagging of solar-like oscillations, ensuring a considerable computational speed and efficiency in performing the analysis. 

\section{Bayesian inference}
\label{sec:bayes}
The heart of the \diamonds\,\,code is Bayes' theorem:
\begin{equation}
p\left( \boldsymbol{\theta}  \mid D, \mathcal{M} \right) = \frac{\mathcal{L} \left( \boldsymbol{\theta} \mid D, \model \right) \, \prior}{p \left( D \mid \mathcal{M} \right)} \, ,
\label{eq:bayes}
\end{equation}
where $\dtheta = \left( \theta_1, \theta_2, \dots, \theta_k \right)$ is the parameter vector, that is, the $k$-dimensional vector containing the free parameters that formalize a given model $\model$, and $D$ is the dataset used for the inference. $\mathcal{L} \left( \boldsymbol{\theta} \mid D, \model \right)$ (hereafter, $\likel$ for simplicity) is the likelihood function, which represents the way we sample the data, while $\prior$ is the prior probability density function (PDF) that reflects our knowledge about the model parameters (see Sect.~\ref{sec:likelihood_priors}). The left-hand side of Eq~(\ref{eq:bayes}) is the posterior PDF, which has a key role in the parameter estimation problem as we shall discuss more in detail in Sect.~\ref{sec:parameter_estimation}. 

\subsection{Model comparison}
\label{sec:model_comparison}
The denominator on the right-hand side of Eq~(\ref{eq:bayes}) is instead a normalization factor, generally known as the Bayesian evidence (or marginal likelihood), which is defined as
\begin{equation}
\evid \equiv p \left( D \mid \model \right) = \int_{\Sigma_\mathcal{M}} \likel \prior d\dtheta
\label{eq:evidence}
\end{equation}
with $\Sigma_\mathcal{M}$ the $k$-dimensional parameter space set by the prior PDF. The Bayesian evidence $\evid$ --- namely,  the likelihood distribution averaged over the parameter space set by the priors --- plays no role in the parameter estimation because it does not depend upon the free parameters by definition, but it is nevertheless central for solving model selection problems as we shall show   in Sect.~\ref{sec:application} for the application of the peak bagging analysis. 

To perform the model comparison one considers the so-called odds ratio, that is the ratio of the posterior probabilities of the different models, which is defined as
\begin{equation}
\mathcal{O}_{ij} \equiv \frac{p \, (\model_i \mid D )}{p\,( \model_j \mid D )} = \frac{\evid_i}{\evid_j} \frac{\pi\,(\model_i )}{\pi\,( \model_j )} = \mathcal{B}_{ij} \frac{\pi \,(\model_i )}{\pi\,( \model_j )} \, ,
\label{eq:odds}
\end{equation}
which comprises both the Occam's razor --- consisting in a trade-off between fit quality of the model to the observations and the number of free parameters involved in the inference and represented by the Bayes factor $\mathcal{B}_{ij}$ --- and our prior knowledge about the models investigated, $\pi \left(\model \right)$, which is given as prior odds ratio here. When $\mathcal{O}_{ij} > 1$ the model $\model_i$ is favored over the model $\model_j$, and vice versa when $O_{ij} < 1$.

In most astrophysical applications, the model selection problem is addressed by setting $\pi \left( \model \right) = \mbox{const}$ for all models \citep[e.g. see][]{Corsaro13}, which means that we assign the same chance of being eligible a priori to all models. This assumption neutralizes the effect of the prior odds ratio and reduces Eq.~(\ref{eq:odds}) to $\mathcal{O}_{ij} = \mathcal{B}_{ij}$ only, i.e. the ratio of the Bayesian evidences of the two models. Occasionally, the prior odds ratio may not be negligible and requires further consideration, see also \cite{Scott10}. We also refer to the so-called Jeffreys' scale of strength \citep{Jeffreys61} for comparing the Bayes' factor and conclude on whether a model ought to be preferred over its competitor.

\section{Nested sampling}
\label{sec:nested}
Since Eq.~(\ref{eq:evidence}) is a multi-dimensional integral, as the number of dimensions increases, its evaluation becomes quickly unsolvable both analytically and by numerical approximations. The NSMC algorithm was first developed by SK04, having not only the aim of an efficient evaluation of the Bayesian evidence for any dimensions but also the sampling of the posterior probability distribution (PPD) for parameter estimation as a straightforward by-product.

We follow SK04 by introducing the \textit{prior mass} (or prior volume) $dX = \prior \, d\dtheta$ such that
\begin{equation}
X \left( \constr \right) = \int_{\mathcal{L} \left( \dtheta \right) > \constr} \prior d\dtheta \, ,
\label{eq:prior_mass}
\end{equation}
with $\constr$ being some fixed value of the likelihood distribution. Clearly, $0 \leq X \leq 1$ because $\prior$ is a PDF. Equation (\ref{eq:prior_mass}) is therefore the fraction of volume under the prior PDF that is contained within the hard constraint $\likel > \constr$, hence the higher the constraining value and the smaller the prior mass considered. In other words, for a given $\constr$, we are considering the parameter space delimited by the iso-likelihood contour $\mathcal{L} \left( \dtheta^* \right) = \constr$, which also includes the maximum value $\mathcal{L}_\mathrm{max}$. 

Considering the inverse function $\mathcal{L} (X)$, i.e. $\mathcal{L} (X (\constr)) = \constr$, Eq.~(\ref{eq:evidence}) becomes
\begin{equation}
\evid = \int^1_0 \mathcal{L} (X) dX \, ,
\label{eq:1d_evidence}
\end{equation}
reducing it to a one-dimensional integral. Assuming one has a set of $\nnest$ pairs $\{ \constr_i, X_i \}$, where $X_i = X(\constr_i)$, with $X_{i+1} < X_i$ and $\constr_{i+1} > \constr_i$, Eq.~(\ref{eq:1d_evidence}) can be evaluated as
\begin{equation}
\evid = \sum^{\nnest - 1}_{i=0} \constr_i w_i \,
\label{eq:sum_evidence}
\end{equation}
with either $w_i = \left(X_i - X_{i+1}\right)$ for a simple rectangular rule or $w_i = \frac{1}{2} \left( X_{i-1} - X_{i+1} \right)$ for a trapezoidal rule (see SK04 for some simple examples). For the sake of clarity, due to numerical reasons related to the implementation of the equations above, quantities such as evidence, prior mass, likelihood and prior probability density values are more conveniently considered in a logarithmic scale.

\subsection{Drawing from the prior}
What happens in practice is that we set a new likelihood constraint that is higher than the previous one at each iteration, and we peel off a thin shell of prior mass defined by the new iso-likelihood contour. This allows us to collect the evidence from that thin shell and cumulate it to the final value given by Eq.~(\ref{eq:sum_evidence}).
Still following SK04, the right-hand side of Eq.~(\ref{eq:sum_evidence}) is computed as follows: 
\begin{enumerate}
\item $\nlive$ ‘live’ points are drawn uniformly from the original prior PDF $\prior$ by setting the initial prior mass $X_0 = 1$.
\item At the first iteration of the nested sampling, $i = 0$, the live point having the lowest likelihood is removed from the sample, its coordinates and its likelihood stored, the latter as the first likelihood constrain $\constr_0$.
\item The missing point of the live sample is then replaced by a new one uniformly drawn from the prior PDF, having likelihood higher than the first constraint $\constr_0$.
\item The prior mass is reduced by a factor $\exp \left(-1 / \nlive \right)$ according to the standard rule defined by SK04
\begin{equation}
X_{i} = \exp \left(-i / \nlive \right) \, .
\label{eq:standard_reduction}
\end{equation}
\end{enumerate}
The entire process from point 2 to point 4 is repeated in the subsequent iterations until some stopping criterion is met. 
While the computation of the evidence is simple, the NSMC has the challenging drawback of requiring drawings from the prior with the hard constraint of the likelihood. This process becomes much more computationally demanding in the case of multi-modal PPDs, high-dimensional problems, and pronounced curving degeneracies. For these reasons, the drawing problem has been widely investigated and different solutions have been proposed, such as single ellipsoidal sampling (M06); clustered and simultaneous ellipsoidal sampling and improved versions with either k-means, X-means (S07, FH08), or an expectation-maximization algorithm (\textsc{MultiNest} code, by F09); Metropolis nested sampling \citep[][FH08]{Sivia06}, artificial neural networks \citep{Graff12}, and more recently Galilean Monte Carlo (GMC, by FS13). 

In Sect.~\ref{sec:ellipsoidal}, we describe the details of another version of the simultaneous ellipsoidal sampling that adopts X-means, which was adopted in this work.

\section{The DIAMONDS code}
\label{sec:code}
The \diamonds\,\,(high-DImensional And multi-MOdal NesteD Sampling) code presented in this work was developed in {\ttfamily C++11} and structured in classes to be as much flexible and configurable as possible. The working scheme from a main function is as follows:
\begin{enumerate}
\item reading an input dataset;
\item setting up model, likelihood, and priors to be used in the Bayesian inference;
\item setting up a drawing algorithm;
\item configuring and starting the nested sampling;
\item computation and printing of the results.
\end{enumerate}

The code can be used for any application involving Bayesian parameter estimation and/or model selection in general. Users can supply new models, likelihood functions, and prior PDFs whenever needed by taking advantage of {\ttfamily C++} class polymorphism and inheritance. Any new model, likelihood, and prior PDFs can be defined and implemented upon a basic template. 

In addition, it is possible to feed the basic nested sampler with drawing methods based on different clustering algorithms (see section below). 

\subsection{Simultaneous ellipsoidal sampling}
\label{sec:ellipsoidal}
Simultaneous ellipsoidal sampling (SES) is a drawing algorithm based on a preliminary clustering of the set of live points at a given iteration of the nested sampling. The clustering is obtained in our case by using X-means \citep{Pelleg2000} with a number of clusters $\nclust$ ranging from a minimum to a maximum value allowed. A good choice for most applications is given by $1 \leq \nclust \leq 6$. The SES was developed by S07 and FH08, based on the first idea by M06, for gaining efficiency when dealing with multi-modal posteriors and pronounced curving degeneracies. 

The SES algorithm proceeds as follows: once a number of clusters has been identified --- i.e. the set of live points has been partitioned into subsets --- ellipsoidal bounds for each of the clusters are constructed, and a new point is drawn from the inside of one of the ellipsoids exploiting a fast and exact method (see S07 for more details). Ellipsoids are intended to approximate the iso-likelihood contours, thus reducing the effective prior volume where the drawing has to take place. This considerably improves the speed of the NSMC because sampling from uninteresting regions of the PPD is in general avoided. This algorithm is in principle repeated at every iteration of the nested sampling (see also FH08).

Below, we present the two main changes applied in the SES algorithm from the version described in Method 1 of FH08, the closest in spirit to the one adopted here. These changes were done to improve speed and efficiency of the drawing process.

The first difference is that we introduced two additional configuring parameters: (i) the number of nested iterations before executing the first clustering of the live points, $\minit$ and (ii) the number of nested iterations with the same clustering, $\msame$, which is the number of iterations that do not involve any clustering of the live points. It is not required and conversely much more computationally expensive to perform the clustering at each iteration of the nested sampling. Point (i) addresses the problem of having X-means identifying one cluster during the early stages of the nesting process, where we do not yet expect any clustering of the live points to happen, and (ii) allows us to speed up the computation by a factor $\msame$, which is able to significantly reduce the run time of the NSMC, since X-means represents the bottleneck of this approach.

While the two additional parameters $\minit$ and $\msame$ complicate the configuration of the code to some extent, their tuning is not tricky because they do not critically affect the efficiency of the sampling. Intuitively, the more live points are being used, the more iterations can be treated with the same clustering.
In general, adopting $\minit$$\,\sim\,$$\nlive$ and $\msame$$\,\sim\,$$0.02-0.05 \minit$ have provided a valuable choice for all the demos presented in Sect.~\ref{sec:demos} and for the peak bagging analysis discussed in Sect.~\ref{sec:application}. Ellipsoids are recomputed at each iteration because this process is not significantly influencing the speed of the NSMC.

The second relevant change applied to the original SES algorithm consists in another additional input parameter, $\matt$, which represents the maximum number of attempts in drawing from the prior with the hard constraint of the likelihood, and it is typically set to $10^4$. On one hand, this parameter allows for a safer control of the drawing process, which can therefore be stopped if the number of attempts exceeds the given limit (meaning that the drawing is not efficient anymore). This avoids situations in which the sampling gets stuck in a flat region of the PPD, resulting in a prohibiting large number of drawing attempts. On the other hand, increasing $\matt$ up to values, such as $10^5$, can sometimes be useful to force more nested iterations and achieve more precision on the final value of the evidence. However, the larger the $\matt$, the slower the computation becomes toward the final iterations of the NSMC. It is important to note that the parameter $\matt$ directly impacts the number of total likelihood computations done during the process. This is because every attempt done involves the drawing of a new point according to the prior PDF and the subsequent assessing of the likelihood constraint. As a result, the final number of sampling points only accounts for those attempts that were successful (hence useful according to the NSMC working criterion) and will always be smaller than the total number of likelihood evaluations that were practically done by the code.

Lastly, the SES implemented in \diamonds\,\,incorporates the dynamical enlargement of the ellipsoids as introduced by FH08, that is, for a given $i$-th iteration and the $k$-th ellipsoid
\begin{equation}
f_{i,k} = f_0 X^{\alpha}_i \sqrt{\frac{\nlive}{n_k}} \, ,
\label{eq:dynamic_enlargement}
\end{equation}
where $n_k$ is the number of live points falling in the $k$-th ellipsoid, while $f_0 \geq 0$ and $0 \leq \alpha \leq 1$ are the two additional configuring parameters, which represent the initial enlargement fraction and the shrinking rate of the ellipsoids, respectively. 

\subsection{Likelihood and prior distributions}
\label{sec:likelihood_priors}
The code includes different likelihood functions, which are all implemented as a log-likelihood, $\lnlikel \equiv \ln \likel$. The application exposed in this paper includes the \emph{exponential likelihood}, which is required for describing Fourier power spectra as introduced by  \cite{Duvall86,Anderson90} for data distributed according to a $\chi^2$ with two degrees of freedom. The log-likelihood reads
\begin{equation}
\Lambda \left( \dtheta \right) = - \sum^{\ndata}_{i=1} \left[ \ln E_i \left( \dtheta \right) + \frac{O_i}{E_i \left( \dtheta \right)} \right] \, ,
\label{eq:exponential_likelihood}
\end{equation}
where the functional form for $E_i \left( \dtheta \right)$ is described in Sects.~\ref{sec:bkg} and \ref{sec:pb}.

\begin{figure*}
   \centering
   \includegraphics[width=6.0cm]{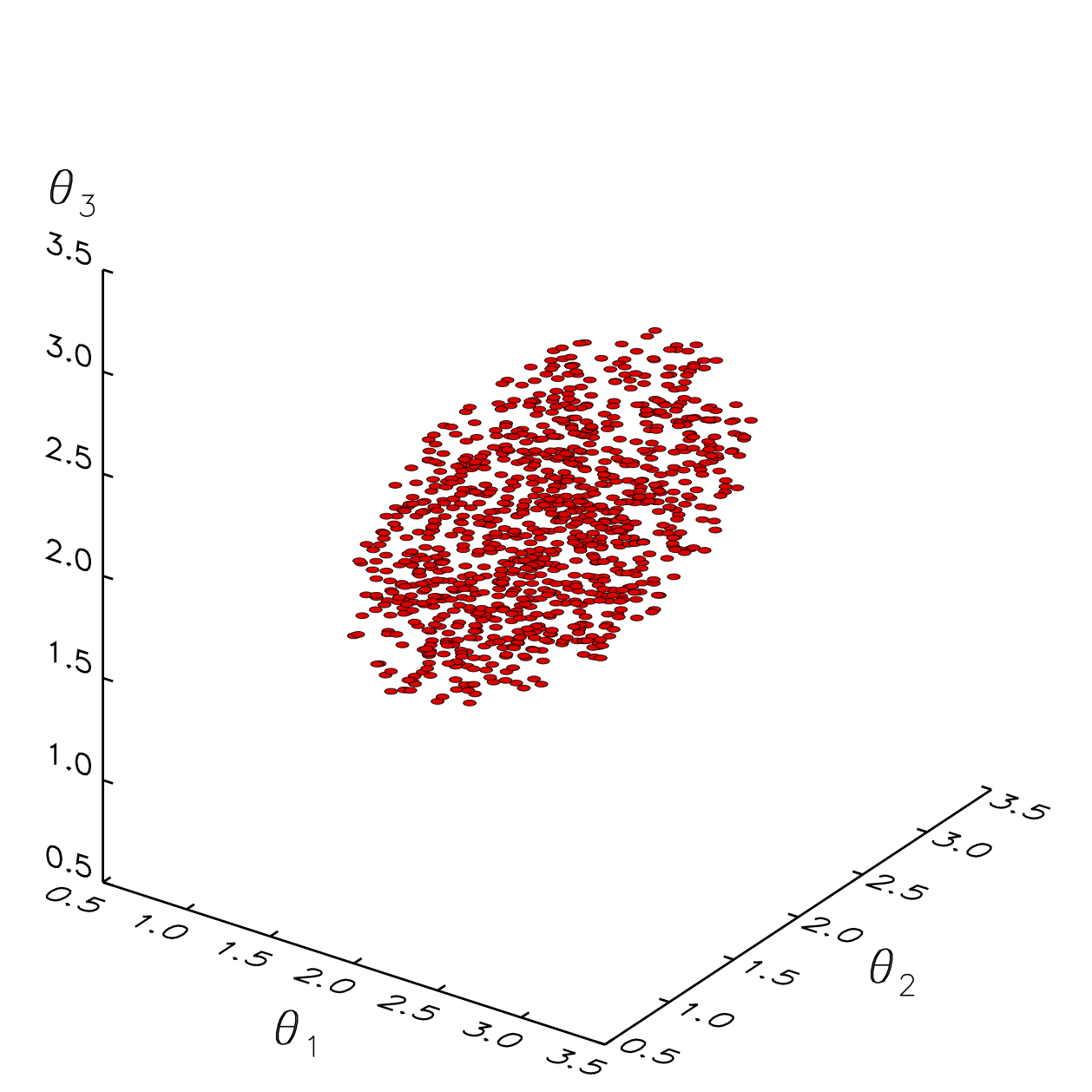}\includegraphics[width=6.0cm]{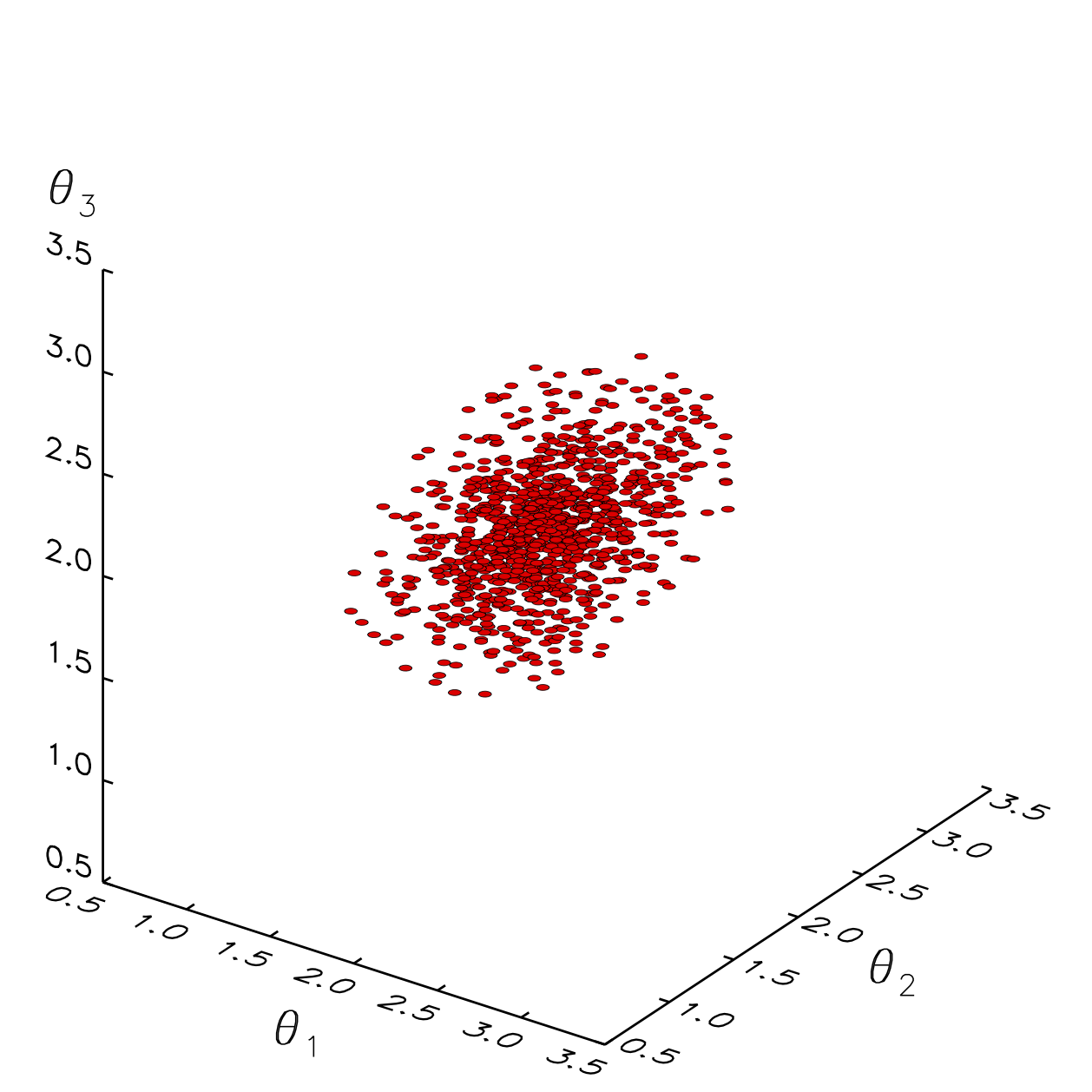}\includegraphics[width=6.0cm]{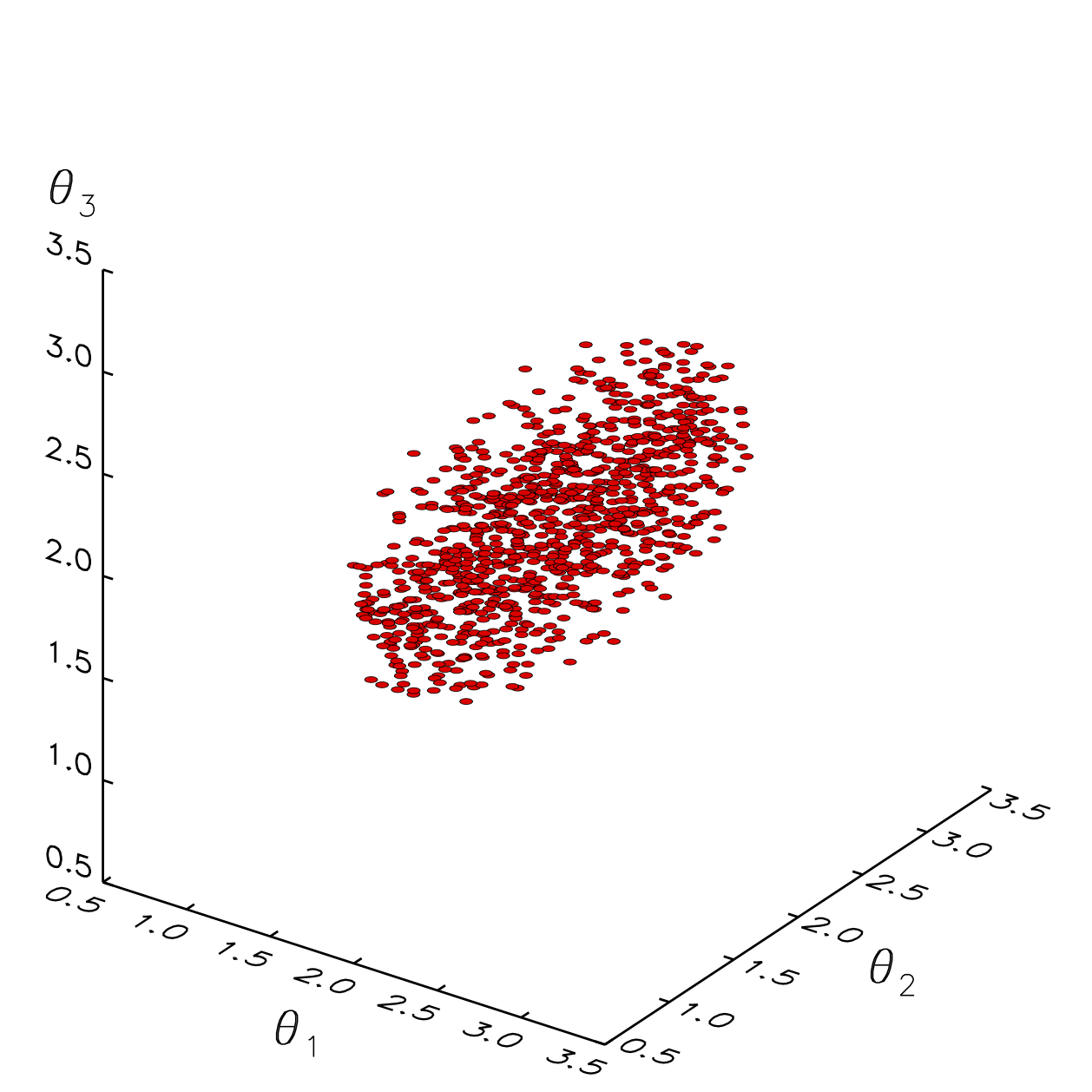}
      \caption{Examples of 1000 points (red dots) drawn according to different prior distributions from a three-dimensional ellipsoid centered at coordinate $(2.0,2.0,2.0)$ in the parameter space defined by $\theta_1, \theta_2, \theta_3 \in [0.5,3.5]$. \textit{Left panel}: case of a uniform prior along each coordinate, $\mathcal{U}_1 (0.5,3.5) \cdot \mathcal{U}_2 (0.5,3.5) \cdot \mathcal{U}_3 (0.5,3.5)$. The points uniformly fill the entire volume of the ellipsoid. \textit{Middle panel}: case of a normal prior along each coordinate, $\mathcal{N}_1 (2.0,0.3) \cdot \mathcal{N}_2 (2.0,0.6) \cdot \mathcal{N}_3 (2.0,0.3)$. The points are more concentrated toward the center of the ellipsoid while having a doubled spread along the direction of $\theta_2$. \textit{Right panel}: case of normal priors along the two coordinates $\theta_1, \theta_3$ and a uniform prior along the coordinate $\theta_2$, $\mathcal{N}_1 (2,0.4) \cdot \mathcal{U}_2 (0.5,3.5) \cdot \mathcal{N}_3 (2,0.4)$. The points are now more concentrated along the coordinate $\theta_2$ since the spread only occurs over the orthogonal directions.}
    \label{fig:prior_demo}
\end{figure*}

\begin{figure*}
   \centering
   \includegraphics[width=6.0cm]{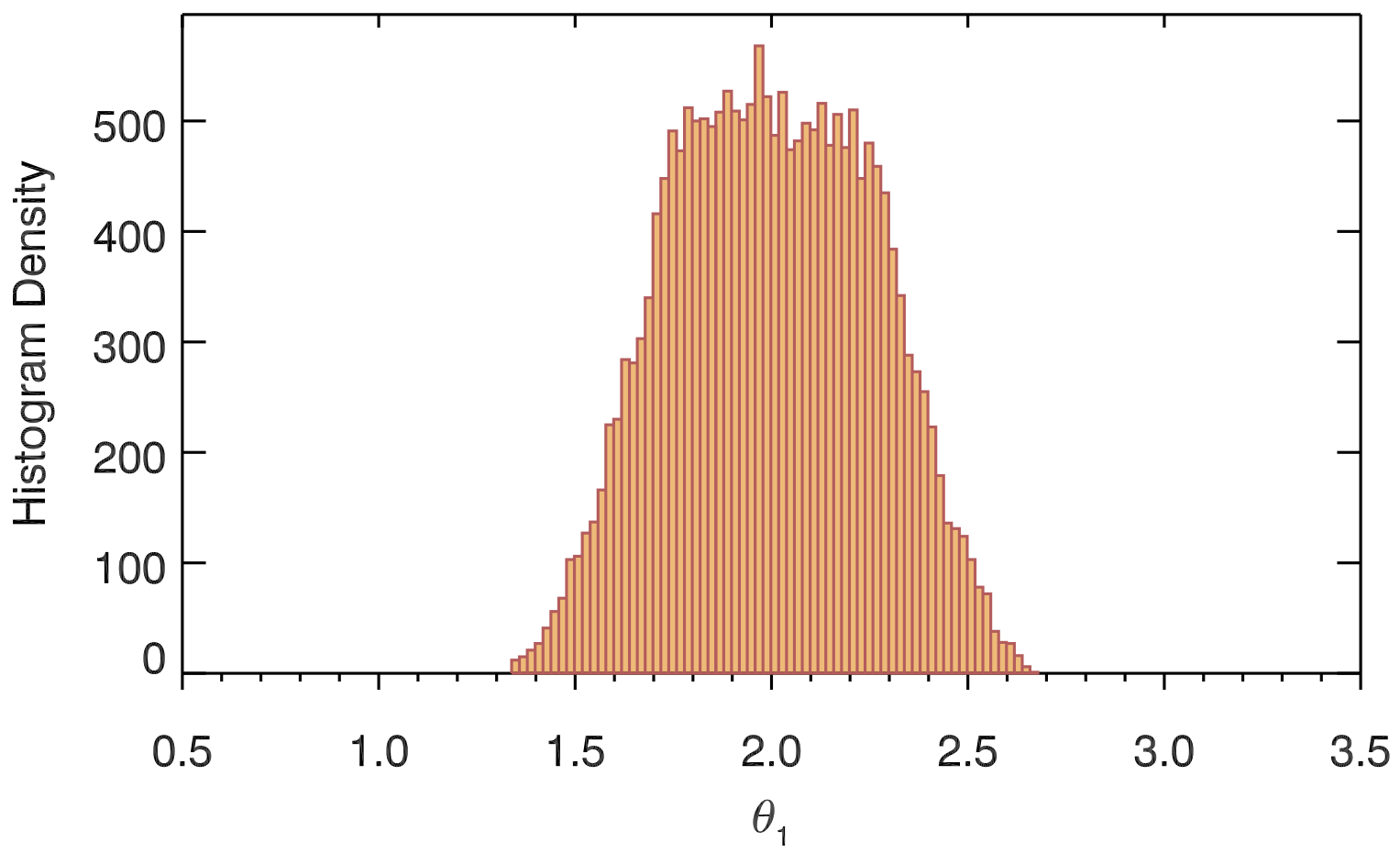}
   \includegraphics[width=6.0cm]{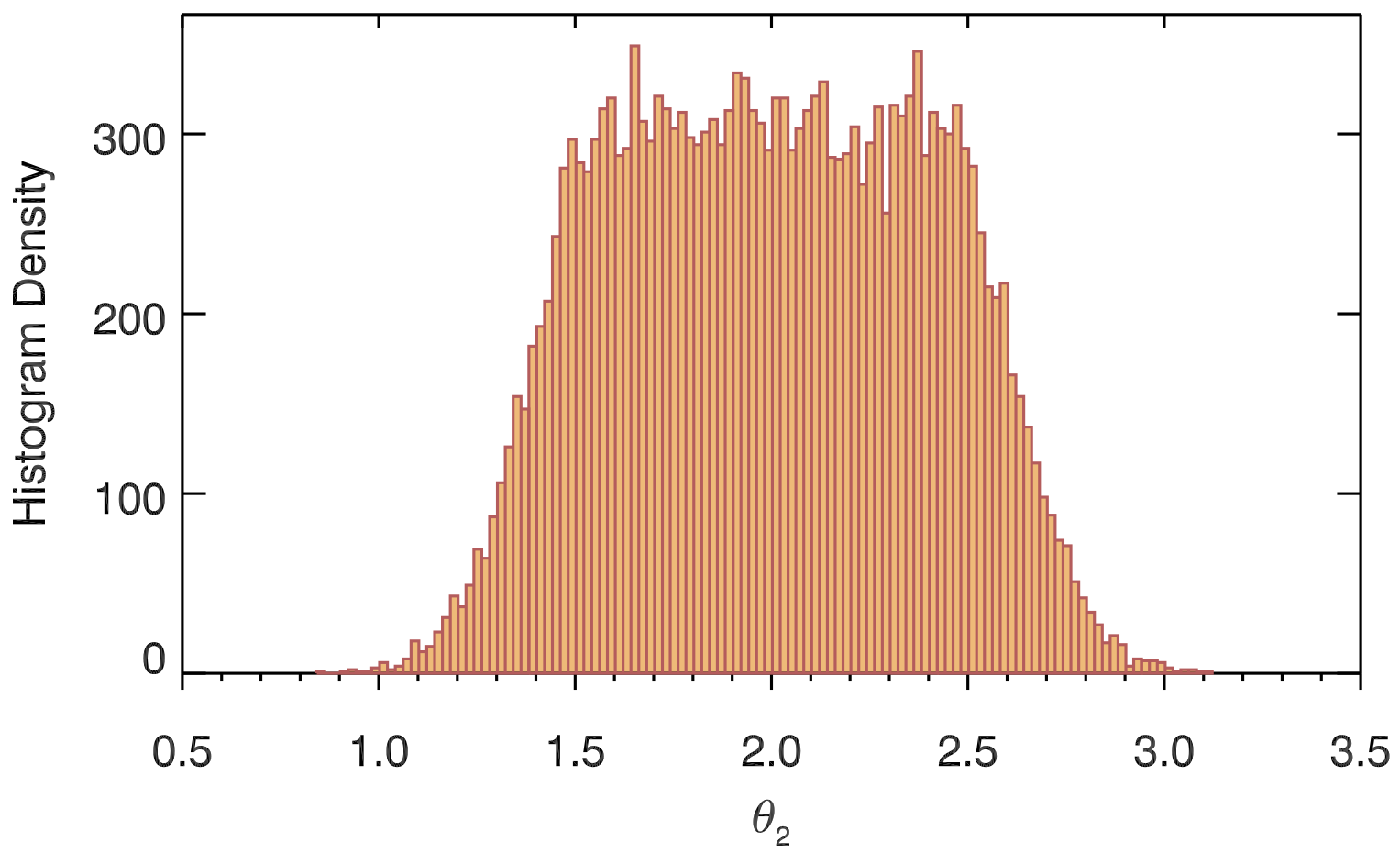}
   \includegraphics[width=6.0cm]{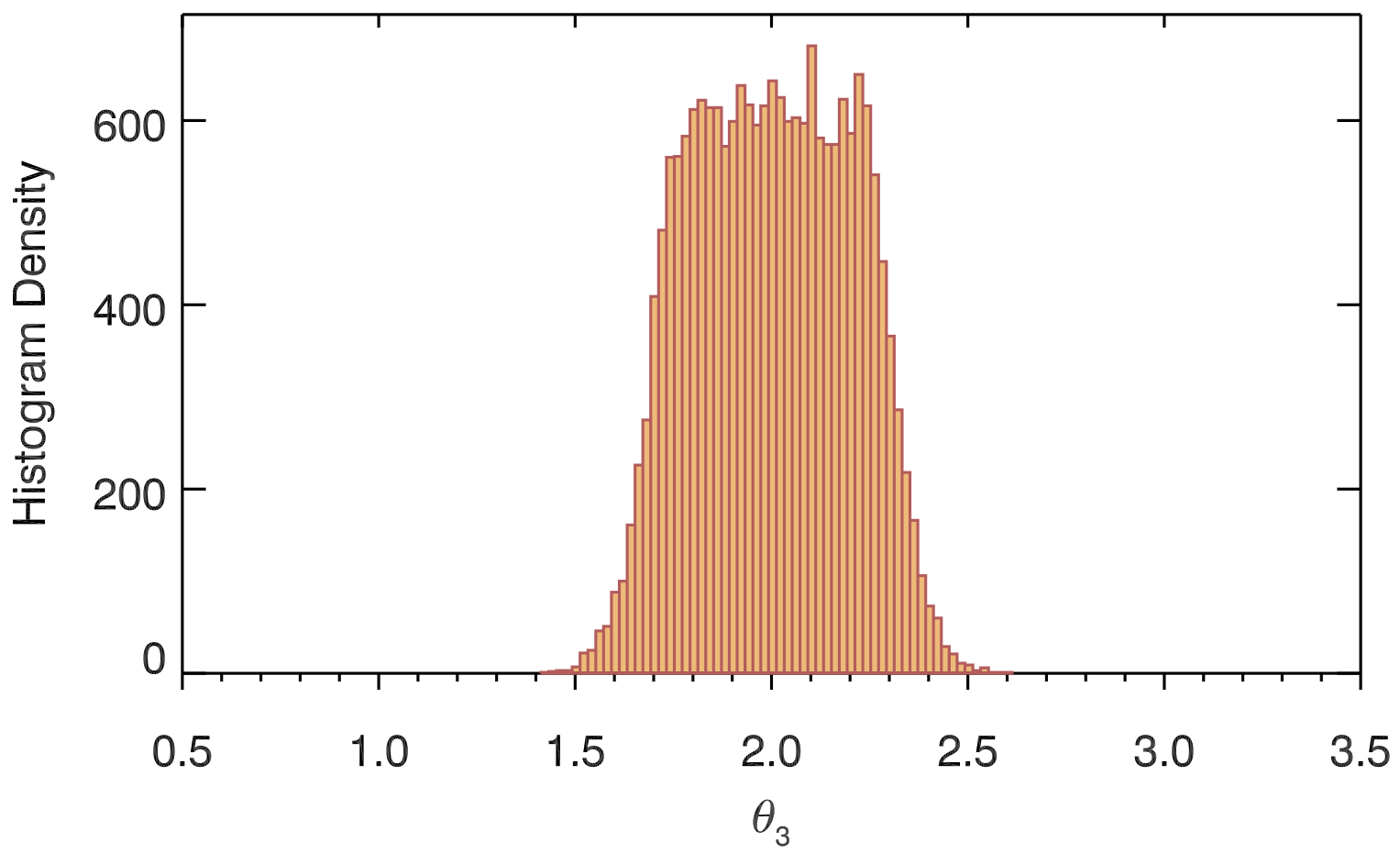}
      \caption{Examples of 20000 points drawn from the 3D ellipsoid used in Fig.~\ref{fig:prior_demo} but now according to the super-Gaussian priors $\mathcal{S}_1 (2.0,0.5,0.2) \cdot \mathcal{S}_2 (2.0,1.0,0.2) \cdot \mathcal{S}_3 (2.0,0.5,0.1)$. Panels from left to right show the histogram densities of the number of drawn points per dimension. By definition, the histogram corresponding to the coordinate $\theta_2$ has a plateau with twice the width of the plateau of the other two coordinates, while we see the same spread in the tails of the histograms for $\theta_1$ and $\theta_2$ and a smaller one for that of $\theta_3$. All the distributions are also centered in the given input center position.}
    \label{fig:prior_demo2}
\end{figure*}

The SES algorithm does not put any particular restriction on the prior PDF that can be used when drawing a new point from an ellipsoid. Prior PDFs, as introduced in Eq.~(\ref{eq:bayes}), allow us to draw a point more frequently from those regions inside the ellipsoid having higher prior probability density. This clearly encompasses our knowledge about the inferred parameters, and it is one of the key points of the Bayesian approach. 

Differently from the implementation adopted by F09 in \textsc{MultiNest}, when using \diamonds\,\,prior distributions can be defined by the user by means of a separate module that implements a general template for any proper, or normalizable, prior. 

The present code package comes with three different prior PDFs with each of them requiring input hyper parameters --- i.e. the parameters defining the  shape of the prior distribution --- for their set-up. In the following, we briefly introduce them as defined for a single free parameter $\theta_j$ (hence in one dimension of the parameter space): 
\begin{itemize}
\item the uniform prior $\uniform$, where $h_l$ and $h_u$ are the hyper parameters defining lower and upper bounds, respectively, for the free parameter for which the prior is defined.
\item the normal prior $\normal$ with $h_\mu$ and $h_\sigma$ being the hyper parameters mean and standard deviation of the normal distribution for the free parameter considered, respectively.
\item the super-Gaussian prior $\supergauss$, consisting a plateau (flat prior) with symmetric Gaussian tails, as defined by the three hyper parameters $h_c$ as the center of the plateau region, $h_w$ the width of the plateau, and $h_\sigma$ the standard deviation of the Gaussian tails. 
\end{itemize}

Each type of prior PDF can also be defined for set of free parameters, requiring an input vector of hyper parameters in which each element corresponds to one different dimension. The overall $k$-dimensional prior PDF is simply given as the product of the $k$ prior PDFs defined for each coordinate.

Three examples for demonstrating the drawing process from a single ellipsoid in a three-dimensional parameter space according to different combinations of the prior PDFs $\uniform$ and $\normal$ are shown in Fig.~\ref{fig:prior_demo} with 1000 points drawn in each demo. In the first plot, the drawn points are uniformly distributed within the entire volume of the ellipsoid because a uniform prior was adopted for each coordinate. In the second plot, the points are concentrated around the center of the ellipsoid, occurring in the position $(2,2,2)$, and are spread over the three directions since a normal prior was used for each coordinate. In the third plot,  the samples are more concentrated along the direction of $\theta_2$ because of a uniform prior, while they are spread over the other two directions because two normal priors were used.

For demonstrating the super-Gaussian prior PDF $\supergauss$ instead, we show the histograms of the cumulated counts in each of the three directions, as seen in Fig.~\ref{fig:prior_demo2}. For this demo, we used the same ellipsoid of Fig.~\ref{fig:prior_demo} but drew 20000 points from it to provide a more clear result in the histogram density. More details are mentioned in the figure caption.

The drawing from the ellipsoids is by default uniform, hence uniform priors ensure the most efficient drawing process. When using normal  and/or super-Gaussian priors instead, it is recommended to put reasonably large standard deviations if one is not appreciably confident about the possible outcome of the free parameters involved in the inference. Moreover, drawing a new point using super-Gaussian priors is more computationally expensive than with normal priors. Especially in higher dimensions, it is often the slowest drawing among the three prior types considered.

\subsection{Stopping criterion and total evidence}
\label{sec:stop_criterion}
For the stopping criterion implemented in \diamonds, we considered the so-called \emph{mean live evidence} \citep[][hereafter K11]{Keeton11}, defined at a given $i$-th nested iteration as
\begin{equation}
\evid^\mathrm{live}_i = \bar{\mathcal{L}}^{(i)} \left( \frac{\nlive}{\nlive +1} \right)^i \, ,
\label{eq:meanlive_evidence}
\end{equation} 
where the product of the average likelihood estimated from the existing set of live points, $\bar{\mathcal{L}}^{(i)}$, by the remaining prior mass is expressed here as a simple power law of the number of live points because it is averaged over all possible realizations of prior mass distribution (see K11 for more details). 

Once the nested sampling is terminated, we compute the total evidence as $\evid^\mathrm{tot} = \evid + \evid^\mathrm{live}$, where $\evid^\mathrm{live}$ is the remaining mean live evidence at the last iteration. This correction ensures that we have a more accurate estimate of the real evidence even in the case the algorithm is stopped prematurely. For achieving the same level of accuracy on the final evidence, fewer iterations might be required than if $\evid$ only is considered, thus representing an additional advantage for the computation.

The final uncertainty on $\ln \evid^\mathrm{tot}$ can still be taken as the classical $\sigma_{\ln \evid} = \sqrt{H/\nlive}$, which is  suggested by SK04, where $H$ is the final information gain \citep[see][for more details]{Sivia06} since the difference with respect to the total statistical uncertainty derived by K11 is negligible. In the remainder part of the paper, we shall refer to $\evid^\mathrm{tot}$ only and adopt the symbol $\evid$ for simplicity.

At this stage, it is important to introduce a change related to the stopping criterion of the NSMC implemented in \diamonds\,\,with respect to other existing codes using NSMC. As shown in the statistical work by K11, the evolving ratio $\delta_i \equiv \evid^\mathrm{live}_i / \evid_i$ between the mean live evidence and the cumulated evidence at the $i$-th iteration can be used as a criterion for terminating the NSMC. A ratio $\delta_\mathrm{final} \equiv \evid^\mathrm{live} / \evid < 1$ is normally enough for obtaining both an accurate estimate of the total evidence and a good sampling of the PPD. This condition turns out to correspond reasonably well to that provided by SK04 by using the information theory; the latter gives an optimal number of iterations $N_\mathrm{opt} = H \nlive + \sqrt{k} \nlive$, where $k$ is once again the number of dimensions in the problem and $H$ the final information gain. The optimal number of iterations suggested by SK04 is also computed at the end of the process so that it can be used as a reference for the total number of iterations defined by the new stopping condition. 

When a dense sampling of the modes in the PPD ought to be preferred, especially for multi-modal distributions, the stopping threshold can be lowered to a value $\delta_\mathrm{final} < 0.1$. Conversely, the threshold can also be set to higher values ($> 10$) in case we have no much knowledge about the inferred parameters and, therefore, intend to test the validity of the prior boundaries. For the application of \diamonds\,\,presented in Sect.~\ref{sec:application}, we fix this parameter to $\delta_\mathrm{final} = 0.01$.

\begin{figure*}[ht]
   \centering
   \includegraphics[width=4.5cm]{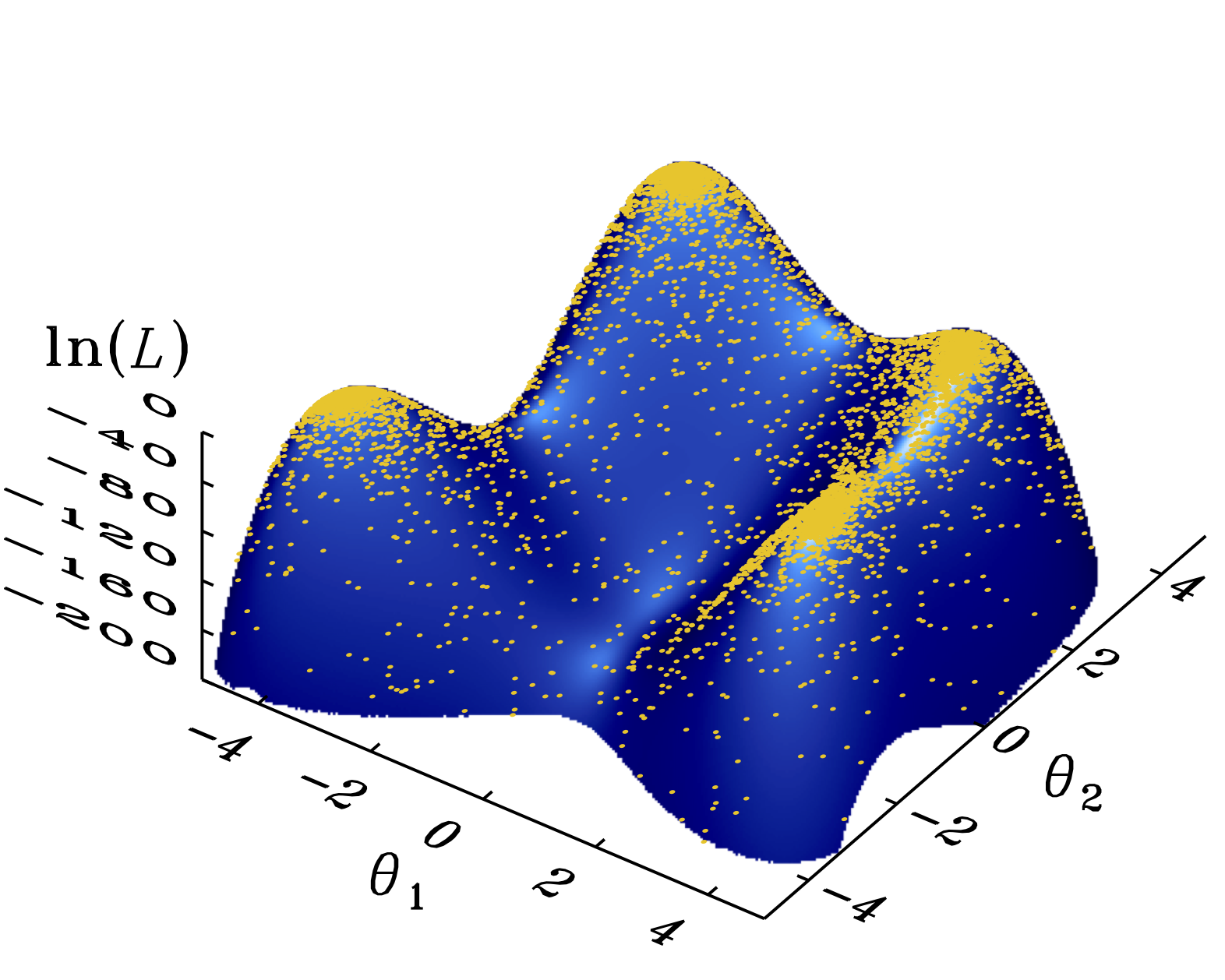}\includegraphics[width=4.5cm]{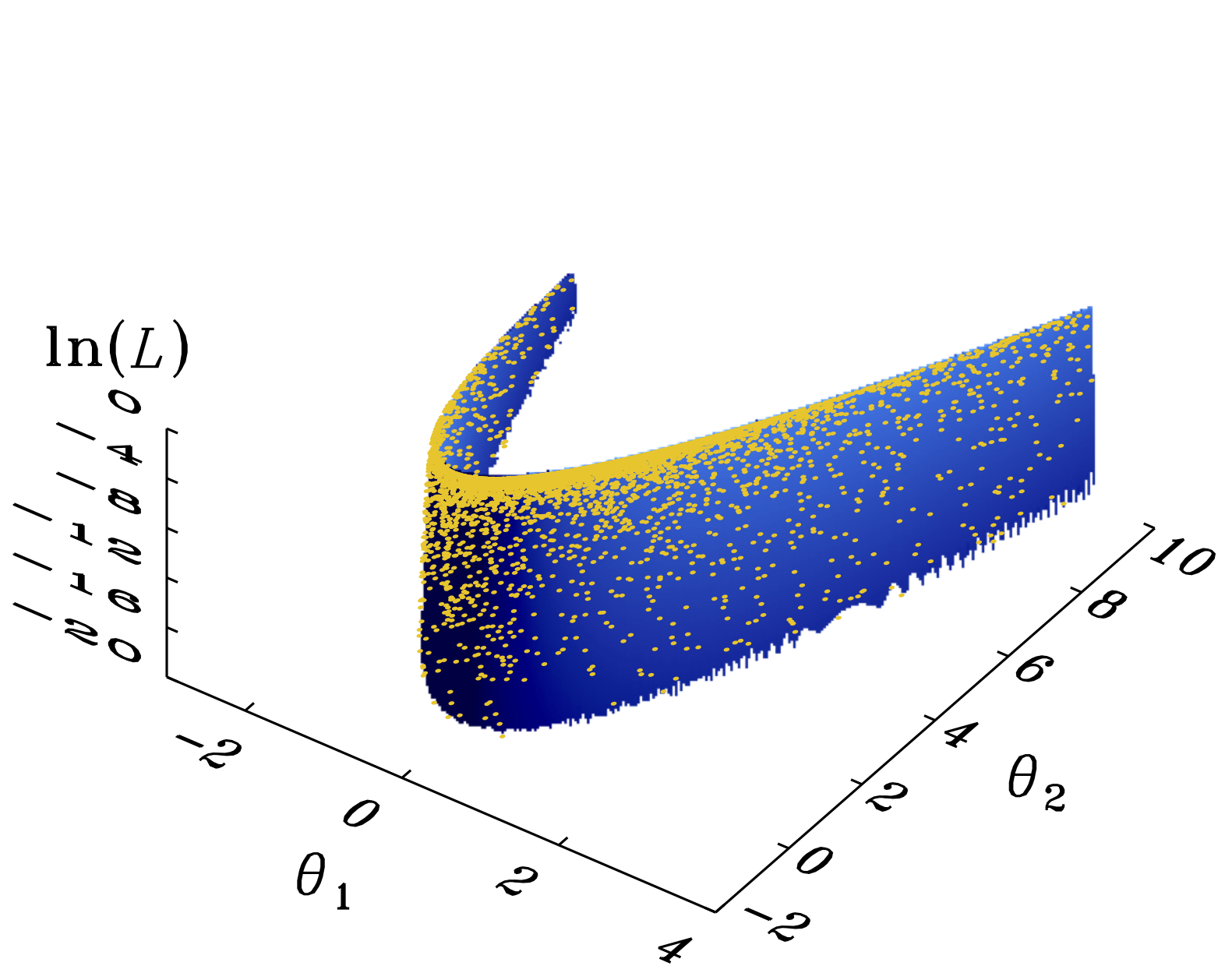}\includegraphics[width=4.5cm]{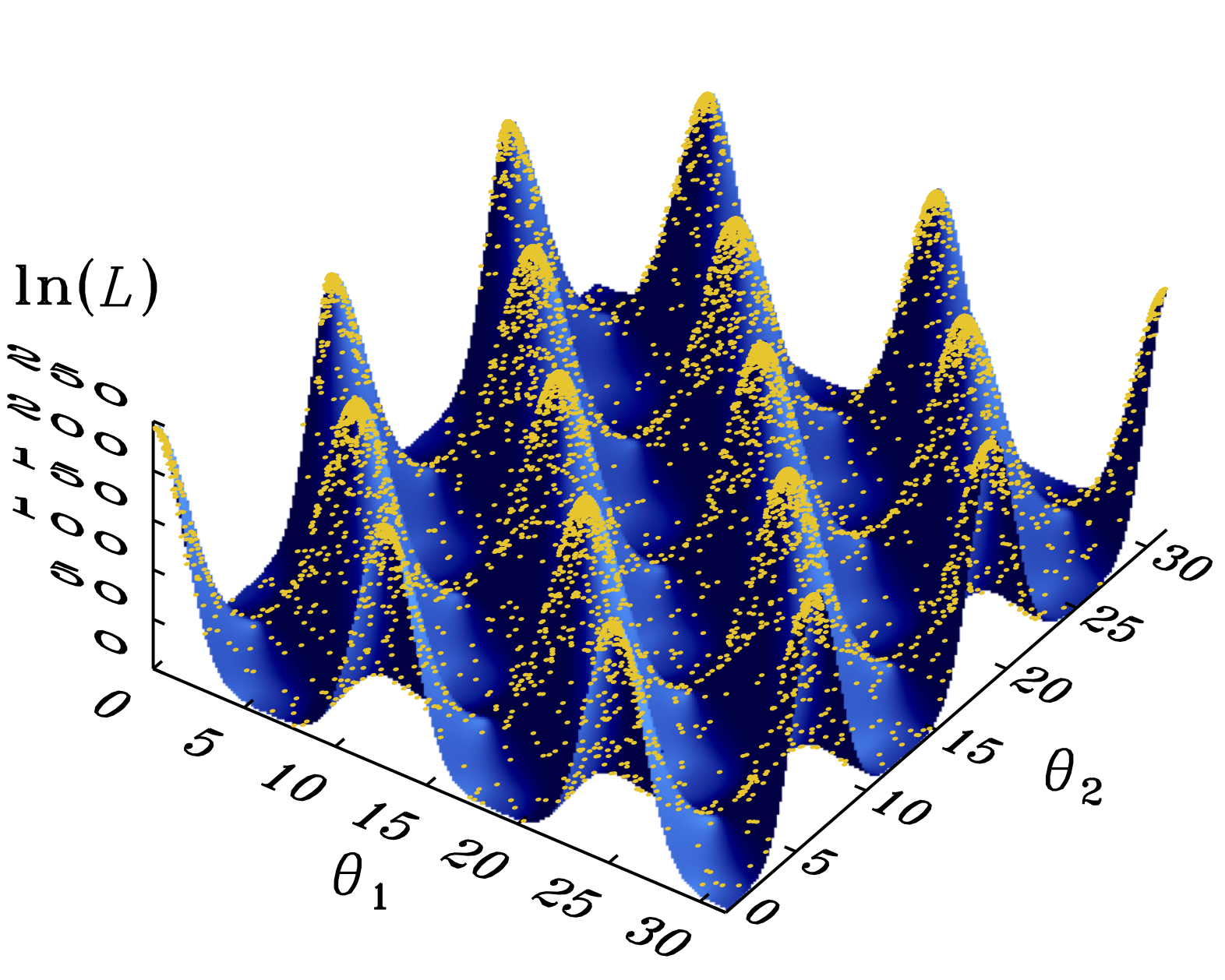}\includegraphics[width=4.5cm]{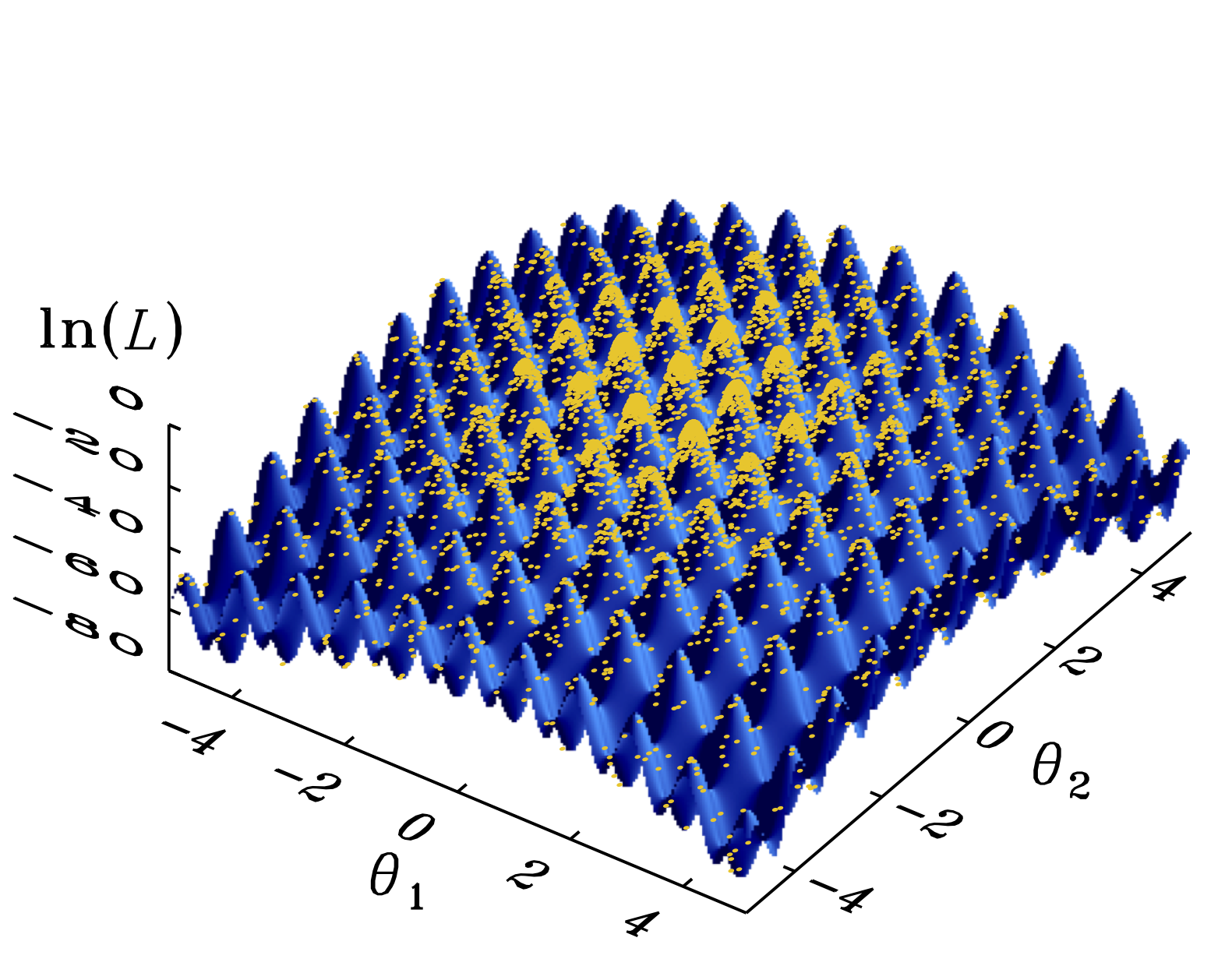}
      \includegraphics[width=4.5cm]{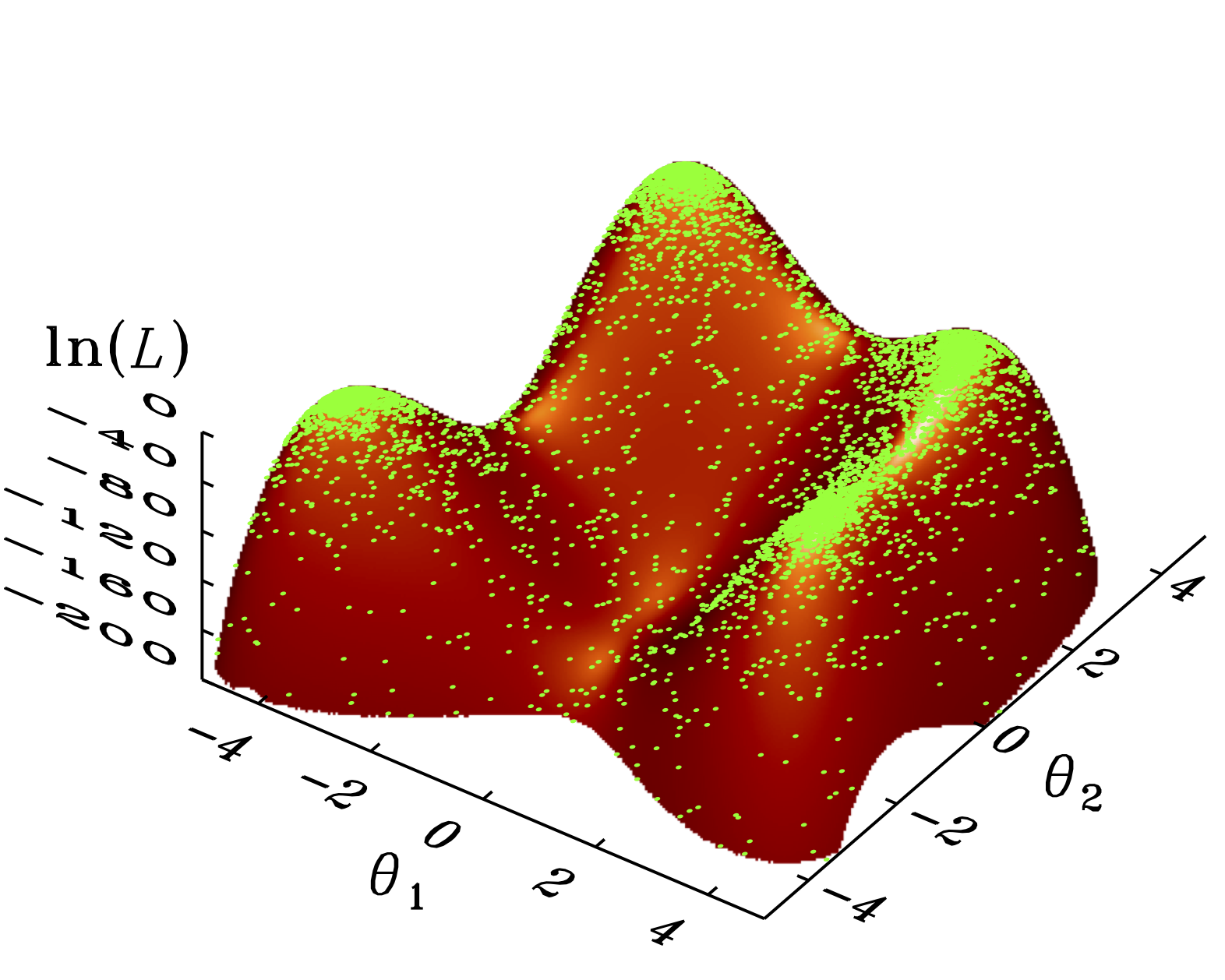}\includegraphics[width=4.5cm]{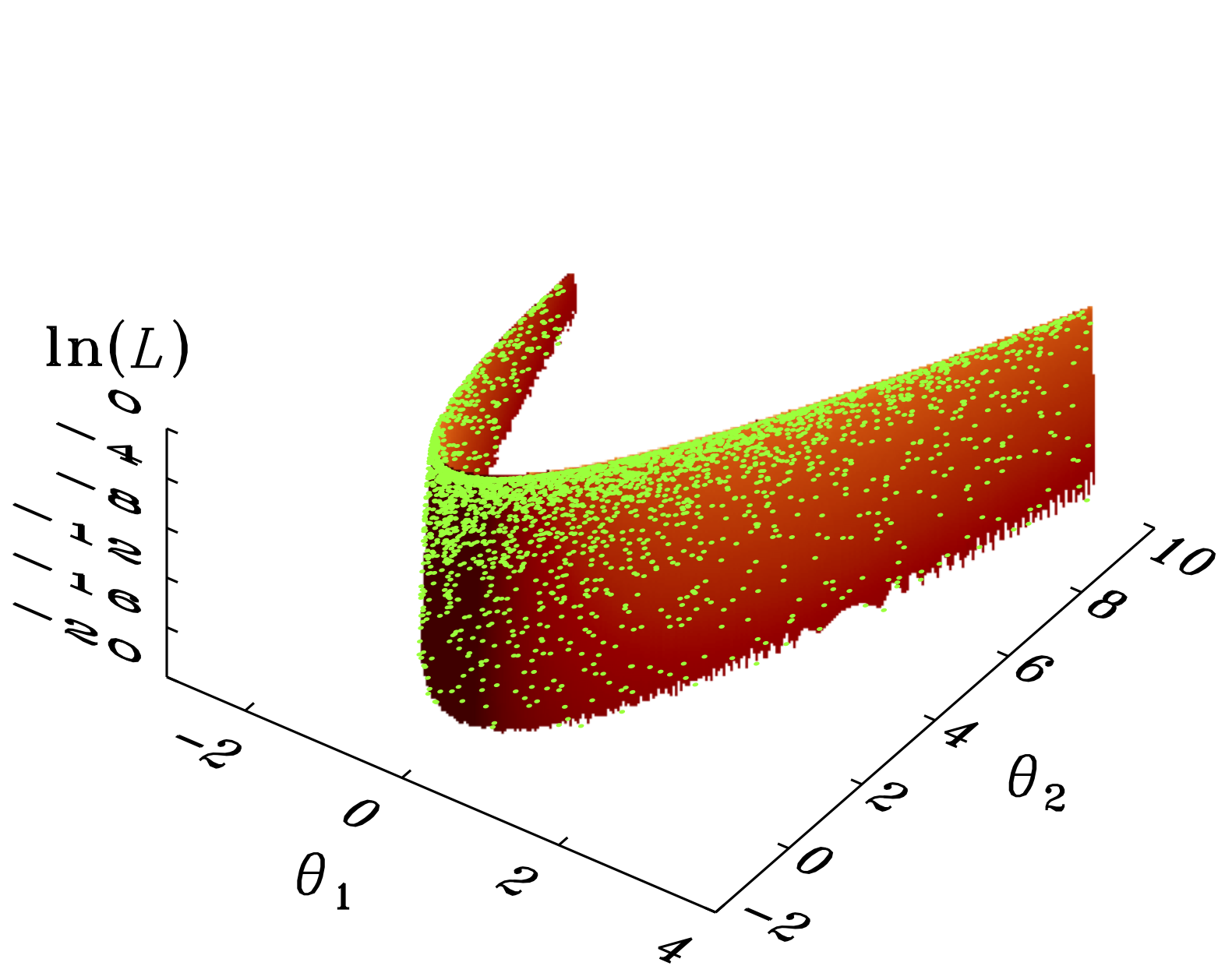}\includegraphics[width=4.5cm]{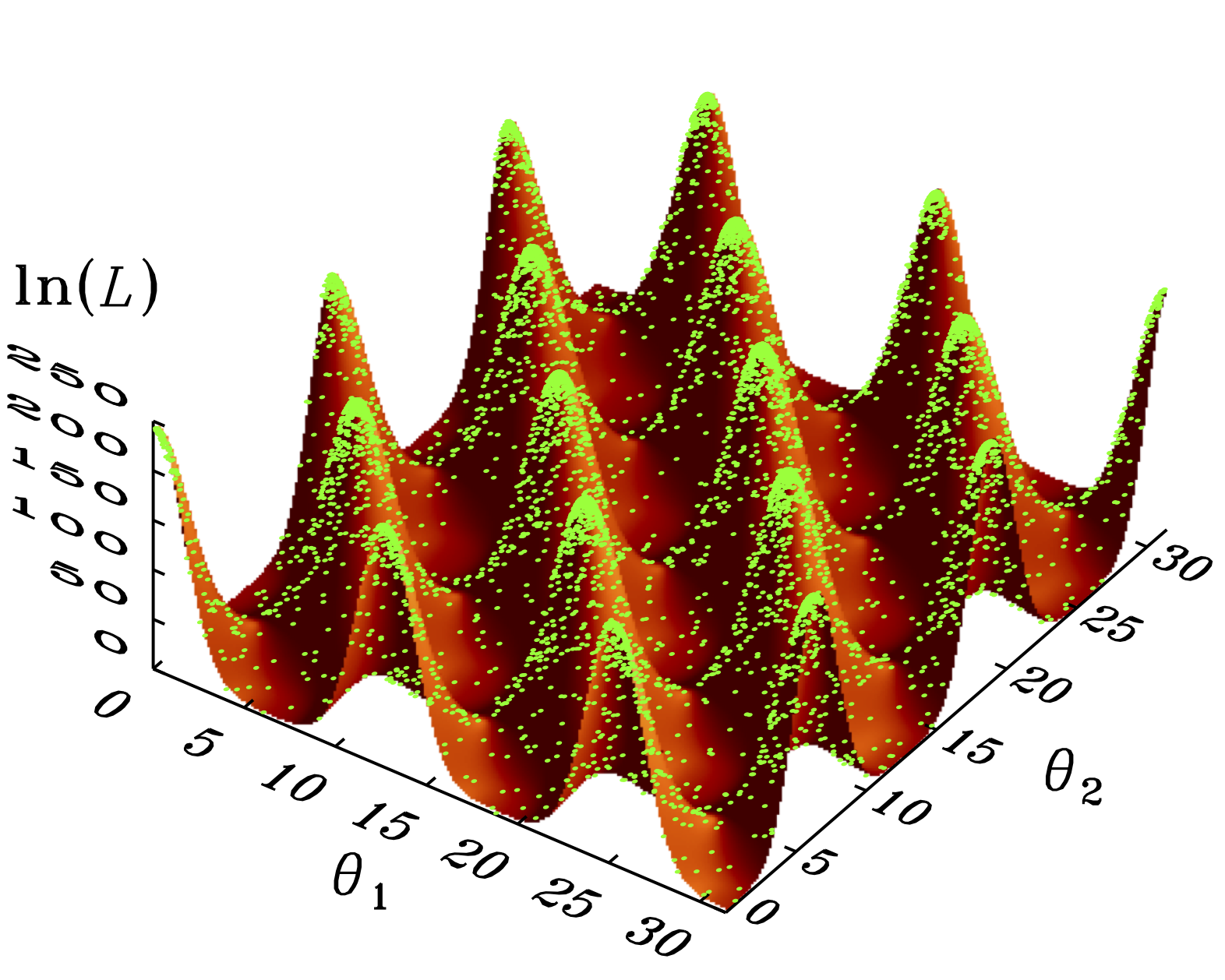}\includegraphics[width=4.5cm]{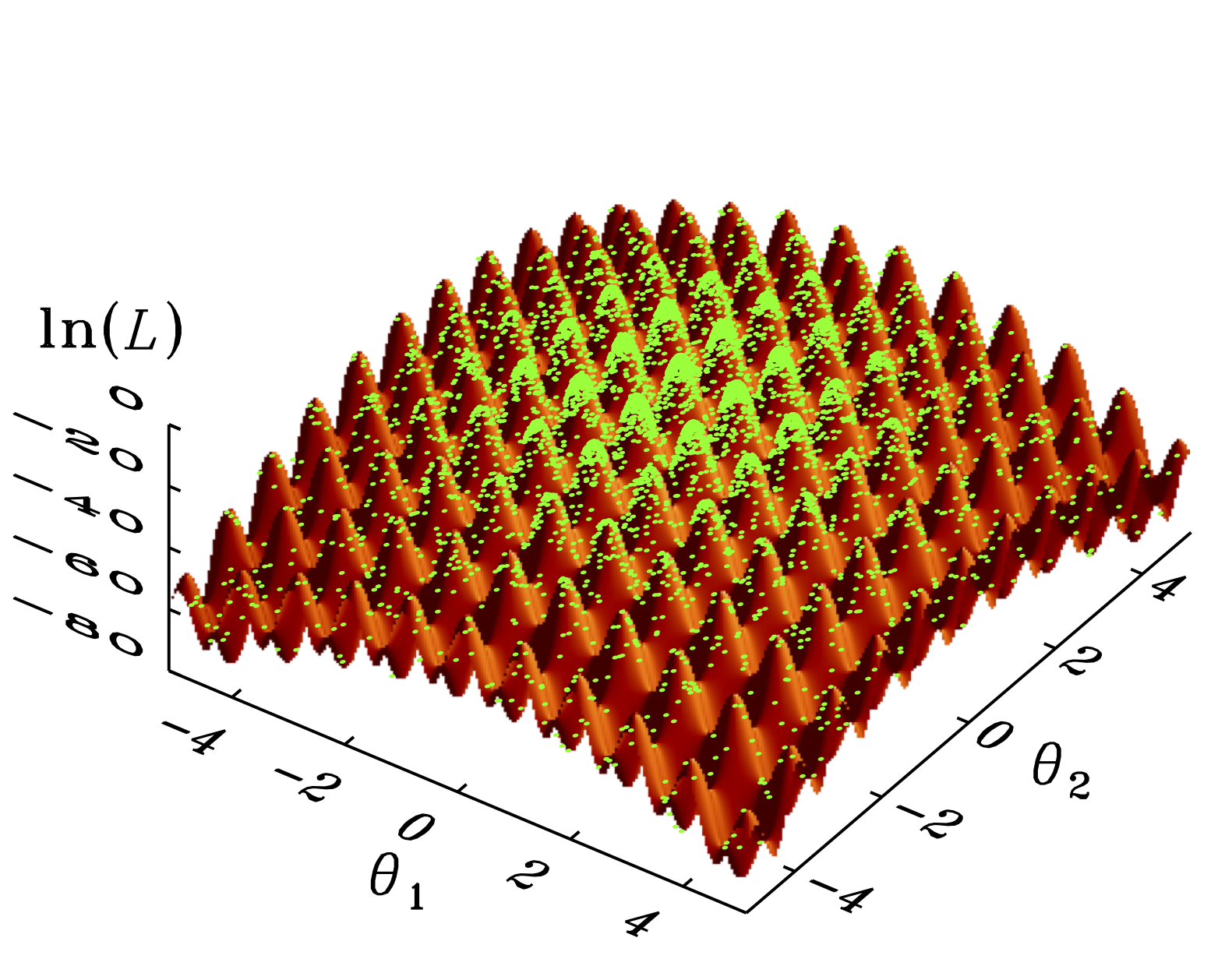}
      \caption{Shaded surfaces show Himmelblau's function in the range $\theta_1, \theta_2 \in [-5,5]$ (left), Rosenbrock's function in the range $\theta_1 \in [-3,4]$ and $\theta_2 \in [-2,10]$ (middle left), Eggbox function in the range $\theta_1, \theta_2 \in [0, 10\pi]$ (middle right) and Rastrigin's function in the range $\theta_1, \theta_2 \in [-5.12,5.12]$ (right). Uniform priors over each coordinate were used for all the demos with stopping thresholds $\delta_\mathrm{final} = 0.05, 0.05, 0.5$, and $0.05$, respectively. \textit{Upper panels}: yellow dots represent (from left to right) the resulting $\nnest = 8485$, $8558$, $8207$, $10648$ samples obtained with the code presented in Sect.~\ref{sec:code} by using $\nlive = 1000$ points for each demo, as presented by FS13. \textit{Lower panels}: green dots represent (from left to right) the resulting $\nnest = 5286, 5151, 5874, 6174$ samples derived by additionally applying the reduction law given by Eq.~(\ref{eq:powerlaw_reducer}) with tol $= 100$, $\gamma = 0.4$, $N_0 = 1000$ and $\nmin = 400$ live points.}
    \label{fig:test}
\end{figure*}

\subsection{Reduction of the live points}
\label{sec:reduction}
As suggested by F09, reducing the number of live points as the NSMC process evolves may help in speeding up the whole computation, since fewer live points also imply fewer nested iterations for the algorithm to converge. This reduction could be needed especially in highly multi-modal problems, where a large number of live points is required at the beginning to ensure all the modes are properly detected. Nonetheless, the reduction of the prior mass with an evolving number of live points cannot be done with the standard rule given by Eq.~(\ref{eq:standard_reduction}), which assumes $\nlive$ to be constant throughout the process, and the new approach requires some thoughts. This case was neither treated by SK04 for the basic algorithm nor explicitly discussed by F09, who proposed an empirical rule for reducing the live points based on the largest evidence contribution estimable at each iteration. 

We explain below how we obtained the reduction rule for the prior mass with an evolving number of live points. We prove the result for one step only, as the principle can easily be generalized to an arbitrary number of reductions. Suppose we start with $\nlive = N_0$ live points at the first nested iteration, $i = 0$. Applying Eq.~(\ref{eq:standard_reduction}), the remaining prior mass at the first iteration is given by 
\begin{equation}
X_{1,N_0} = X_0 \exp \left(-1/N_0 \right) \, \,
\end{equation}
with the subscript $N_0$ indicating that it is based on $N_0$ live points ($X_0 = 1$ independently of the number of live points, hence no subscript is used).
At the second iteration, $i = 1$, we reduce the number of live points to $\nlive = N_1$ and once again reduce the prior mass. According to the standard reduction rule adopted for the case of $N_0$ live points, we simply have for $N_1$ that 
\begin{equation}
X_{2,N_1} = X_{1,N_1} \exp \left(-1/N_1 \right) \, ,
\end{equation}
where $X_{1,N_1}$ is the remaining prior mass from the first iteration given the number of live points is $N_1$. Since we do not know $X_{1, N_1}$ a priori, we need to derive its relation to the old $X_{1, N_0}$, the latter being known already because it was computed at the previous iteration, $i = 0$. Without losing in generality, we can write
\begin{equation}
X_{1, N_1} = \beta^{(N_0, N_1)}_1 X_{1, N_0} \, ,
\end{equation}
where the factor $\beta^{(N_0, N_1)}_1$ depends on both the previous and the new number of live points. By replacing the reduction rule for deriving $X_1$ from $X_0$, one obtains
\begin{equation}
\beta^{(N_0, N_1)}_1 = \exp \left( \frac{1}{N_0} -\frac{1}{N_1} \right) \, ,
\end{equation}
which clearly shows that if $N_1 < N_0$ then $\beta^{(N_0, N_1)}_1< 1$, thus resulting in a new remaining prior mass that is lower than the old one, as one would expect intuitively by adopting fewer live points. Iterating the result yields the generalized reduction rule; that is
\begin{equation}
X_{i+1, N_i} = \beta^{(N_{i-1},N_i)}_i X_{i,N_{i-1}} \exp \left(-1 / N_i \right) \, ,
\label{eq:generalized_reduction}
\end{equation}
where
\begin{equation}
\beta^{(N_{i-1}, N_{i})}_i = \exp \left( \frac{i}{N_{i-1}} - \frac{i}{N_i} \right) \, ,
\label{eq:stretching_factor}
\end{equation}
$i$ is the iteration in which the prior mass is updated and $N_i$ is the number of live points to be used for the next iteration. Clearly, Eq.~(\ref{eq:generalized_reduction}) reduces to the standard reduction rule expressed by Eq.~(\ref{eq:standard_reduction}) for $N_{i} = N_{i-1} = \nlive$.

Equations (\ref{eq:generalized_reduction}) and (\ref{eq:stretching_factor}) are implemented in the code, and two input parameters are therefore required: the initial number of live points, $N_0$, and the minimum number of live points allowed in the computation, $\nmin$. In case the two values coincide, the reduction of the live points is turned off automatically.

For \diamonds, in addition to the empirical rule proposed by F09 we also implemented another one that allows for a different behavior of the reduction process. As shown from our testing phase the function adopted by F09 appears to reduce the number of live points only at the beginning of the computation. It could be convenient instead to start reducing live points at a later stage, especially if the modes in the PPD are difficult to detect. This choice is also supported by the slowing down of the computation when approaching the termination condition described in Sect.~\ref{sec:stop_criterion}. This happens because it becomes more difficult to draw a new point that satisfies the likelihood constraint as we further rise up to the top of the likelihood distribution. Hence, the whole process would benefit more from removing live points at a later stage than in an early one since it ensures all the modes have been sampled efficiently in the previous steps.

Following the notation used above, our new relation for reducing live points can be expressed as
\begin{equation}
N_i = N_{i-1} - \left( \frac{\mbox{tol}}{\delta_i / \delta_\mathrm{final}}\right)^\gamma \, ,
\label{eq:powerlaw_reducer}
\end{equation}
where the final and evolving ratios of the live to the cumulated evidence introduced in Sect.~\ref{sec:stop_criterion} are adopted. The configuring parameter tol, which is the tolerance on the ratio $\delta_i / \delta_\mathrm{final}$, determines the initiating nested iteration for the reduction process. The lower the tolerance, the later the stage at which the live points start to be reduced. The minimum value allowed is tol~$= 1$, meaning that the reduction is not taking place. The exponent $\gamma$ instead controls the speed of the reduction process. The default value is $\gamma = 1$ for a linear reduction. For $\gamma > 1$ the reduction is super-linear, hence faster, while it is sub-linear for $0 \leq \gamma < 1$, hence slower. In the case of $\gamma = 0$, Eq.~(\ref{eq:powerlaw_reducer}) reduces to the simple form $N_i = N_{i-1} -1$, which implies that the sample of live points is constantly reduced by one at each iteration.

Some caution when using the reduction process is nevertheless needed. In this case, properties such as prior mass, density of the sampling, and evidence collection, change considerably during the computation. Deviations from the standard method introduced by SK04 may hamper the goodness of the final result. This happens mostly when too many live points are removed over very few iterations. This bad condition can generally be caused by a strong reduction rate. In the testing phase, we could note some side effects of a bad reduction process, which we list below:
\begin{enumerate}
\item The final sampling of the PPD may not correctly resemble the density of the probability function. This happens because when live points are removed, ellipsoids undergo an additional enlargement according to Eq.~(\ref{eq:dynamic_enlargement}), hence causing the sampling to occur in a region of the parameter space that is larger than expected.
\item The additional enlargement of the ellipsoids caused by having fewer live points also implies a loss in efficiency for the drawing algorithm.
\item The evidence collection is affected by additional (systematic) uncertainties, since reducing the live points decreases $\nnest$, hence the number of contributing terms used in Eq.~(\ref{eq:sum_evidence}). This effect produces a significant underestimation of the final evidence.
\end{enumerate}
Therefore, we recommend using the reduction of the live points with care and possibly only when it is really needed to speed up the inference analysis. We also advise not to use the reduction for computing evidences. Some examples on how to apply Eq.~(\ref{eq:powerlaw_reducer}) for speeding up the computation are shown in Fig.~\ref{fig:test} (see Sect.~\ref{sec:demos} for more discussion). We refer to Sect.~\ref{sec:parallelization} for a more suitable way of decreasing the computational time without directly affecting the number of live points during the process for both the models investigated.

\subsection{Parameter estimation}
\label{sec:parameter_estimation}
Parameter estimation is addressed by a separate module of \diamonds. The module uses the sample of nested points, which are the points found to have the lowest likelihood at each iteration and collected during the computation. The sample includes the $k$ coordinates of each nested point, and the corresponding likelihood value $\constr_i$ and weight $w_i$, as defined in Sect.~\ref{sec:nested} for the case of the trapezoidal rule.

Posterior probability values (not densities) for each sampled point are calculated by $P_i = \constr_i w_i / \evid$, as described by SK04. Since each free parameter $\theta_j$ of the $k$-dimensional parameter space $\Sigma_\mathcal{M}$ has $\nnest$ sampled values, one can marginalize the posterior probability --- i.e. integrate the posterior probability over the remainder $(k-1)$ coordinates --- by simply sorting the $\nnest$ sampled probabilities according to the ascending order of sampled values of the free parameter we want to estimate. Mean, median, and modal values of each free parameter $\theta_j$, and the second moment (variance) of the marginalized distribution are then computed.

Upper and lower credible limits for the shortest Bayesian credible intervals (CI, e.g. see \citealt{Sivia06} for a definition) are also calculated and provided as an output with all the parameter estimation values discussed above. For the computation of the CI, a refined marginal probability distribution (MPD) is obtained for each free parameter $\theta_j$ by rebinning its $\nnest$ sampled values according to the Scott's normal rule and by adopting the averaged shifted histogram \citep[ASH,][]{Hardle04}. The ASH is then interpolated by a grid that is ten times finer by means of a cubic spline interpolation. An example of this result is shown in Fig.~\ref{fig:marginal} for the analysis presented in Sect.~\ref{sec:bkg}. The final interpolated ASH of the marginal distribution can also be stored in output ASCII files for possible usage after the computation.

\subsection{Demos and comparison with \textsc{MultiNest}}
\label{sec:demos}
For demonstrating the capability of \diamonds\,\,to sample challenging likelihood surfaces, we tested it with the four two-dimensional examples used by FS13 with the GMC algorithm applied to \textsc{MultiNest}. These two-dimensional surfaces prove to be difficult to explore with standard Markov Chain Monte Carlo (MCMC) methods and they are: 
\begin{itemize}
\item the Himmelblau's function
\begin{equation}
f \left(\theta_1, \theta_2 \right) = ( \theta^2_1 + \theta_2 - 11 )^2 + ( \theta_1 + \theta^2_2 - 7 )^2 \, ,
\end{equation}
which has four local minima at $(3.0, 2.0)$, $(-2.81, 3.13)$, $(-3.78, -3.28)$, and $(3.58, -1.85)$; 
\item the Rosenbrock's function
\begin{equation}
f \left(\theta_1, \theta_2 \right) = ( 1 - \theta_1 )^2 + 100 \, (\theta_2 - \theta^2_1 )^2 \, ,
\end{equation}
having a global minimum at $(1,1)$ hidden in a pronounced thin curving degeneracy; 
\item the Eggbox function
\begin{equation}
f \left(\theta_1, \theta_2 \right) = - \left[ 2 + \cos \left( \frac{\theta_1}{2} \right) \cos \left( \frac{\theta_2}{2} \right) \right]^5 \, ,
\end{equation}
which presents identical local minima all equally spaced along each coordinate; 
\item the Rastrigin's function
\begin{equation}
f \left(\theta_1, \theta_2 \right) = 20 + \theta^2_1 + \theta^2_2 - 10 \left[ \cos \, (2 \pi \theta_1) - \cos \, (2 \pi \theta_2) \right] \, ,
\end{equation}
having a global minimum at $(0,0)$ hidden among a large number of local minima.
\end{itemize}
Following FS13, we adopted a log-likelihood $\ln \mathcal{L} \left( \theta_1, \theta_2 \right) = - f \left(\theta_1, \theta_2 \right)$ with $\theta_1$ and $\theta_2$ the coordinates identifying the two-dimensional parameter space. The results of the tests are shown in Fig.~\ref{fig:test}, where a fixed number of $\nlive = 1000$ points and uniform priors over each coordinate were adopted for all the demos for a more reliable comparison. We used uniform priors in the ranges specified in the caption of the figure. The code required $\nnest < 10^4$ samples for each demo to identify all the global maxima of the distributions. The number of iterations was about ten times fewer than in the case presented by FS13 for all the demos.

As already announced in Sect.~\ref{sec:reduction}, we also tested the same distributions by additionally applying a reduction of the live points according to Eq.~(\ref{eq:powerlaw_reducer}). The samples, as visible in the green dots of Fig.~\ref{fig:test}, lower panels, resemble well the shape of the distributions and also allow for a correct identification of all the global maxima. In this case, the configuration adopted (see the caption of the figure for more details) yielded a final number of nested iterations reduced by up to 40\,\%, resulting in a significant increase of speed in the computation. The evidence collected led to final values reduced by about 48\,\% with respect to that obtained without the reduction process for all the demos considered.

\subsection{Parallelization}
\label{sec:parallelization}
Based on the suggestion by K11 and references therein, one could improve the goodness of the results by unifying different and independent runs having $N_1, N_2$, etc live points with each, hence obtaining a joint run equivalent to a single one having $N_\mathrm{tot} = N_1 + N_2 +$, etc live points. The computation can be made parallel by running the split processes in several CPUs, hence merging the results in the end. The merging for the likelihood can be done by simply sorting the likelihood values from each independent run into a global ascending order. The merging of the parameters is done according to the sorting order of the corresponding likelihood values. For the prior mass, an easy re-computation based on Eq.~(\ref{eq:standard_reduction}) with a number of live points given by $N_\mathrm{tot}$ is required instead. The final evidence of the joint process can therefore be recomputed according to Eq.~(\ref{eq:sum_evidence}). 

This simple parallelization allows us in principle to gain the same level of accuracy on the final result of that obtained by a single run with $N_\mathrm{tot}$ live points, while significantly reducing the run tine of the process at the same time. Another advantage of this parallelization is that the sampling of the PPD can be rendered much finer than that of a single process, even in a shorter computational time. 
As a side note, one should keep in mind that there the number of live points that can be used in each of the split processes has a lower limit, which is directly related to the complexity of the PPD and the number of dimensions for the given problem.

\section{Observations and data}
\label{sec:data}
The \textit{Kepler} satellite has monitored thousands of pulsating stars among the 150,000 observed in its field of view, exploiting a very high duty cycle and sampling time in two different modes, short cadence \citep[SC,][]{Gilliland10}, and long cadence \citep[LC,][]{Jenkins10}. 

Among the initial $\sim$550 stars showing solar-like oscillations and observed in SC during the first year of operation with \Kepler, 61 of them were then selected for an extended observing time because they are bright and have higher SNR than other stars. The 61 final targets comprise G-type and F-type main-sequence (MS) and sub-giant stars, which were investigated by \citet[][hereafter A12]{App12Freq}, who measured their single p-mode frequencies.

For the purpose of our paper, we chose KIC~9139163 (HIP~92962), known also as \emph{Punto}, one F-type MS star of the final sample. The selected star was studied in further detail in a subsequent analysis concerning oscillation linewidths and heights of a sub-sample of 23 targets, done by \cite{App14} (hereafter A14). In addition, \cite{Campante14} also investigated the star concerning the effect of stellar activity on the amplitudes of solar-like oscillations. The object KIC~9139163 shows the largest number of individual oscillations ($> 50$) among all the stars of the final sample. This peculiarity makes the star even more suitable for a high-dimensional and multi-modal problem. 

A revised \textit{Hipparcos} parallax for KIC~9139163, $\pi = 9.49 \pm 0.83\,$mas is provided by \cite{Van07} and can be useful for accurately derive the stellar luminosity. We refer to \cite{Bruntt12} for an estimate of the temperature from spectroscopy, $T_\mathrm{eff} = 6375 \pm 70\,$K, which is largely compatible within the error bars to the value $T_\mathrm{eff} = 6405 \pm 44\,$K derived by \cite{Pin12} from SDSS photometry.
Furthermore, the rotation and activity level of the star have been studied by \cite{Karoff13Act}, who found a rotation period $P_\mathrm{rot} = 6.5 \pm 0.2$\,days by means of a periodogram analysis of the \kepler light curve, combined to asteroseismic and spectroscopic measurements.

\begin{figure}
   \centering
   \includegraphics[width=9.2cm]{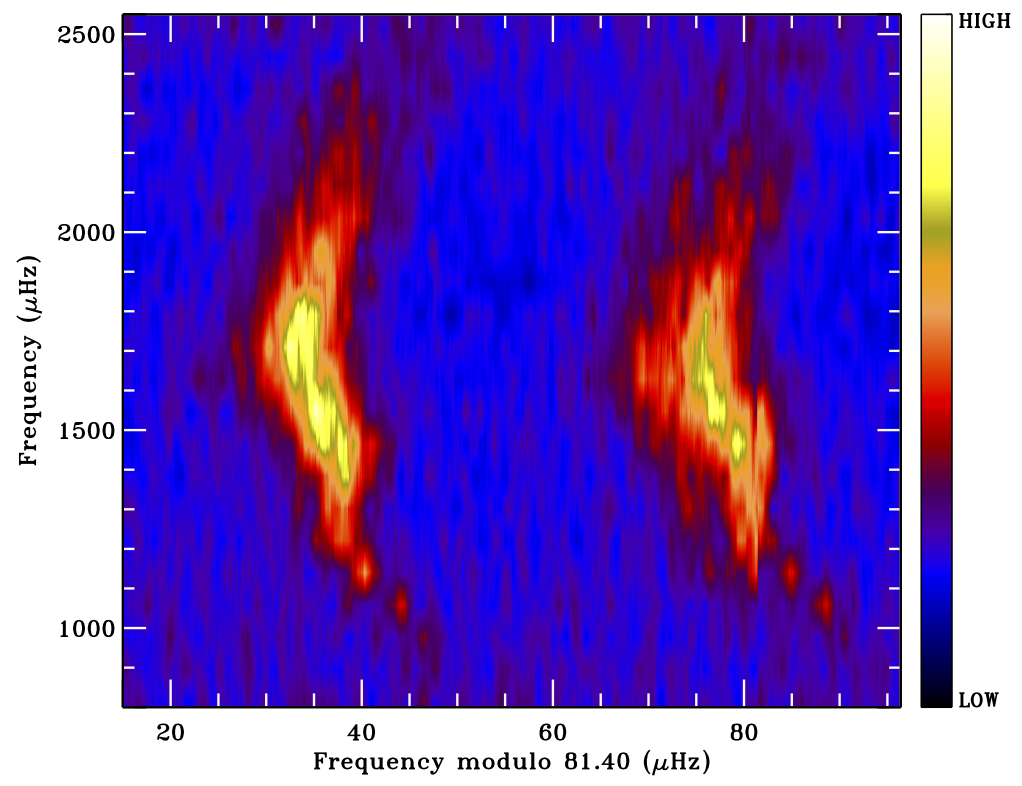}
      \caption{\'{E}chelle power spectrum of KIC~9139163 on a colored scale for $\Dnu = 81.4\,\mu$Hz and smoothed by $1 \,\mu$Hz. On the left, we find the $\ell = 1$ ridge of oscillation, while we have those corresponding to $\ell = 2, 0$ on the right. The plot makes the presence of a curvature of the ridges clear along the entire frequency range and the strong blending between quadrupole and radial peaks.}
    \label{fig:echPS}
\end{figure}

Referring to the studies by A12 and A14, oscillation frequencies, linewidths, and heights already derived for several oscillations of the given star allow for a more fruitful comparison of the results presented in Sect.~\ref{sec:results}. We now exploit a more recent and larger dataset available for this star, which includes the \kepler observing quarters (Q) from 5 to 17, namely 1147.5 days of observations. These data were stitched together and corrected from instrumental instabilities and drifts following the procedures explained in \cite{Garcia11data}. Moreover, we have high-pass filtered the final light curve with triangular smoothing with a cut-off frequency at $\sim$\,4\,days. For minimizing the impact of the quasi-regular gaps due to the angular momentum desaturation of the \kepler spacecraft, we have interpolated the gaps of less than 1 hour using a third order polynomial interpolation algorithm \citep{Garcia14}. 

The new light curve used in this work is about 14 months longer than the one adopted by A14, who used Q5-Q12 light curves, hence ensuring higher accuracy and precision that allows for further constraint of the free parameters of the models investigated. For the inference problem presented in Sect.~\ref{sec:application}, we use a power spectral density (PSD) computed by means of a Lomb-Scargle algorithm \citep{Scargle82} applied to the \kepler light curve. The new PSD has a frequency resolution of $0.01\,\mu$Hz and contains a total of more than 840,000 bins. This remarkable amount of data points makes the peak bagging analysis even more challenging in terms of computational effort.

\section{Application to peak bagging analysis}
\label{sec:application}

\subsection{Introduction to solar-like oscillations}
\label{sec:pmodes}
Before describing the details of the peak bagging analysis, it is useful to briefly introduce the physical quantities that we investigate. For a detailed description of the theory of solar-like oscillations, we refer the reader to \cite{CD04} for more insightful discussions. To avoid any ambiguity in terminology from now on, we shall refer to the individual oscillation mode as ‘peak’, while using the term ‘mode’ to indicate the modal value of the outcome coming from the Bayesian inference analysis.

According to the asymptotic theory of solar-like oscillations \citep[e.g.][]{Tassoul80}, acoustic standing waves (also known as pressure modes or simply \p modes) with a low angular degree $\ell$, the number of nodal lines on the stellar surface, and high radial order $n$, the number of nodes along the radial direction of the star, show a characteristic regular pattern in frequency, expressed by the asymptotic relation approximated at the first order,
\begin{equation}
\nu_\mathrm{\ell,n} \simeq \Dnu \left( n + \frac{\ell}{2} + \epsilon \right) - \delta\nu_\mathrm{0\ell} \, ,
\label{eq:asymp}
\end{equation}
where $\Dnu$ is the main characteristic frequency spacing of p modes having different radial order, known as the large frequency separation, which scales roughly as the square root of the mean stellar density \citep{Ulrich86}. The phase shift $\epsilon$ is instead sensitive to the physics of the near-surface layers of the star \citep{CD92}. The small frequency spacing $\delta_\mathrm{0\ell}$ is related to the sound speed gradient in the stellar core, and it is defined for $\ell = 1, 2$ as
\begin{align}
\dnua &\equiv \frac{1}{2} \left( \nu_\mathrm{n,0} + \nu_\mathrm{n+1,0} \right) - \nu_\mathrm{n,1} \, , \label{eq:d01}\\
\dnub &\equiv \nu_\mathrm{n,0} - \nu_\mathrm{n-1,2} \, ,\label{eq:d02}
\end{align}
respectively. 

By plotting Eq.~(\ref{eq:asymp}) as a function of the frequency modulo $\Dnu$, we obtain an \'echelle diagram where the oscillations having different angular degree align vertically to form separate ridges. In practice, it often happens that the ridges are curved, since the observed frequencies may depart from the first order approximation given by Eq.~(\ref{eq:asymp}). Figure~\ref{fig:echPS} shows an example of such a curvature effect with the \'echelle power spectrum of KIC~9139163 in a color-coded scale, computed in the same way as by \cite{Corsaro12}. For this plot, the PSD was normalized by the background level derived in Sect.~\ref{sec:bkg}, hence smoothed by a boxcar filter having width $1\,\mu$Hz. Oscillation ridges for dipole peaks ($\ell = 1$, left), quadrupole, and radial peaks ($\ell = 2, 0$, right) are visible over a frequency range of more than $1500\,\mu$Hz. A large frequency separation $\Dnu = 81.4\,\mu$Hz, as derived by A12, was adopted.

\subsection{Background modeling}
\label{sec:bkg}
A preliminary step for performing the peak bagging analysis consists in estimating the background level in the star's PSD. Although fitting a background is a relatively low-dimensional problem, there is no universal model that can be used for all the stars as it closely depends on how many physical phenomena are involved, namely granulation \citep{Harvey85,Aigrain04,Michel09}, and the more recently investigated bright spots activity (faculae) \citep{Chaplin10,Karoff10,Karoff12,Karoff13}. As a consequence (since the asteroseismic analysis of the oscillations is sensitive to the stellar background components) assuming different models may sensibly change the final results (see the discussion by A14 concerning the impact of the background on the oscillation characteristic parameters). 

For these reasons, it is essential to properly address this part of the analysis. With Bayesian statistics, this is achieved by exploiting the Bayesian evidence computed by \diamonds\,\,for testing different hypotheses through Eq.~(\ref{eq:odds}), as discussed in Sect.~\ref{sec:bayes} already.

\begin{figure}
   \centering
   \includegraphics[width=9.0cm]{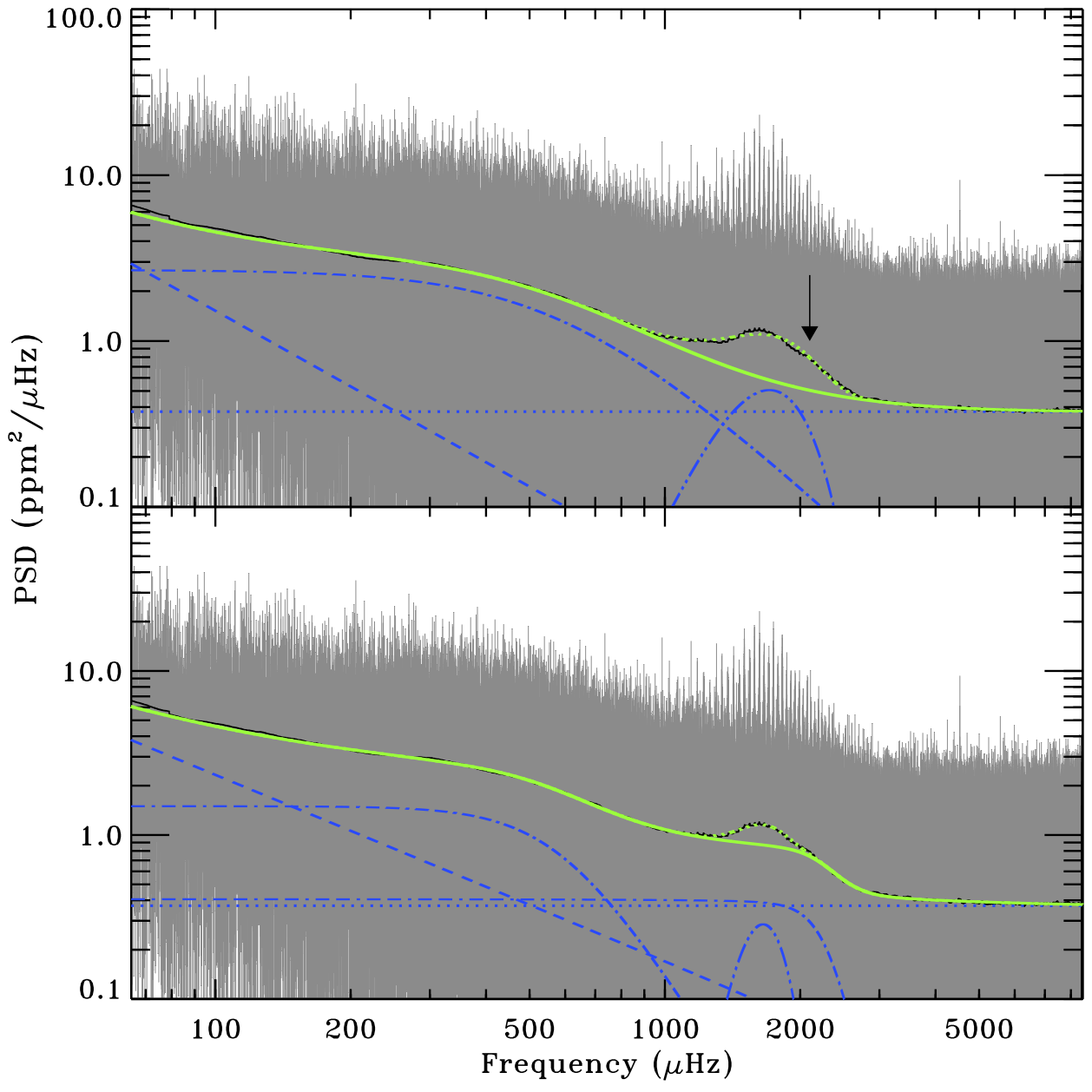}
      \caption{PSD of KIC~9139163 (gray) with overall background from Eq.~(\ref{eq:overall_bkg}) and median values reported in Table~\ref{tab:bkg} (thick green line) with the additional Gaussian envelope included (dotted green line). The solid black line represents the smoothed PSD by $81.4\,\mu$Hz. The single background components of constant photon noise (dotted), power law (dashed), granulation and faculae (dot-dashed), and Gaussian envelope (double-dot-dashed) are shown in blue. \textit{Upper panel}: The model $\model_1$ accounting for one Harvey-like profile. The arrow indicates the presence of a kink that is not reproduced by the model. \textit{Lower panel}: the model $\model_2$ accounting for two Harvey-like profiles. The winning model $\model_2$ is strongly favored as it yields a Bayes' factor $\ln \mathcal{B}_{21} = 58.2 \pm 0.2$ over its competitor. }
    \label{fig:bkg}
\end{figure}

\begin{figure*}
   \centering
   \includegraphics[height=4.3cm]{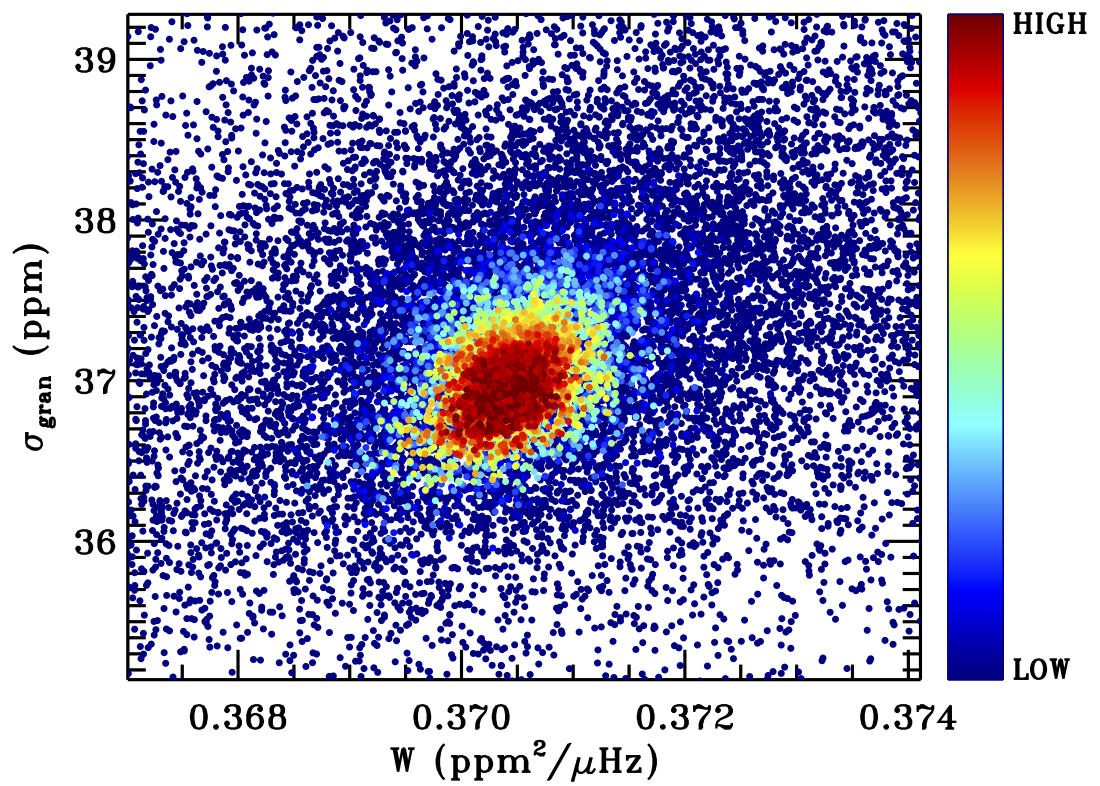}\includegraphics[height=4.3cm]{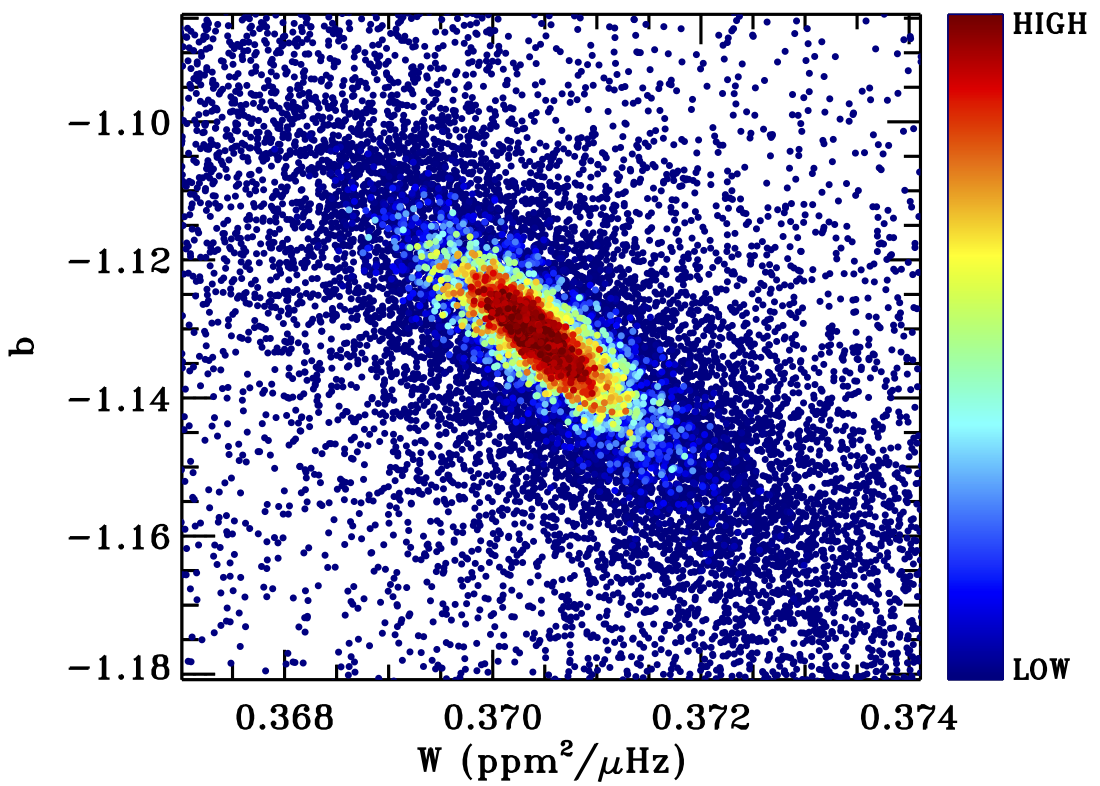} \includegraphics[height=4.3cm]{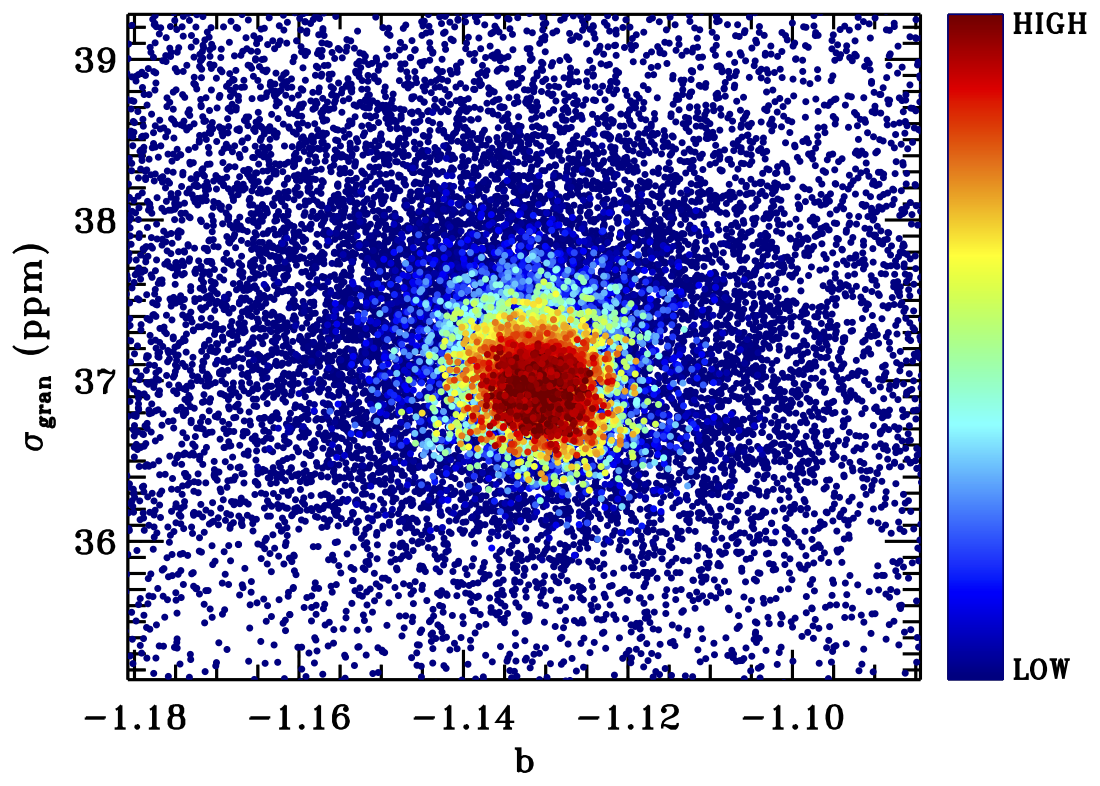}
   \includegraphics[width=4.8cm]{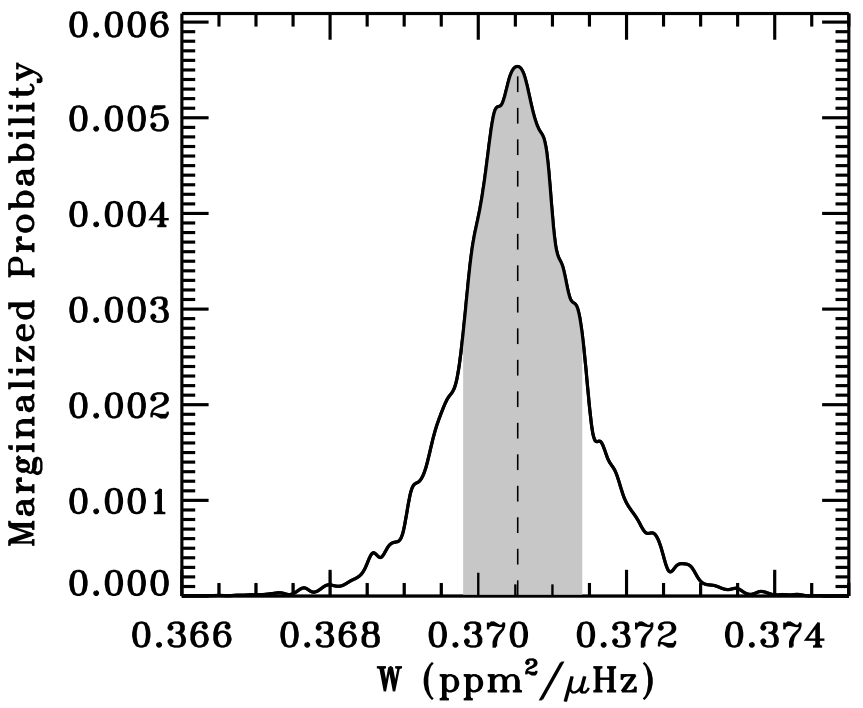}\includegraphics[width=4.8cm]{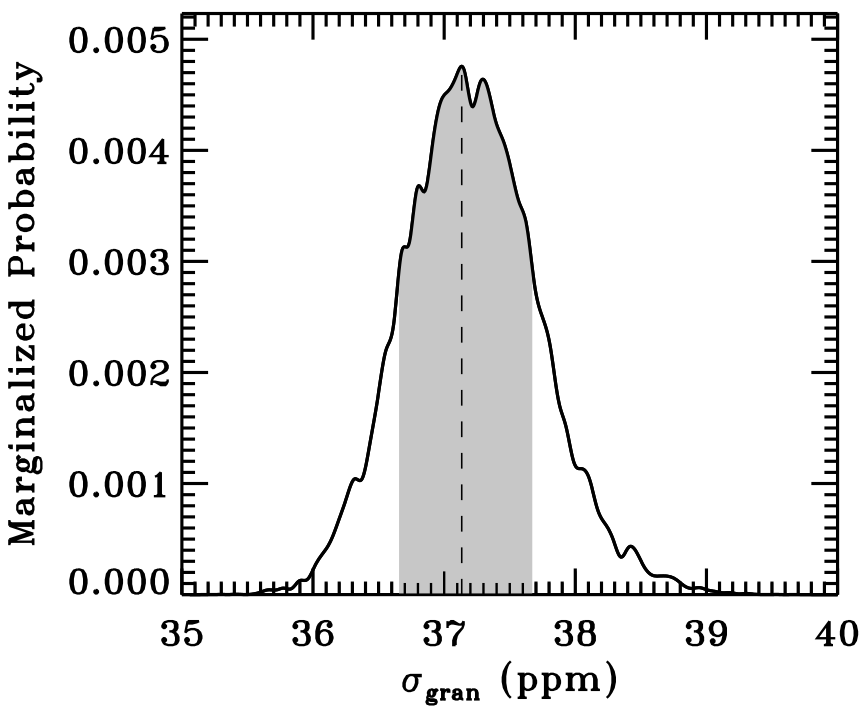}\includegraphics[width=4.8cm]{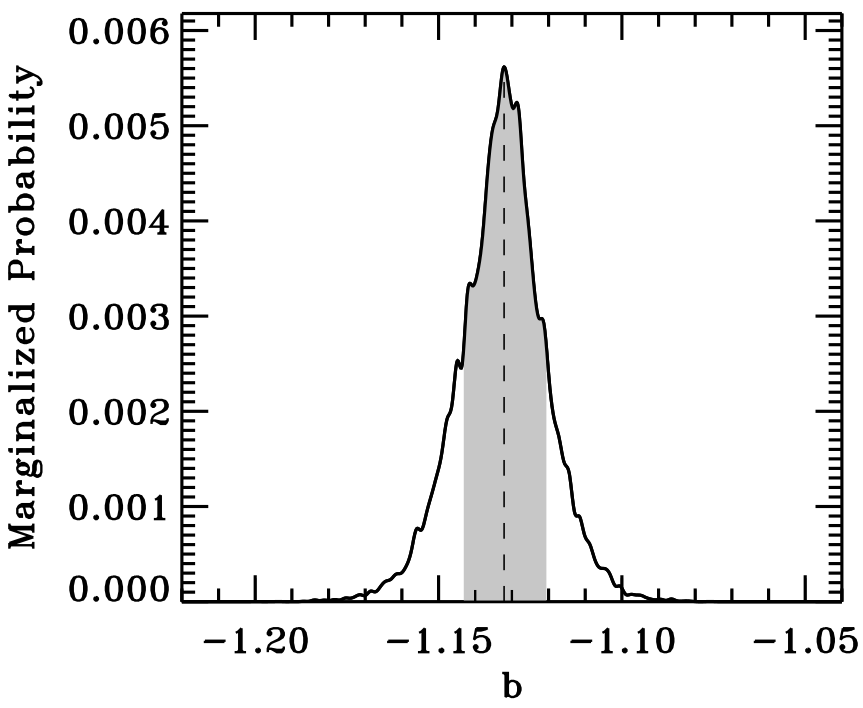}
      \caption{\textit{Upper panels}: examples of correlation maps of the three free parameters $W$, $\sigma_\mathrm{gran}$, and $b$ used in Eq.~(\ref{eq:bkg}) using two Harvey-like profiles with color-coded likelihood values. Each point in the diagram is a sampling point that stems from the NSMC process. The plotted realization consists of $\sim$\,27,000 samples. \textit{Lower panels}: corresponding MPDs of the free parameters as computed by means of \diamonds. The shaded region indicates the portion of the distribution containing 68.3\,\% of the total probability, defining the shortest credible intervals listed in Table~\ref{tab:bkg}. The dashed line indicates the mode of the distribution. }
    \label{fig:marginal}
\end{figure*}

\begin{figure*}
   \centering
   \includegraphics[height=4.3cm]{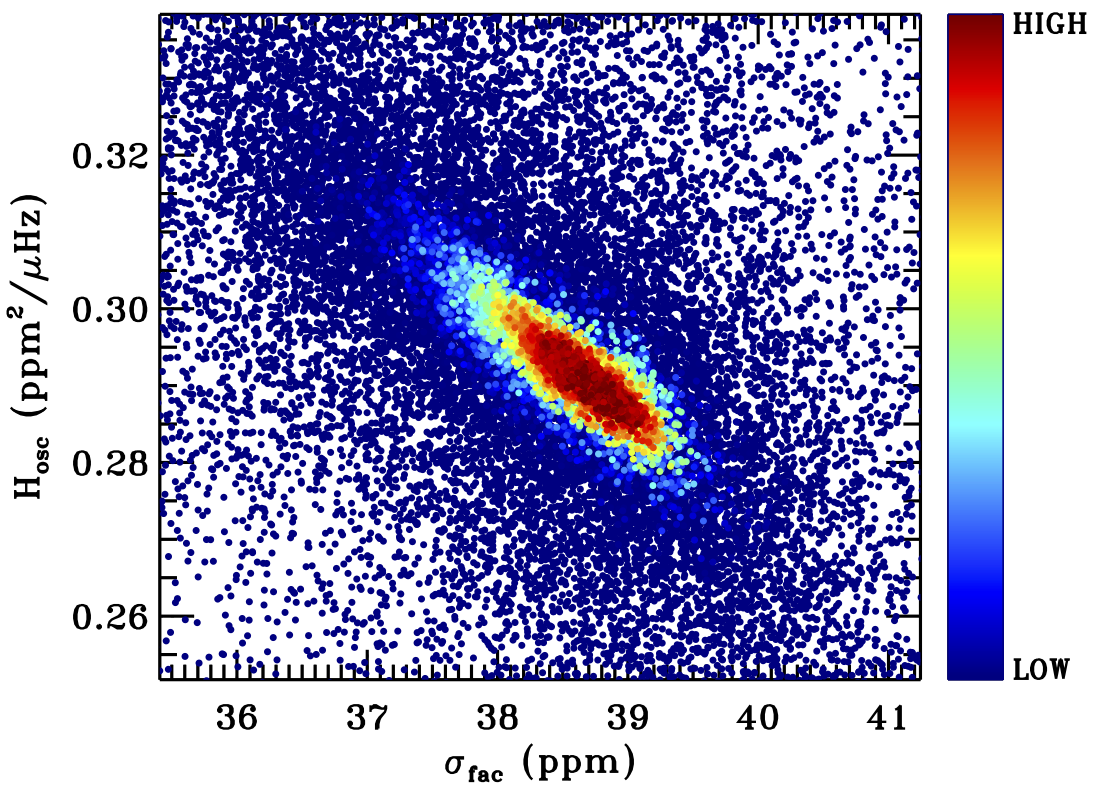}\includegraphics[height=4.3cm]{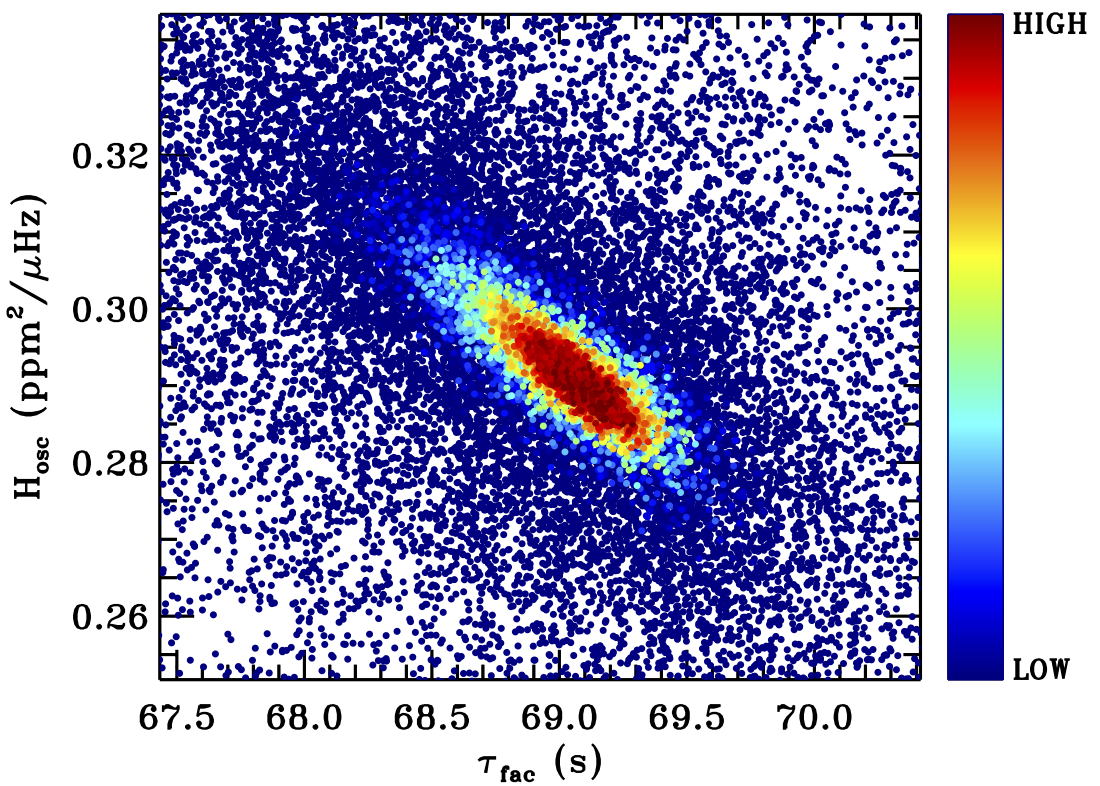} \includegraphics[height=4.3cm]{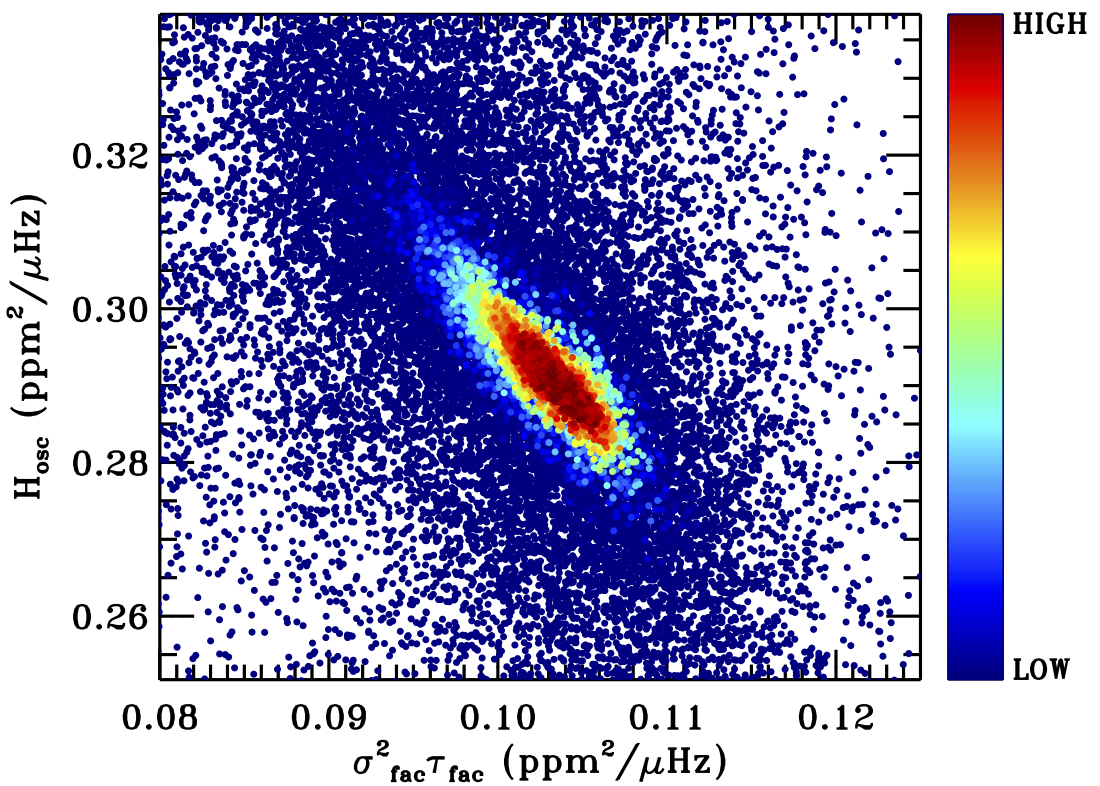}
      \caption{Correlation maps for the parameters describing the Harvey-like profile related to the faculae activity --- namely $\sigma_\mathrm{fac}$, and $\tau_\mathrm{fac}$, their combination $\sigma^2_\mathrm{fac} \tau_\mathrm{fac}$ in PSD units, and the height of the oscillation envelope $H_\mathrm{osc}$ with color-coded likelihood values plotted similarly to Fig.~\ref{fig:marginal} --- using the same sample of points.}
    \label{fig:corr_faculae}
\end{figure*}
We therefore considered a general model based on those presented by \cite{Mathur11} and by \cite{Karoff13}, which can be expressed as
\begin{equation}
P \left(\nu \right) = W + R \left( \nu \right) \left[ B\left( \nu \right) + G \left( \nu \right) \right] \, ,
\label{eq:overall_bkg}
\end{equation}
where $W$ is a constant noise (mainly photon noise), $R\left(\nu\right)$ is the response function that considers the sampling rate of the observations,
\begin{equation}
R\left( \nu \right) = \mbox{sinc}^2 \left( \frac{\pi \nu}{2 \nu_\mathrm{Nyq}} \right) \, ,
\label{eq:resp}
\end{equation}
where $\nu_\mathrm{Nyq} = 8496.36\,\mu$Hz is the Nyquist frequency for \kepler SC data in our case. The background components are given by
\begin{equation}
B\left(\nu\right) = a \nu^{-b} + \sum^h_{i=1} \frac{4 \tau_i \sigma^2_i}{1 + \left( 2 \pi \nu \tau_i \right)^{c_i}} \, ,
\label{eq:bkg}
\end{equation}
and the power excess is described with
\begin{equation}
G\left(\nu\right) = H_\mathrm{osc} \exp \left[ - \frac{ \left( \nu - \nu_\mathrm{max} \right)^2}{2 \sigma_\mathrm{env}^2} \right] \, . 
\label{eq:env}
\end{equation}
The first term on the right-hand side of Eq.~(\ref{eq:bkg}) is a power law that models slow variations caused by stellar activity, while the second term is a summation of $h$ (either $1$ or $2$) as the Harvey-like profiles \citep{Harvey85}, $\tau_i$ being the characteristic timescale, $\sigma_i$ the amplitude of the signature, and $c_i$ the corresponding exponent related to the decay time of the physical process involved \citep{Harvey93,Karoff12,Karoff13}.
The Gaussian component appearing as the last term to the right-hand side of Eq.~(\ref{eq:bkg}) models the power excess caused by solar-like oscillations with $H_\mathrm{osc}$ as the height of the oscillation envelope, and $\numax$ and $\sigma_\mathrm{env}$ as the corresponding frequency of maximum power and width, respectively. The component modeling the oscillation envelope is replaced afterwards by the global peak bagging model, as seen in Sect.~\ref{sec:pb}, to fit the individual oscillation peaks.

As already introduced in Sect.~\ref{sec:likelihood_priors}, the log-likelihood to be adopted for the inference analysis that involves a PSD is the exponential log-likelihood given by Eq.~(\ref{eq:exponential_likelihood}). In the context of the peak bagging analysis, the data point is now the observed PSD at a single frequency bin $\nu_i$, namely $O_i \equiv P_\mathrm{obs} \left(\nu_i \right)$, while  the corresponding prediction is given as  $E_i \left( \dtheta \right) \equiv P\left(\nu_i; \dtheta \right)$ for the specific case of the background fitting, as expressed by Eq.~(\ref{eq:overall_bkg}). We used uniform priors for all the parameters of the model except for the parameter $a$, the amplitude of the slow variation in stellar activity. The indeterminacy on $a$ is larger than that of the other parameters because $a$ is mostly constrained by the PSD at very low frequencies. For this reason, we set up a Jeffreys' prior $\pi\left( a \right) \propto a^{-1}$ \citep{Kass96}, giving equal weight to different orders of magnitude. This was done in practice by adopting the new parameter $\ln a$, since the Jeffreys' prior becomes uniform in logarithmic scale.

Figure~\ref{fig:bkg} shows the final results of the Bayesian inference for the background modeling of KIC~9139163 done by means of \diamonds\,\,(background without and with Gaussian envelope shown in the Figure and overlaid to the observed PSD and its smoothing by $\Dnu$. The figure presents both the case of the model including only one Harvey-like profile, which is the granulation signal component ($h = 1$), and the most likely model accounting for two Harvey-like profiles ($h = 2$), which represent the granulation signal and faculae activity of the star (see the caption of the figure for more details and Table~\ref{tab:bkg} for a list of all the values derived from the Bayesian parameter estimation). By comparing the Bayesian evidence from the two competing models ($\model_1$ for $h = 1$ and $\model_2$ for $h = 2$), the resulting natural logarithm of the Bayes' factor $\ln \mathcal{B}_{21} = \ln \evid_2 - \ln \evid_1 = 58.2 \pm 0.2$ in favor of model $\model_2$ suggests that an additional source of background signal coming from the faculae activity is strongly decisive for the model comparison. The model with two Harvey-like profiles thus ought to be preferred (strong evidence for $\ln \mathcal{B}_{21} \simeq 5$ in the Jeffreys' scale of strength). More reliability is added to this conclusion since an error bar on the Bayes' factor is also included, which is computed through error propagation of the statistical uncertainty on the final evidence given by the NSMC algorithm. The result of the model selection also agrees with the presence of a kink in the smoothed PSD (indicated by an arrow in the upper panel of Fig.~\ref{fig:bkg}) at the high-frequency side of the oscillation envelope. In Fig.~\ref{fig:marginal}, we plot an example of three correlation maps with likelihood values in a color-coded scale for the background parameters $W$ and $\sigma_\mathrm{gran}$ and $b$ of the winning model with the corresponding MPDs computed by \diamonds. We further provide three interesting cases of correlation maps for the background parameters characterizing the Harvey-like profile of the faculae activity, which are its amplitude $\sigma_\mathrm{fac}$, timescale $\tau_\mathrm{fac}$, and height at zero frequency $\sigma^2_\mathrm{fac} \tau_\mathrm{fac}$ in PSD units, relative to the height of the oscillation envelope $H_\mathrm{osc}$. As clearly visible, amplitude, timescale, and height of the faculae background are anti-correlated to $H_\mathrm{osc}$, as one would expect, since the faculae component rises inside the oscillation region, while no significant correlation is found for the exponent $c_\mathrm{fac}$ of the corresponding Harvey-like profile.

For the configuration of the code, the following set of parameters was used: $f_0 = 1.5$, $\alpha = 0.02$, $\nlive = 1000$, $1 \leq \nclust \leq 6$, $\matt = 10^4$, $\minit = \nlive$, and $\msame = 50$. This has allowed us to maintain a good efficiency throughout the sampling process for both the models investigated.

\begin{table}
\caption{Median values with corresponding 68.3\,\% shortest credible intervals of the background parameters for KIC~9139163, given by Eq.~(\ref{eq:bkg}) with $h = 2$, as derived by \diamonds.}             
\centering                         
\begin{tabular}{l l r}       
\hline\hline
\\[-8pt]         
Parameter & Median & Units\\    
\hline
\\[-8pt]
   $W$ & $0.3705^{+0.0009}_{-0.0007}$ & ppm$^2$ $\mu$Hz$^{-1}$\\[1pt]
   $a$ & $428\pm25$ & ppm$^2$ $\mu$Hz$^{-1+b}$\\[1pt]
   $b$ & $-1.13\pm0.01$ &\\[1pt]
   $\sigma_\mathrm{gran}$ & $37.2\pm0.5$ & ppm \\[1pt]
   $\tau_\mathrm{gran}$ & $271^{+3}_{-2}$ & s\\[1pt]
   $c_\mathrm{gran}$ & $4.3\pm0.2$ & \\[1pt]
   $\sigma_\mathrm{fac}$ & $38.4^{+0.8}_{-0.6}$ & ppm \\[1pt]
   $\tau_\mathrm{fac}$ & $69.0\pm0.4$ & s\\[1pt]
   $c_\mathrm{fac}$ & $12.5^{+0.5}_{-0.3}$ & \\[1pt]
   $H_\mathrm{osc}$ & $0.294^{+0.010}_{-0.011}$ & ppm$^2$ $\mu$Hz$^{-1}$\\[1pt]
   $\numax$ & $1655^{+5}_{-4}$ & $\mu$Hz\\[1pt]
    $\sigma_\mathrm{env}$ & $193^{+13}_{-12}$ & $\mu$Hz\\[1pt]
\hline                                
\end{tabular}
\label{tab:bkg}
\end{table}

\subsection{Characterization of \p modes}
\label{sec:pb}
It can be shown \citep[e.g. see][]{Kumar88,Anderson90} that the limit PSD for a series of $N_\mathrm{peaks}$ independent oscillations can be expressed by means of a Lorentzian mixture
\begin{equation}
P_\mathrm{osc} \left( \nu \right) = \sum^{N_\mathrm{peaks}}_{i=1} \frac{A_i^2 / \left( \pi \Gamma_i \right)}{1 + 4 \left( \frac{\nu - \nu_{0,i}}{\Gamma_i} \right)^2} \, ,
\label{eq:general_pb_model}
\end{equation}
where $A_i$ is the amplitude of the $i$-th oscillation peak in the PSD (expressed in ppm), $\nu_{0,i}$ its central frequency, and $\Gamma_i$ the linewidth, which is related to the oscillation lifetime $\tau_i$ by $\Gamma_i = ( \pi \tau_i)^{-1}$. The quantities $A_i$, $\nu_{0,i}$, and $\Gamma_i$ thus represent the free parameters characterizing one oscillation peak profile. We can easily assign uniform priors for each of these quantities by a quick look at the PSD.

For the Bayesian inference, we adopt once again the exponential log-likelihood, but we now restrict the frequency range of the PSD to the interval $900$-$2800\,\mu$Hz; namely, the region containing the oscillation peaks we intend to analyze, following the results by A12. Moreover, similar to the work done by A12 and by A14, we set a fixed background $B\left(\nu\right) = \overline{B} \left(\nu\right)$ and $\overline{W}$ by using the mean values of the corresponding parameters listed in Table~\ref{tab:bkg}. This choice is motivated by the fact that the background fitting is performed over the full-length PSD, thus providing the most precise and accurate result we can stem for the background from the given dataset.
In addition, we adopt mean values since they are the optimal estimators post-data in the context of Bayesian statistics \citep{Bolstad13}, hence the most reliable outcomes of our fit. 

Therefore, for the fitting and characterization of the individual oscillation peaks, the Lorentzian mixture model for solar-like oscillations described by Eq.~(\ref{eq:general_pb_model}) replaces the previous approximation of the power excess given by the Gaussian envelope $G \left( \nu \right)$ defined by Eq.~(\ref{eq:env}), which yields the overall peak bagging model,
\begin{equation}
P \left(\nu \right) = \overline{W} + R \left( \nu \right) \left[ \overline{B} \left( \nu \right) + P_\mathrm{osc} \left( \nu \right) \right]
\label{eq:overall_pb}
\end{equation}
with $R \left( \nu \right)$ once again as the response function given by Eq.~(\ref{eq:resp}).

\subsection{Rotation from $\ell = 1$ peaks}
The presence of rotation in a PSD lifts the $(2 \ell + 1)$-fold degeneracy of the frequency of non-radial peaks, hence directly measuring a mean angular velocity, $\Omega$, of the star \citep{Ledoux51} if the rotationally split peaks are properly resolved. For KIC~9139163, we exclude a priori the possibility of detecting rotation from the dipole (and therefore quadrupole) peaks since its surface rotation rate ${P_\mathrm{rot}}^{-1}\simeq 1.78\,\mu$Hz \citep{Karoff13Act}, as compared to the typical linewidth of the highest SNR $\ell = 1$ peaks, where $\Gamma_{\ell = 1}\,\sim\,4\,\mu$Hz (this work, see Sect.~\ref{sec:results2}), satisfies the condition ${P_\mathrm{rot}}^{-1} < 2 \Gamma_{\ell = 1}$, hence this does not allows us to resolve multiplets coming from rotation \citep[e.g. see][]{Gizon03}. This is essentially related to the short oscillation lifetime of \p modes occurring in the shallow convective regions of F-type stars. The result is even more enhanced if one considers the projection effect of the estimated inclination angle of the rotation axis of the star. By combining the spectroscopic $v \sin i \simeq 4$\,km\,s$^{-1}$ \citep{Bruntt12} and deriving the radius of the star through asteroseismology $R \simeq 1.52\, R_\odot$ (this work agrees within 2\,\% with that stemmed by \citealt{Karoff13Act}) through the relation
\begin{equation}
\sin i = \frac{v \sin i \, P_\mathrm{rot}}{2 \pi R} \, ,
\end{equation}
one obtains $i \simeq 20^\circ$. This low inclination angle implies a relative height of the central (non-split) component to that of the two split for the dipole oscillations \citep{Gizon03} of $\sim15.4$, meaning that the split frequency peaks are almost undistinguishable from the background level. For all these reasons from now on, we assume there is no rotational effect observable in any of the dipole (and quadrupole) peaks of KIC~9139163.

\subsection{Peak significance and detection criterion}
\label{sec:significance}
\begin{figure}
   \centering
    \includegraphics[height=7.1cm]{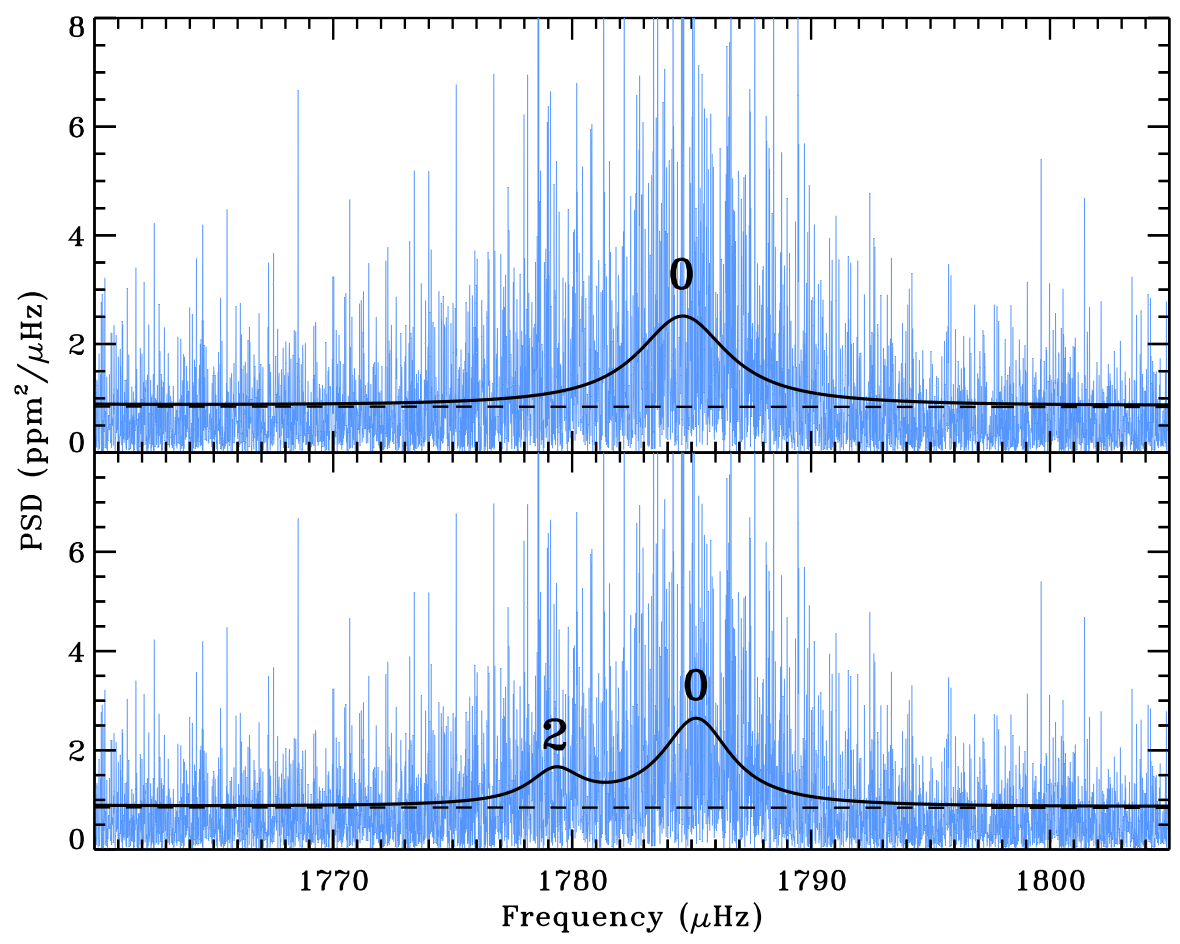}
      \caption{Example of an ambiguous detection of a high SNR duplet $\ell = 2, 0$ from the PSD of KIC~9139163. The solid line shows the resulting best-fit as computed by \diamonds, while the dashed line represents the mean background model as obtained in Sect.~\ref{sec:bkg}, according to Eq.~(\ref{eq:overall_pb}). \textit{Upper panel}: the case of model $\model_B$ where only $\ell = 0$ is fitted. \textit{Lower panel}: the case of model $\model_C$ where both $\ell = 2, 0$ are fitted. The resulting $\ln \mathcal{B}_{CB} =  16.5 \pm 0.1$ strongly favors model $\model_C$, hence the presence of two peaks in the observed structure.}
    \label{fig:duplet}
\end{figure}

One of the main problems arising in the context of a peak bagging analysis consists in assessing the number of significative peaks to be fitted and/or accepted for the final result. In a frequentist-based approach, one would generally adopt a detection threshold based on either the estimated noise-level of the star or the maximum likelihood value of the fitted oscillation peak \citep[e.g. see][]{App98}. In a Bayesian context, one  computes instead a probability stating how reliable the given peak is, which may rely on either the null-hypothesis test or a direct estimate of the odds ratio given by Eq.~(\ref{eq:odds}) (e.g. see A12), as explained in Sect.~\ref{sec:bayes}. In this work, we exploit the Bayesian odds ratio, hence the Bayes' factor, since its computation is straightforward within the NSMC process, as already shown earlier in the text. This allows us to statistically weight the peak detection in terms of both goodness-of-fit and model complexity, hence penalizing those peaks that have lower SNR. We apply this criterion in three different scenarios:
\begin{itemize}
\item A single low-SNR peak arising from a background level. This detection process involves the model comparison between two competing models, which we indicate as $\model_\mathrm{A}$ when we exclude the peak in the fitting process and $\model_\mathrm{B}$ when we include the peak instead. This case is typically that of dipole peaks occurring in the wings of the oscillation envelope. The corresponding natural logarithm of the Bayes' factor $\ln \mathcal{B}_\mathrm{BA} = \ln \evid_\mathrm{B} -  \ln \evid_\mathrm{A}$ in favor of the model with a fitted peak, is therefore included in the final list of the results for all the ambiguous peak detections, to provide prompt confirmation of the outcome of the model comparison process. In addition, since assigning a quantitative probability value for the detection of an individual peak could be useful for weighting the reliability of the different detections, the Bayesian evidences $\evid_\mathrm{A}$ and $\evid_\mathrm{B}$ can be used to compute the detection probability as
\begin{equation}
p_\mathrm{B} \equiv \frac{\evid_\mathrm{B}}{\evid_\mathrm{A} + \evid_\mathrm{B}} \, ,
\label{eq:p_B}
\end{equation}
or equivalently, the non-detection probability,
\begin{equation}
p_\mathrm{A} \equiv 1 - p_\mathrm{B} = \frac{\evid_\mathrm{A}}{\evid_\mathrm{A} + \evid_\mathrm{B}} \, .
\label{eq:p_A}
\end{equation}

\item Two high-SNR peaks appearing in a blended structure. This situation makes it ambiguous to distinguish between one or two different oscillation peaks. The issue is well represented by the duplet $\ell = 2, 0$ of F-type stars because their large oscillation linewidths can produce very strong blending, as shown in Fig.~\ref{fig:duplet} for two oscillation peaks of KIC~9139163. For this case, we can once again assume two competing models, $\model_\mathrm{B}$ when only one peak ($\ell = 0$) is fitted and $\model_\mathrm{C}$ when two peaks (both $\ell =2$ and $0$) are fitted. We can thus compute the natural logarithm of the Bayes' factor, $\ln \mathcal{B}_\mathrm{CB} = \ln \evid_\mathrm{C} - \ln \evid_\mathrm{B}$ in favor of the model with two peaks. Following Eqs.~(\ref{eq:p_B}) and (\ref{eq:p_A}), we also define the probability of detecting two peaks, $p_\mathrm{C}$ (or equivalently only one, $p_\mathrm{B} \equiv 1 - p_\mathrm{C}$). 

\item Two low-SNR peaks appearing in a blended structure. The typical example is given by the duplets $\ell = 2, 0$, which fall in the wings of the oscillation envelope. To properly address this event, one has to consider three possible competing models, namely $\model_\mathrm{A}$, $\model_\mathrm{B}$, and $\model_\mathrm{C}$, as previously defined. We therefore compute the natural logarithm of the Bayes' factors $\ln \mathcal{B}_\mathrm{BA}$ for checking whether a single peak ($\ell = 0$) is detected or not, and $\ln \mathcal{B}_\mathrm{CB}$ for assessing the presence of two peaks ($\ell = 2, 0$) from the blended structure. The corresponding detection probabilities can be defined as 
\begin{equation}
p_\mathrm{C} \equiv \frac{\evid_\mathrm{C}}{\evid_\mathrm{A} + \evid_\mathrm{B} + \evid_\mathrm{C}}
\label{eq:p_C3}
\end{equation}
for detecting two peaks; 
\begin{equation}
p_\mathrm{B} \equiv \frac{\evid_\mathrm{B}}{\evid_\mathrm{A} + \evid_\mathrm{B} + \evid_\mathrm{C}} \, ,
\label{eq:p_B3}
\end{equation}
for detecting one peak; and
\begin{equation}
p_\mathrm{A} \equiv 1 - p_\mathrm{B} - p_\mathrm{C} \, ,
\end{equation}
for the case of just the background level.
\end{itemize}

To apply the peak significance method and perform the Bayesian parameter estimation at the same time, we fit the PSD of KIC~9139163 as follows: \begin{enumerate}
\item When assessing and fitting a dipole oscillation, we consider a chunk of PSD that contains a series of five consecutive oscillation peaks having an angular degree $\ell = 2,0,1,2,0$ in the order from left to right (hence with the dipole peak in the center of the selected PSD window). This allows the fit in the PSD to be stable for the central peak, hence comparing the models (cases  $\model_\mathrm{A}$ and $\model_\mathrm{B}$) accurately, and stemming parameter estimates more conveniently, since we adopt a maximum of $k = 15$ dimensions in the fit.
\item When assessing and fitting a duplet of blended quadrupole and radial oscillations, we consider a chunk of PSD containing a series of four consecutive oscillation peaks having an angular degree $\ell = 1,2,0,1$ in the order from left to right (hence with the duplet falling in the center of the selected PSD window). The reasoning is equivalent to that of the first case, allowing us to set the number of dimensions in the inference problem not beyond $k = 12$.
\end{enumerate}
The choice of fitting windows in the PSD does not practically affect the parameter estimation process, since the oscillation peaks of the angular degrees $\ell = 2,0$ are separated by those of angular degree $\ell = 1$ by $\sim40\,\mu$Hz, a value about ten times larger than the typical linewidth of the oscillations. For the blended $\ell = 2,0$ oscillations, we instead always fit a duplet of Lorentzian profiles, since the two peaks cannot be separated from each other. 

We note that in this work linewidths and amplitudes are fit independently for each individual oscillation peak, as expressed by Eq.~(\ref{eq:general_pb_model}), in contrast to a classical fitting method to peak bagging (e.g. see A12, A14). This implies that we drop both the assumption of the height ratios for oscillation peaks having a different angular degree (therefore using a single height per radial order) and the use of a common linewidth within a single radial order. A direct consequence of this approach is that the resulting PPD presents a uni-modal solution for any number of fit oscillation peaks, as we discuss in more detail in Sect.~\ref{sec:multi_modal_pb}.

\subsection{Results}
\label{sec:results}
All the results performed according to the approach based on the model as given by Eq.~(\ref{eq:general_pb_model}) and the peak significance as explained in Sect.~\ref{sec:significance} (hereafter Approach 1 for shortness) are listed in Table~\ref{tab:frequencies} and documented in Appendix~\ref{sec:results2}. From a computational point of view, the configuration of \diamonds\,\,required the tuning of the parameters for the dynamical enlargement of the ellipsoids, as introduced in Sect.~\ref{sec:ellipsoidal}, and used values in the ranges $2.0 \leq f_0 \leq 3.2$ and $0.01 \leq \alpha \leq 0.04$. We adopted a fixed number of live points $\nlive = 2000$, as it ensured enough sampling points for the given problem. We allowed a number of clusters $1 \leq \nclust \le 4$, which proved to be enough to sample possible degeneracies in the PPD. Moreover, we set $\matt = 10^4$, $\minit = \nlive$, and $\msame = 50$, which is similar to the case of the background fitting presented in Sect.~\ref{sec:bkg}. The typical amount of final sampling points when analyzing each chunk of PSD settles to $\sim\,40,000$.

\begin{figure*}
   \centering
    \includegraphics[height=6.9cm]{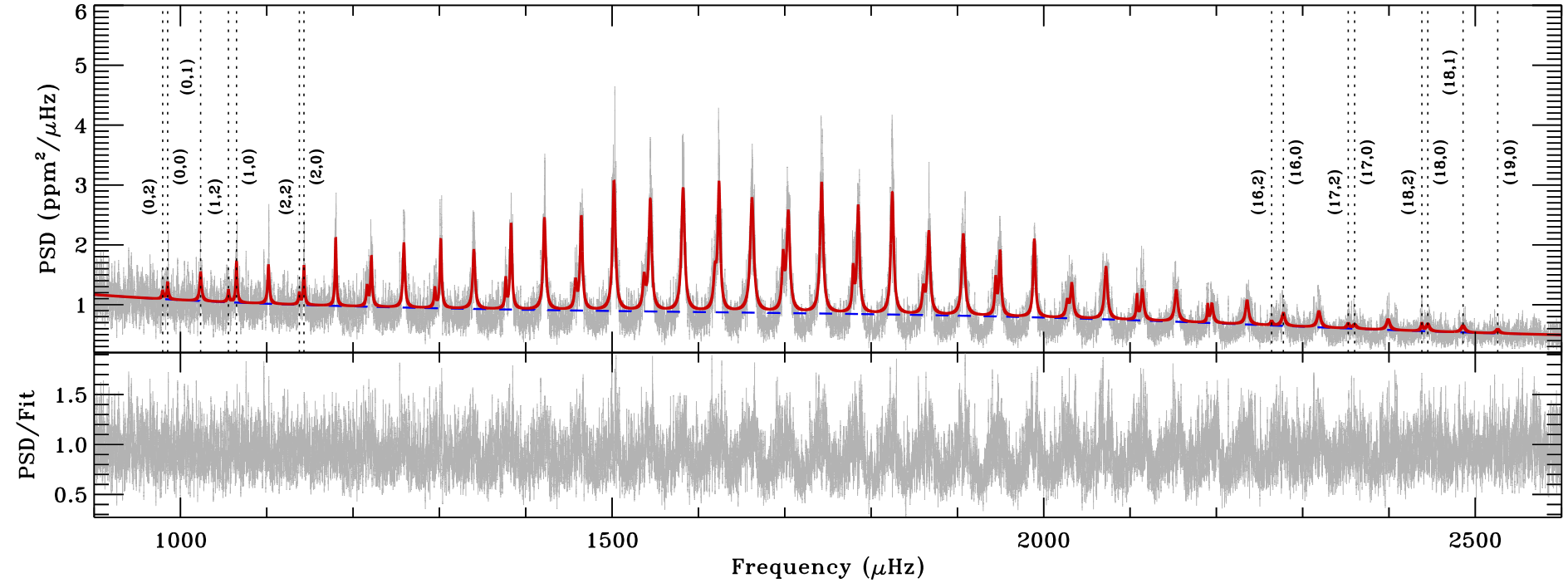}
      \caption{ \textit{Upper panel}: resulting peak bagging best-fit for KIC~9139163 as derived by means of \diamonds\,\,by using Approach 1 based on the background that is estimated in Sect.~\ref{sec:bkg} (red thick line) overlaid on the PSD smoothed by $0.25\,\mu$Hz (gray). The mean background level is shown as a dashed blue line. Dotted vertical lines mark the oscillation peaks for which the detection probability is below 99\,\% with labels indicating the corresponding peak identification, as reported in Table~\ref{tab:frequencies} and as explained in Appendix~\ref{sec:results2}. \textit{Lower panel}: ratio between the smoothed PSD and the resulting red line fit that is shown in the upper panel.}
    \label{fig:results1}
\end{figure*}

\begin{figure}[tb]
   \centering
    \includegraphics[height=7.5cm]{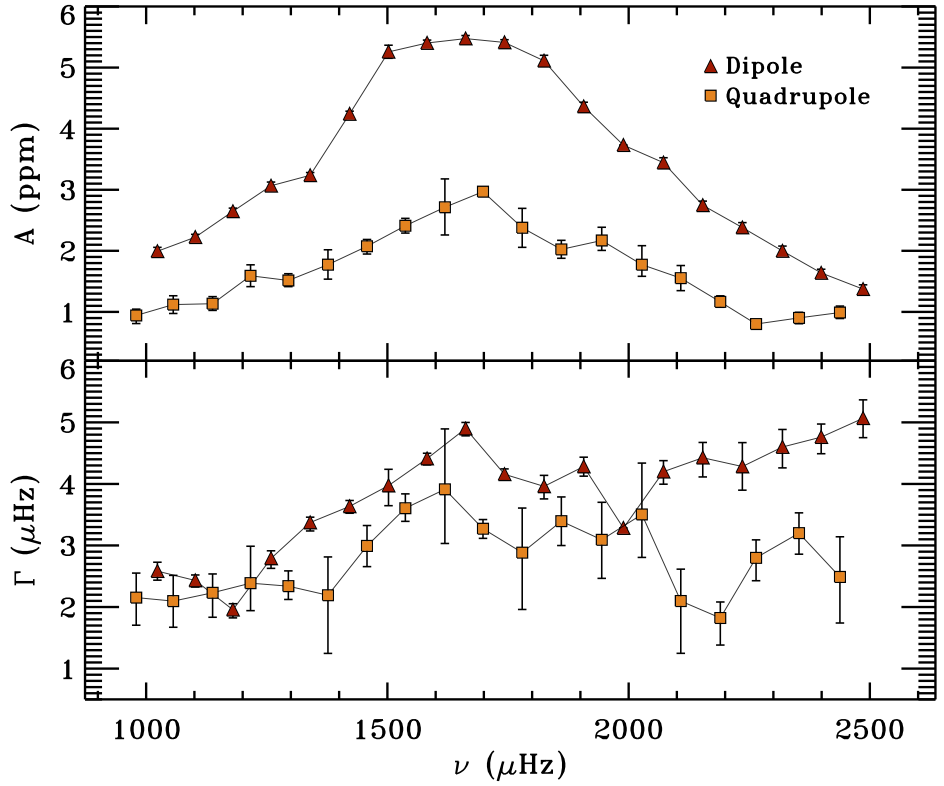}
      \caption{Oscillation parameters for the dipole (red triangles) and quadrupole (orange squares) peaks of KIC~9139163 as derived by means of \diamonds\,\,by using Approach 1 based on the background derived in Sect.~\ref{sec:bkg}, for both amplitudes $A$ (upper panel) and linewidths $\Gamma$ (lower panel), as a function of the frequency position $\nu$. All the results shown are the mean estimates from the Bayesian inference with corresponding 68.3\,\% credible intervals overlaid.}
    \label{fig:results2}
\end{figure}

The best-fit model based on Approach 1 is plotted in the upper panel of Fig.~\ref{fig:results1}, which is overlaid to the PSD of KIC~9139163 and smoothed by $0.25\,\mu$Hz for a visual comparison. The ratio between the smoothed PSD and the resulting fit is shown in the lower panel of the same figure.
For clarity, the oscillation peaks that have a detection probability lower than 99\,\% are indicated with mode identification $(\mbox{Peak}\,\#, \ell)$ (see Appendix~\ref{sec:results2}). The background level shown corresponds to the mean background, as given in Eq.~(\ref{eq:overall_pb}). Single dipole, and quadrupole oscillation parameters are shown in Fig.~\ref{fig:results2} for both amplitudes and linewidths. As it appears clear from the plot, the linewidths of the dipole peaks increase with the frequency position, while those of the quadrupole peaks are higher toward the center of the oscillation envelope, although they have a much larger scatter. Conversely, amplitudes visibly resemble the Gaussian envelope of the oscillation pattern for both angular degrees with a maximum occurring around $\numax$. The $\ell = 1$ and 2 frequencies derived by means of Approach 1 agree within 0.1\,\% and 0.3\,\%, respectively, with those given by A12 for the full set of peaks. 

\begin{figure}
   \centering
    \includegraphics[height=7.5cm]{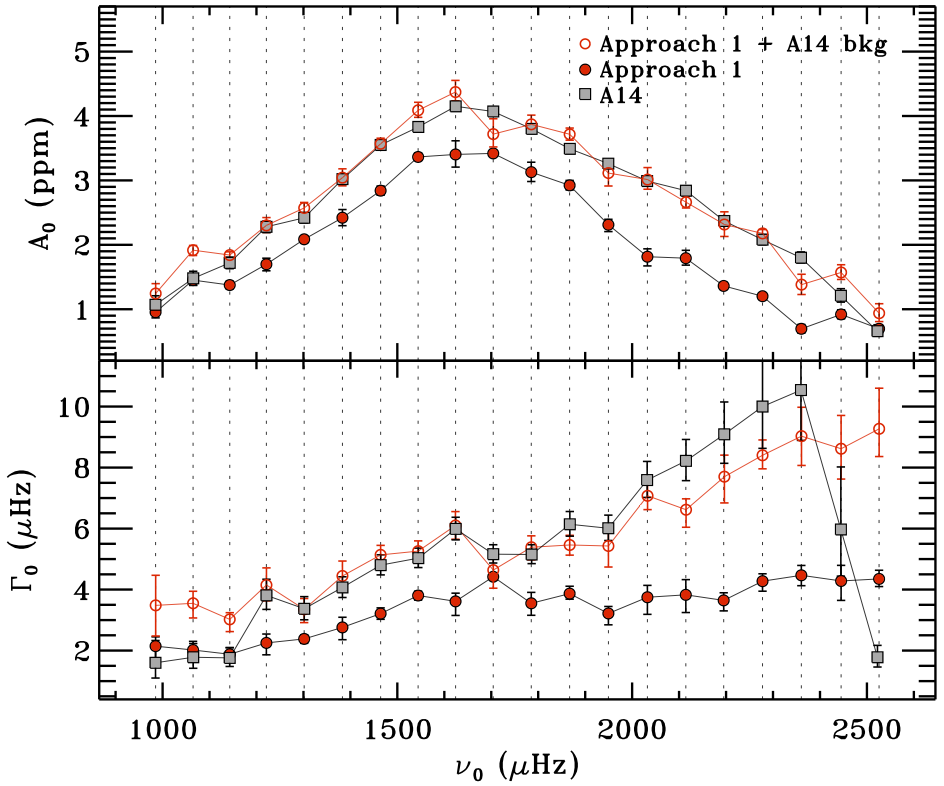}
      \caption{Comparison of the oscillation parameters for the radial peaks of KIC~9139163 between the results derived in this work by means of \diamonds\,\,(filled red circles, Approach 1 based on the background derived in Sect.~\ref{sec:bkg} and, open red circles, Approach 1 based on A14 background) and those provided by A14 (gray squares) for both amplitudes $A_0$ (upper panel), and linewidths $\Gamma_0$ (lower panel), as a function of the frequency position $\nu_0$. The red circles represent the mean estimates coming from the Bayesian inference with their corresponding 68.3\,\% credible intervals, while gray squares are the mean estimates from the MLE fit done by A14 and plotted with their error bars. The $A_0$ values from Approach 1 shown in the plot are obtained by scaling down by a factor $\sqrt{2}$ those reported in Tables~\ref{tab:frequencies} and \ref{tab:frequenciesApp} to be consistent with the definition of amplitude indicated by A14. Dotted vertical lines mark the frequency position of the red circles for a better visual comparison with that of the gray squares.}
    \label{fig:comparison_App}
\end{figure}

For the case of the radial peaks derived with Approach 1, which is based on the background derived in Sect.~\ref{sec:bkg}, we instead refer to Fig.~\ref{fig:comparison_App}, where their amplitudes $A_0$ and linewidths $\Gamma_0$ as a function of their frequency position $\nu_0$ are directly confronted to those obtained by A14. The comparison underlines the presence of a significant difference in amplitude (on average $\sim$24\,\%) and in linewidth (on average $\sim42$\,\%), especially in the high-frequency side of the oscillation envelope ($\nu > 1900\,\mu$Hz, see Sect.~\ref{sec:discussion} for more discussion), which is caused by the different background adopted in this work with respect to that used by A14. The frequencies of the radial peaks estimated in this work and marked in Fig.~\ref{fig:comparison_App} instead agree within 0.2\,\% with those by A14 and within 0.7\,\% with the previous set by A12 throughout all the frequency range.

To test the effect of a different background and compare our results to existing ones in a more favorable way, we have additionally performed the peak bagging of KIC~9139163 by adopting the background used in A14, thus applying Approach 1 to the same oscillation peaks as those presented in Table~\ref{tab:frequencies}. The results are presented in Appendix~\ref{sec:results3} and listed in Table~\ref{tab:frequenciesApp}. As shown in Fig.~\ref{fig:comparison_App}, amplitudes and linewidths are now in good agreement with the previous results by A14 (up to 0.5\,\%), which is well within the reported error bars for most of the cases.

\subsection{Exploiting the multi-modality of Peak Bagging}
\label{sec:multi_modal_pb}
In the method termed Approach 1, described in Sect.~\ref{sec:pb} and Sect.~\ref{sec:significance}, the number of dimensions of the parameter space scales linearly with the number of peaks in the PSD of the star. For example, fitting one peak in a narrow part of the PSD implies searching for one maximum of the PPD in a 3D parameter space ($\Gamma$, $\nu_0$, $A$). Fitting two peaks  simultaneously implies searching for one maximum of the PPD in a 6D parameter space ($\Gamma_1$, $\nu_{0,1}$, $A_1$, $\Gamma_2$, $\nu_{0,2}$, $A_2$) and so on. In principle, one could try fitting one Lorentzian peak using a large part of the PSD and hope that the individual oscillation peaks pop up as several local maxima in the PPD. The advantage would clearly be a strongly reduced dimensionality of the parameter space to only 3D, irrespective of the number of oscillation peaks in the PSD. In practice, such an approach hardly works, because of a number of problems:

\begin{enumerate}
\item The fitting algorithm can have the tendency to consider the entire spectrum as one peak with an extremely large width.
\item The approach is prone to false maxima as it may consider local noise peaks also as an oscillation peak. 
\item Sampling a multi-modal parameter space and not missing local maxima is hard, especially with MCMC algorithms. 
\item Even if we do sample all the local maxima in the parameter space, it is still unclear how to extract the fit parameters for each oscillation peak individually. 
\end{enumerate}
In this section, we show how each of these problems can be solved, and we refer to the resulting method as \emph{Approach 2}. This approach can be very useful in the process of automating the peak bagging analysis, thus providing a possible solution for performing the analysis on a large number of targets.

The first problem is easily solved by imposing a physically reasonable prior on the linewidth, so that the algorithm never diverges. Although the third problem, is hard to handle with standard MCMC algorithms, it is a lot easier with \diamonds, as shown in the demos presented in Sect.~\ref{sec:demos}. We solved the second and fourth problems by letting the algorithm search for (i.e. fit) a pattern of three Lorentzian profiles, instead of fitting a single Lorentzian profile. Concretely, the model now becomes 
\begin{equation}
P_\mathrm{osc} \left( \nu \right) = M_1 \left( \nu \right) + M_2 \left( \nu \right) + M_0 \left( \nu \right)\, 
\label{eq:islands_pb_model}
\end{equation}
where $M_i$ corresponds to a Lorentzian profile of an oscillation peak with degree $\ell = i$. This corresponds to nine fit parameters: ($\Gamma_1$, $\nu_{1}$, $A_1$, $\Gamma_2$, $\nu_{2}$, $A_2$, $\Gamma_0$, $\nu_{0}$, $A_0$). We replace, however, the parameters $\nu_0$ and $\nu_2$ with the frequency spacings $\delta_{10} \equiv \nu_{n,0} - \nu_{n-1,1}$, which is analogous to the definition given in Eq.~(\ref{eq:d01}), and $\delta_{02} \equiv \nu_{n,0} - \nu_{n,2}$, as introduced in Eq.~(\ref{eq:d02}), hence having $\nu_0 = \nu_1 + \delta\nu_{10}$ and $\nu_2 = \nu_0 - \delta\nu_{02}$. The reason is that it is much easier to physically specify constraining priors for these frequency spacings within one radial order than it is for the individual frequencies $\nu_0$ and $\nu_2$. In summary, the frequency of the dipole peak, $\nu_1$, is the fit parameter used to locate the 3-peak pattern in the power spectrum, and the spacings $\delta\nu_{10}$ and $\delta\nu_{02}$ are used as fit parameters to determine the distances between the peaks inside the pattern. Each oscillation peak inside the 3-peak pattern has a separate fit parameter for the amplitude and the width, resulting in still nine fit parameters. The 9D posterior distribution shows local maxima, one for each 3-peak pattern identified in the PSD, which is one for each radial order. We found that fitting such a pattern is much less susceptible to noise than fitting a single Lorentzian profile.

We note what we did \emph{not} impose: we did not specify the number of radial orders in the PSD. This is an output of the algorithm: each local maximum detected in the PPD corresponds to an observed radial order in the PSD. We also did not impose the condition that the radial (or dipole) peaks in the PSD should be equidistant nor that they may show an oscillatory behavior due to acoustic glitches. This also will be an output of the algorithm: the different maxima found in the 9D PPD will show up as equidistant along the $\nu_1$ axis, hence disentangling the different radial orders without ambiguity. 

\begin{figure}
   \centering
    \includegraphics[height=8.4cm]{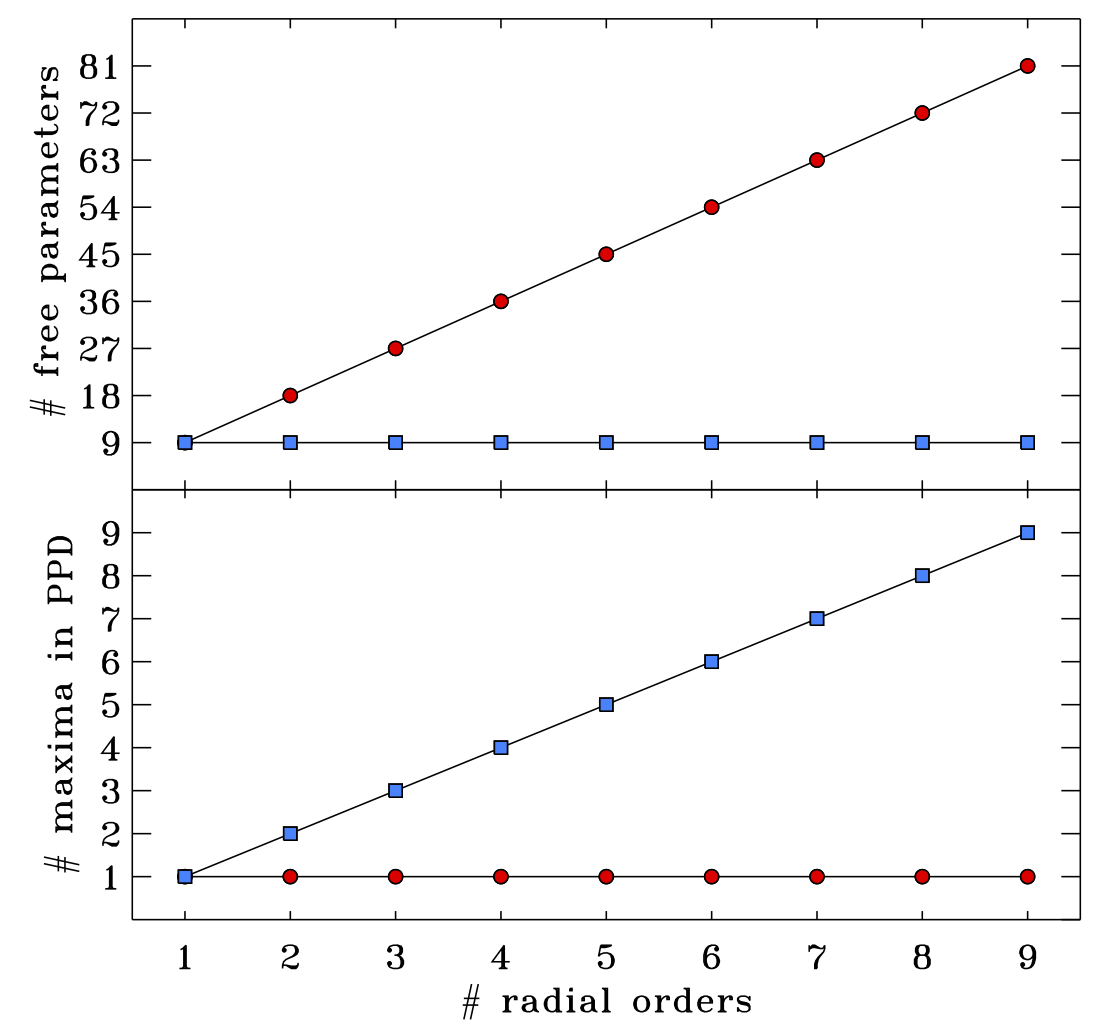}
      \caption{Illustrative comparison between Approach 1 (red circles) and Approach 2 (blue squares) in terms of the number of free parameters used in the fit (top panel) and the number of local maxima identified in the final PPD (bottom panel), as a function of the number of radial orders considered.}
    \label{fig:a1_v_a2}
\end{figure}

To highlight two substantial differences between Approach 1 and Approach 2, as discussed in this section, we refer to an illustrative comparison shown in Fig.~\ref{fig:a1_v_a2}. In the top panel, we can observe the linear increasing trend of the number of fit parameters when using Approach 1 over an increasing number of radial orders of the star (nine new free parameters added for each new radial order included), as opposed to the constant number of nine free parameters used by Approach 2. In the bottom panel, we see that the number of local maxima obtained in the PPD increases linearly with Approach 2 when used over an increasing number of radial orders (one local maximum more for every additional radial order), leading to a multi-modal distribution when at least two radial orders are considered. The PPD remains instead uni-modal when using Approach 1, irrespective of the number of radial orders included in the fit. It is then clear that the two approaches have a reversed balance in terms of the number of free parameters and number of local maxima in the PPD, while they become essentially identical for the limiting case of one radial order.

\begin{figure*}
   \centering
   \includegraphics[width=18.5cm]{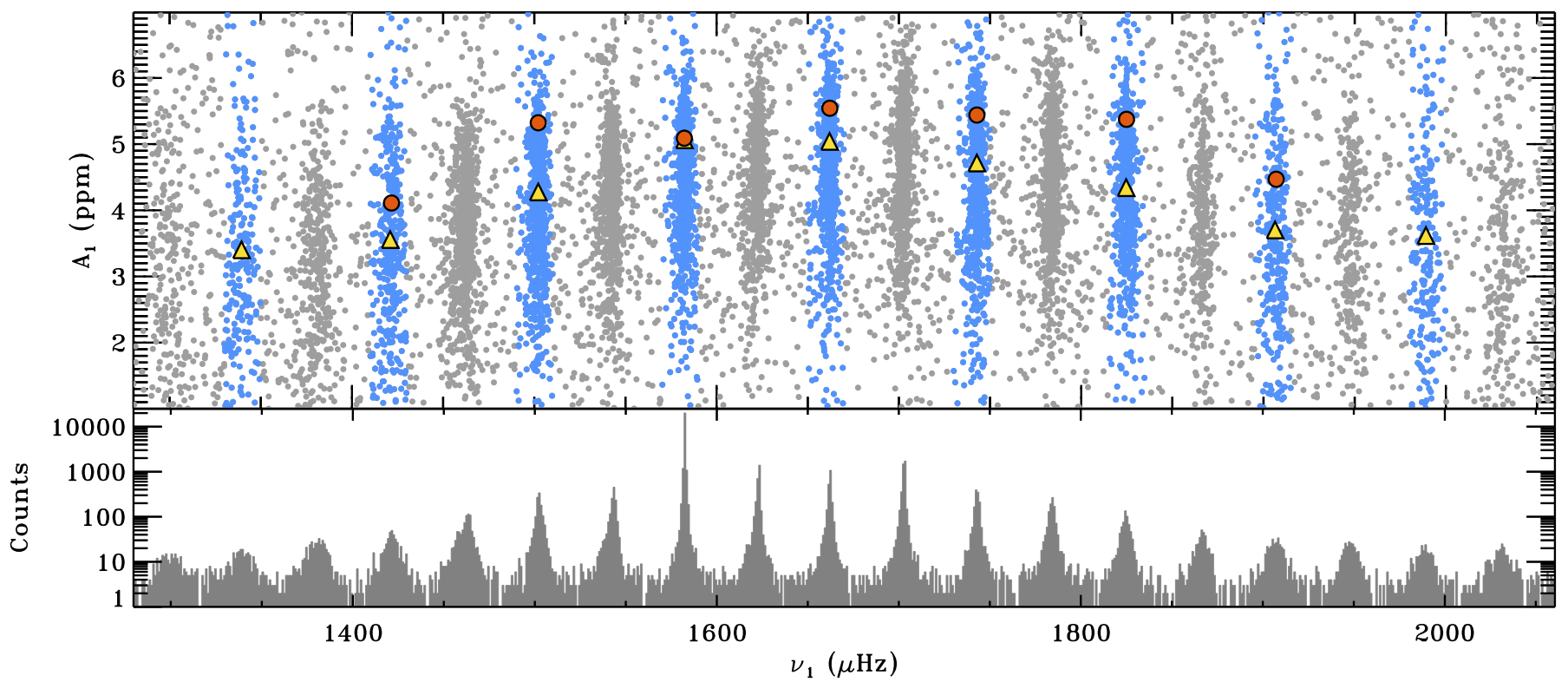}
      \caption{Resulting 43374 samples obtained by \diamonds\,\,by using Approach 2 applied to the PSD of KIC~9139163 in the frequency range $1280$-$2090\,\mu$Hz, covering nine radial orders. \textit{Upper panel}: distribution of the sampling points for the amplitude of dipole peaks, $A_1$ and the corresponding frequency position, $\nu_1$. The islands marked in light blue represent the true positions of the dipole peaks, while those in gray are relative to the positions of the blended quadrupole-radial peaks, which are ignored. The red circles represent the mean Bayesian estimates derived from the MPDs of the light blue islands, while yellow triangles show the corresponding sampling average positions in both coordinates and are not related to the MPDs. \textit{Lower panel}: the corresponding histogram density along the direction of $\nu_1$, showing the number of counts occurring in each island.}
    \label{fig:amplitude_islands}
\end{figure*}

We tested Approach 2 by applying it to the PSD of KIC~9139163 in a frequency range containing nine radial orders (27 peaks), spanning $\sim800\,\mu$Hz and centered around $\numax$. For configuring the code, we adopted the parameters $f_0 = 0.6$, $\alpha = 0.04$, $\nlive = 2000$, $1 \leq \nclust \leq 10$, $\matt = 10^4$, $\minit = 500$, and $\msame = 50$.
Figure~\ref{fig:amplitude_islands} shows a sampling result of this multi-modal application, where we plot the fitted amplitude for dipole peaks, $A_1$, against $\nu_1$, with an histogram density along the coordinate $\nu_1$. As it appears evident from the plot, the different islands sampled by \diamonds\,\,are regularly spaced in frequency, which clearly corresponds to the regular Tassoul comb-pattern of the solar-like oscillations observed in KIC~9139163. In particular, we identify two types of islands, those corresponding to the true position of the dipole peaks and those that occur where the duplets $\ell = 2,0$ arise, in which the latter case is to be discarded, since it does not match the definition of the free parameter $\nu_1$. 
The result that the algorithm confuses $\ell=1$ peaks with $\ell=0$ peaks is particular for KIC~9139163, or for F-type solar-like oscillators in general, where the $\ell=2$ and $\ell=0$ peaks are blended, which makes it more difficult to recognize the typical Tassoul pattern. The position of the light blue islands along the direction of $A_1$ decreases when moving from the center of the plot --- i.e. the maximum power region --- towards the wings, which agrees with the typical Gaussian modulation for the amplitudes of solar-like oscillations. We also notice that the left- and right-side light blue islands do not show an estimated Bayesian mean, though the position of the corresponding oscillation peaks can still be obtained by considering the sampling average position of the associated islands. This occurs because the side MPDs have a probability that is zero, as compared to that of the central MPDs, therefore, no Bayesian estimates could be computed for this case. As one can observe from the bottom panel of Fig.~\ref{fig:amplitude_islands}, the central peaks having higher amplitudes tend to be sampled about 100 times more than those falling in the wings of the range and having a lower amplitude. Since higher amplitude peaks produce higher likelihood regions in the parameter space, the NSMC algorithm tends to sample them more densely than the remainder regions that have lower likelihood values and, consequently for this case, lower posterior probabilities. 

\begin{figure*}
   \centering
   \includegraphics[width=6.0cm]{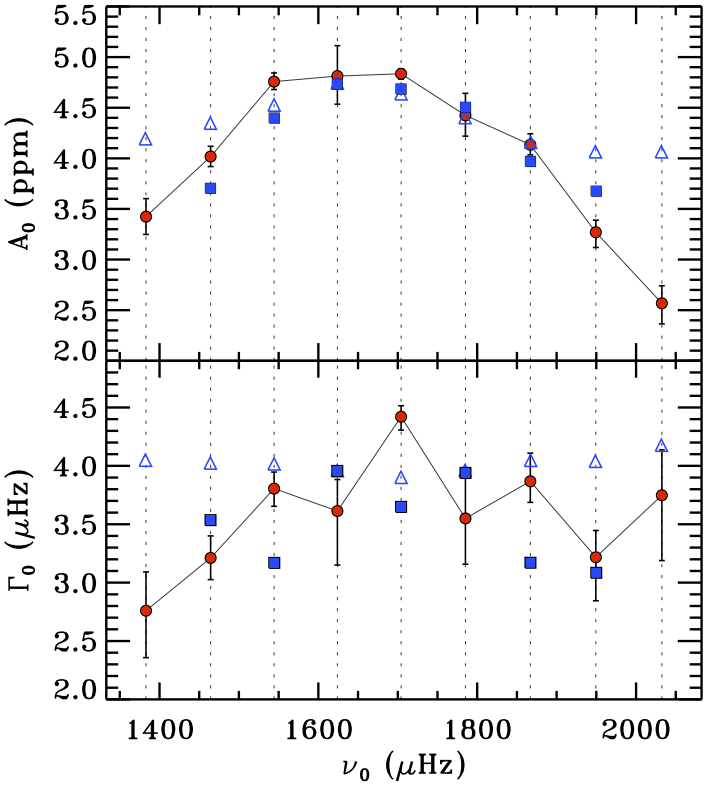} \includegraphics[width=6.0cm]{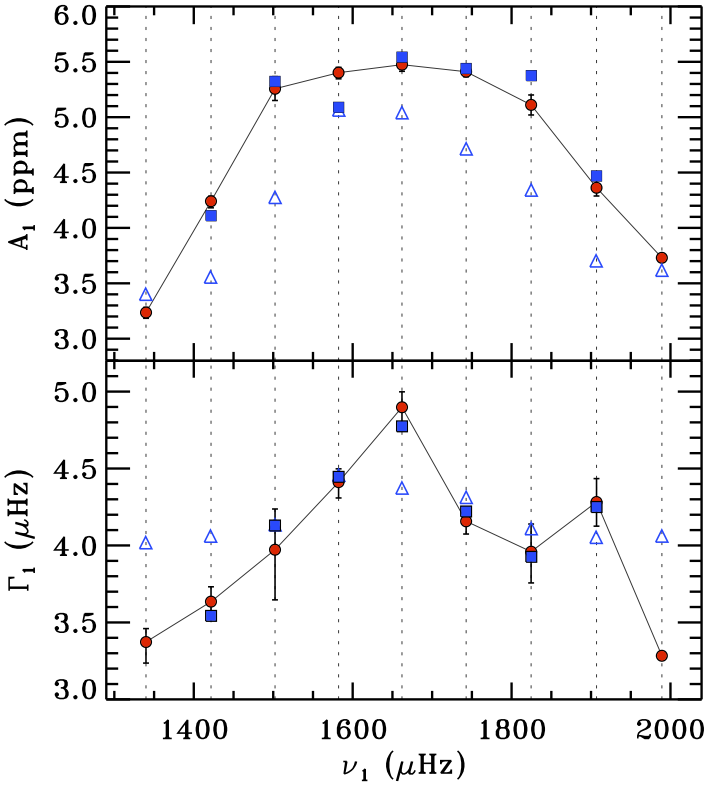} \includegraphics[width=6.0cm]{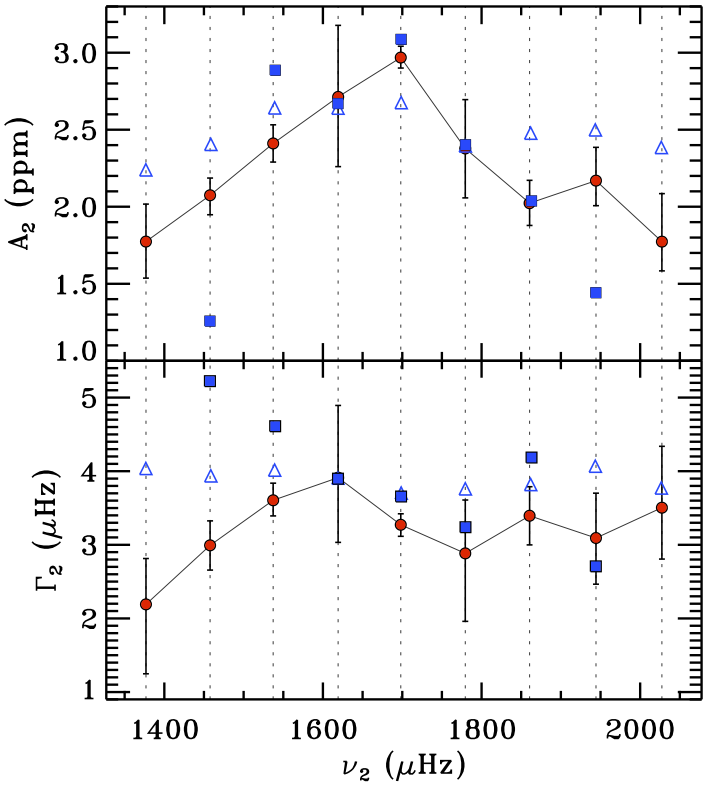}
      \caption{Comparison of the results for KIC~9139163 relative to the nine radial orders in the frequency range $1280$-$2090\,\mu$Hz, obtained by \diamonds\,\,according to Approach 1 (red circles) and Approach 2 (blue squares). Panels from left to right show the amplitudes (upper panels) and linewidths (lower panels) against the frequency position of the full set of radial, $\nu_0$, dipole, $\nu_1$, and quadrupole oscillations, $\nu_2$, respectively. The red circles are the mean values as derived from the MPDs with their corresponding 68.3\,\% shortest credible intervals, while the blue squares are the mean values from the MPDs of each blue island (see Fig.~\ref{fig:amplitude_islands}). The open triangles are the sample average positions (see Fig.~\ref{fig:amplitude_islands}) used for providing a frequency estimate for the two side peaks of each panel. Dotted vertical lines represent the frequency position of the red circles, as shown as a reference for the blue squares.}
    \label{fig:comparison_islands}
\end{figure*}

Figure~\ref{fig:comparison_islands} shows the resulting estimates, obtained by using Approach 2, for all the peaks compared to those derived by means of Approach 1, with sample average positions for each light blue island overlaid for completeness. The frequencies estimated by means of Approach 2 agree within 0.1\,\% (for radial and dipole peaks) and within 0.2\,\% (for quadrupole peaks) to those obtained by using Approach 1. Concerning amplitudes and linewidths instead, the agreement is more evident for the dipole peaks (where amplitudes match up to 0.5\,\% and linewidths up to 0.8\,\%), while it tends to be worse for radial and quadrupole peaks, especially in the wings, which produces discrepancies that can be up to 12\,\% and 40\,\% in amplitude and 18\,\% and 75\,\% in linewidths, respectively. These differences mainly rely on the result that Approach 2 is supposed to recognize the 3-peak pattern $\ell = 1, 2, 0$, which for KIC~9139163 is often ambiguous, as already pointed out before. In addition, the total number of samples obtained for the lowest-amplitude modes in the multi-modal PPD might be too low to ensure accurate Bayesian estimates.

\section{Discussion}
\label{sec:discussion}
The fitting of the background models for KIC~9139163 is a challenging problem in terms of both computational speed and sampling efficiency of the resulting PPD. The estimated set of free parameters for the favored model that incorporates both granulation and faculae activity signal corresponds to a non-trivial combination of the different background components. This is mainly caused by the rising of the faculae component inside the region that contains the solar-like oscillations, which itself has not a prominent height if compared to the background level of the star. This leads to a SNR $\sim$\,1.3 as measured from the PSD around $\numax$. The oscillation signal in the observed PSD is noisier due to the higher temperature of this target if compared to the solar value. The rising of the faculae signal beyond 1000\,$\mu$Hz agrees with the interpretation of the stellar background proposed by \cite{Harvey93} and by \cite{Karoff12}. In addition, the resulting Harvey-like profiles of both granulation and faculae activity are consistent with the findings by \cite{Karoff13} for other pulsating MS stars, for which we expect the faculae component to decay more rapidly with respect to that of the granulation. The decaying coefficients $c_\mathrm{gran}$ and $c_\mathrm{fac}$ estimated in this work, however, exceed by about 23\,\% and 200\,\%, respectively, from those measured for the Sun by \cite{Karoff13}. The detection of a faculae signal in KIC~9139163 is even consistent with the presence of a stellar activity, as detected by \cite{Karoff13Act}, while the estimated $\numax$ from the global background fitting remains compatible within 0.5\,\% with the value indicated by A12. 

The selection of a proper background model plays a crucial role in estimating the asteroseismic parameters that characterize the individual oscillation peaks of the star. As outlined by A14, oscillation amplitudes and linewidths heavily depend on the background model adopted. A clear example of this behavior is depicted in Fig.~\ref{fig:comparison_App} for the effect produced on the radial peaks (see Sect.~\ref{sec:results} and below for more discussion). The model comparison of the different background models, as performed by means of the Bayesian evidence computed by \diamonds\,\,(see Sect.~\ref{sec:bkg}) has led to an unambiguous interpretation of the background. This addressed the problem of the choice of the background model and proved that the method could be applied very straightforwardly with an immediate result.

Concerning the outcomes of the peak bagging analysis based on Approach 1, as described in Sects.~\ref{sec:pb} and \ref{sec:significance}, we stress that all the frequencies derived in this work (see Tab.~\ref{tab:frequencies}) match remarkably well with those obtained in previous works by A12 and A14, showing an agreement level well below 1\,\% (see Sect.~\ref{sec:results} for more details). Significant discrepancies in the fitted amplitudes and linewidths of the radial oscillation peaks with respect to those measured by A14 are observed, especially for the region of the oscillation envelope beyond $\numax$ (see Fig.~\ref{fig:comparison_App}). These new measurements from Approach 1 are on average $\sim$1.36 and $\sim$1.54 times smaller for amplitudes and linewidths respectively, as compared to the values provided by A14. As pointed out already, this relies on the different background adopted for this star, which is that of the granulation signal in the case of A14. The Harvey-like profile related to the faculae activity fitted in this work arises around $2000\,\mu$Hz, as also clearly visible from Fig.~\ref{fig:bkg}. This is confirmed by the results presented in Table~\ref{tab:frequenciesApp}, for which Approach 1 was adopted using a similar background fit to that considered in A14. The approach provided amplitude and linewidth estimates that are in good agreement with those reported by A14 (see Fig.~\ref{fig:comparison_App} and Sect.~\ref{sec:results} for more details). Moreover, as observed in Fig.~\ref{fig:comparison_App}, the increasing trend of the linewidths of the radial peaks with their frequency position tends to flatten in the region around $\numax$. This so-called linewidth depression, expected for radial peaks (e.g. see \citealt{Belkacem11} for a theoretical explanation) agrees with what observed by A14 for the same star (Fig.~\ref{fig:comparison_App}, lower panel). Finally, Approach 1 has led to a final set of 14 radial, 18 dipole, and 13 quadrupole peaks --- as shown in Table~\ref{tab:frequencies} from $(3,0)$ to $(16,0)$, from $(1,1)$ to $(17,1)$ and from $(3,2)$ to $(16,2)$ --- whose detection probability is $p \geq 99$\,\%. The remainder peaks, except for $(2,19)$, which was not detected, are marked in Fig.~\ref{fig:results1} and can still be considered valid, though keeping in mind that their detection probabilities are lower than 99\,\% and even below 50\,\% in some cases. Besides, the error bars on the frequencies appear to be about five times smaller than those provided by A12 and from 1.5 to 4 times smaller than those reported by A14 for amplitudes and linewidths. This is caused by the use of a longer dataset and also by their smaller values obtained in this work as compared to those reported by A14 in the particular case of the linewidths \citep[see][for more details on the behavior of the error bar of the linewidth for a single oscillation]{Toutain94}.

We described a novel peak bagging method (Approach 2) in Sect.~\ref{sec:multi_modal_pb} (see Fig.~\ref{fig:a1_v_a2} and Fig.~\ref{fig:amplitude_islands}, upper panel) based on multi-modality. The capability of the code to sample multi-modal PPDs has proven to be of great help for reducing the dimensionality of the peak bagging analysis, which succeeds in sampling 27 different oscillation peaks by means of only $k = 9$ free parameters. In particular, Approach 2 has allowed us to retrieve Bayesian estimates of the frequency position of 21 out of 27 peaks in the frequency range $1280$-$2090\,\mu$Hz, which result in good agreement (well within 1\,\%) with those derived by means of Approach 1 for all the three different angular degrees. The concordance weakens for the amplitudes and linewidths of radial and quadrupole peaks (see Fig.~\ref{fig:comparison_islands}, left and right panels respectively), as detailed in Sect.~\ref{sec:multi_modal_pb}, while still ensuring accurate estimates for the dipole modes (Fig.~\ref{fig:comparison_islands}, central panels). Nevertheless, the method is sensitive to the configuration of the code parameters related to both the clustering and the ellipsoidal sampling algorithms. The process of identifying clusters in the PPD is much more tapped for Approach 2 (multi-modal) than for Approach 1 (uni-modal). In addition, as already described in Sect.~\ref{sec:multi_modal_pb}, the NSMC algorithm samples more frequently those regions in the PPD having a higher likelihood value, which causes the final sampling to be sparser for those peaks that have a lower amplitude and fall in the wings of the oscillation envelope (see Fig.~\ref{fig:amplitude_islands}, lower panel). One feasible choice to overcome this problem is to avoid sampling peaks having too different amplitudes from one another, hence subdividing the process in chunks of PSD where the different oscillations are similar in terms of asteroseismic properties. In general, however, the merging of several runs, as explained in Sect.~\ref{sec:parallelization}, tends to increase the total number of samples for all the islands, hence giving the possibility to improve the results also for the side peaks.

Importantly, Approach 2 can be further improved, optimized, and used for the automatization of the peak bagging analysis in MS stars exhibiting solar-like oscillations by adopting a two-step approach: (i) quickly estimating the asteroseismic parameters of several oscillation peaks by means of Approach 2, since it requires a low number of free parameters with a low constraint level of their corresponding prior PDFs, which can in turn be derived through simple asteroseismic scaling relations; (ii) fitting the PSD by using Approach 1, or another similar to it, where the prior PDFs are now set by the outcomes of Approach 2, to end up with more accurate results.

\section{Conclusions}
\label{sec:conclusions}
Based on the descriptions given in Sec.~\ref{sec:code}, we can summarize the new features implemented in \diamonds\,\,as follows:
\begin{itemize}
\item A revised SES algorithm based on the clustering of the live points that allows more control and speed in the drawing process of the NSMC
\item The possibility to use different types of prior PDFs (either uniform, normal and super-Gaussian, or new user-defined types) for any free parameter of the inference problem and in any possible combination
\item The inclusion of a flexible stopping criterion for the NSMC based on the amount of evidence coming from the remaining set of live points
\item An improved computation of the final evidence based on the statistical work by K11
\item A new relation for the reduction of the live points that allows better control of the reduction process.
\end{itemize}
As proven with the demos presented in Sect.~\ref{sec:demos} and especially with the application given in Sect.~\ref{sec:application}, the \diamonds\,\,code developed for this work shows great potential for fast and efficient Bayesian inferences of challenging high-dimensional and multi-modal problems, such as the peak bagging analysis of an F-type star, hence allowing for a direct model comparison aimed at measuring the reliability of the peak detection. In addition, we demonstrated that \diamonds\,\,is capable of making a valuable alternative to Approach 1 by sampling multiple peaks simultaneously and in terms of only a few free parameters (Approach 2), thus reducing the number of dimensions involved in the fitting process (Fig.~\ref{fig:a1_v_a2}) with a considerable gain in both computational power and speed. The automatization of the peak bagging analysis for the measurement of detailed asteroseismic parameters in thousands of observed pulsating low-mass MS stars will be crucial for testing theoretical models of stellar evolution and of asteroseismic inversion that is aimed at probing stellar structure. Therefore, \diamonds\,\,gives us the possibility to progress in the analysis of a large ensemble of asteroseismic datasets for the first time.

\begin{acknowledgements}
The research leading to these results has received funding from the European
Research Council under the European Community's Seventh Framework Programme
(FP7/2007--2013) ERC grant agreement n$^\circ$227224 (PROSPERITY) and PEOPLE-IRSES n$^\circ$269194 (ASK), 
from the Fund for Scientific Research of Flanders (G.0728.11), 
and from the Belgian federal science policy office (C90291 Gaia-DPAC).
EC acknowledges Rafael A. Garc{\'{\i}}a for providing the new corrected \kepler data, Tiago Campante and Paul Beck for useful discussions that helped in improving the paper. The authors wish to thank the referee Thierry Appourchaux for the useful comments.
\end{acknowledgements}

\bibliographystyle{aa} 
\bibliography{biblio} 

\appendix
\section{Results for peak bagging of KIC~9139163}
\label{sec:results2}
All oscillation frequencies, amplitudes and linewidths, with the peak significance probabilities are listed in Table~\ref{tab:frequencies}. The peaks are labeled with increasing integer numbers, moving from low to high frequency, for each different angular degree, and use the same labels in both Tables. This, therefore, allows the reader to easily identify the single oscillation and all the corresponding information by using the notation $\left(\mbox{Peak}\,\#, \ell\right)$. The results presented in Table~\ref{tab:frequencies} refer to the use of Approach 1 based on the background derived in Sect.~\ref{sec:bkg}, thus it includes the faculae activity component.

We note that we do not list any detection probability $p_\mathrm{A}$ (which can possibly be assumed zero) for some peaks $\ell = 2, 0$ for which we performed the model comparison between models $\model_\mathrm{C}$ and $\model_\mathrm{B}$. This is because they have a high SNR, and, therefore, assessing their significance over the background level was not needed. In addition, all the peaks labeled with the same number and having angular degrees $\ell = 0$ and $2$ have the same peak significance analysis. This happens since the model comparison is performed for the duplet 2-0 and not for the single peaks, separately.  Whereas the model comparison favored the detection of one peak only (model $\model_\mathrm{B}$), namely the $\ell = 0$ peak, or alternatively just the background component (model $\model_\mathrm{A}$), which has no peaks, it is also important to stress that we decided to show the resulting fit for the model fitting the duplet 2-0 (model $\model_\mathrm{C}$), since the two peaks cannot be disentangled from one another and their characteristic parameters obtained from the fit might still be useful for subsequent investigations. When model $\model_\mathrm{C}$ is favored over $\model_\mathrm{B}$, this results in detection probabilities $p_\mathrm{C} > 0.5$ and $p_\mathrm{B} < 0.5$. Conversely, when the model $\model_\mathrm{B}$ ought to be preferred, the corresponding detection probabilities become $p_\mathrm{C} < 0.5$ and $p_\mathrm{B} > 0.5$. The same rule applies to the detection probabilities $p_\mathrm{A}$ and $p_\mathrm{B}$. The reader can decide whether to consider the quadrupole peak (or even the radial one) or not if the corresponding detection probability $p_\mathrm{C}$ (or even $p_\mathrm{B}$) is too low (e.g. below 50\,\%).

Lastly, for most of the dipole peaks, we did not perform any peak significance analysis thanks to their high amplitude (ranging from $\sim$\,3.5 to $\sim$\,1.5 times that of the lowest amplitude peaks), which makes their detection unambiguous. For the dipole peaks falling in the wings of the oscillation envelope, we instead computed the model comparison between models $\model_\mathrm{B}$ and $\model_\mathrm{A}$, hence listing the detection probabilities $p_\mathrm{A}$ and $p_\mathrm{B}$. With similar arguments to those adopted for the duplets 2-0, we show the result of the model having the dipole peak included, $\model_\mathrm{B}$, even when it is disfavored over its competitor when just the background level is considered.

\begin{table*}
\caption{Median values with corresponding 68.3\,\% shortest credible intervals for the oscillation frequencies, amplitudes, and linewidths of KIC~9139163, as stemmed by \diamonds\,\,for the case of the background model derived in Sect.~\ref{sec:bkg}. The first column represents the peak number in increasing frequency order and is shown for each angular degree (second column). The three last columns correspond to detection probabilities referred to the models $\model_\mathrm{A}$ (only background), $\model_\mathrm{B}$ (one peak), and $\model_\mathrm{C}$ (two peaks), as defined in Sect.~\ref{sec:significance}.}             
\centering                         
\begin{tabular}{l l l l l c c c}       
\hline\hline
\\[-8pt]          
Peak \# & Angular Degree & \multicolumn{1}{c}{Frequency} & \multicolumn{1}{c}{Amplitude} & \multicolumn{1}{c}{Linewidth} & $p_\mathrm{A}$ & $p_\mathrm{B}$ & $p_\mathrm{C}$ \\
 & & \multicolumn{1}{c}{($\mu$Hz)} & \multicolumn{1}{c}{(ppm)} & \multicolumn{1}{c}{($\mu$Hz)}\\
\hline
\\[-8pt]
0 & 0 & $985.10_{-0.37}^{+0.41}$ & $1.35\pm0.12$ & $2.13_{-0.36}^{+0.32}$ & 0.244 & 0.618 & 0.138\\[1pt]
1 & 0 & $1064.94_{-0.14}^{+0.15}$ & $2.06_{-0.12}^{+0.11}$ & $2.00_{-0.29}^{+0.30}$ & 0.000 & 0.982 & 0.018\\[1pt]
2 & 0 & $1143.03_{-0.10}^{+0.09}$ & $1.95_{-0.09}^{+0.07}$ & $1.88\pm0.22$ & 0.000 & 0.291 & 0.709\\[1pt]
3 & 0 & $1221.10\pm0.15$ & $2.40\pm0.14$ &  $2.24_{-0.38}^{+0.30}$ & - & 0.000 & 1.000\\[1pt]
4 & 0 & $1301.54\pm0.09$ & $2.95_{-0.08}^{+0.07}$ & $2.38_{-0.15}^{+0.14}$ & - & 0.002 & 0.998 \\[1pt]
5 & 0 & $1383.01_{-0.19}^{+0.20}$ & $3.43_{-0.18}^{+0.17}$ & $2.74_{-0.38}^{+0.35}$ & - & 0.000 & 1.000\\[1pt]
6 & 0 & $1464.36\pm0.11$ & $4.02\pm0.10$ & $3.21\pm0.19$ & - & 0.000 & 1.000\\[1pt]
7 & 0 & $1544.25\pm0.10$  & $4.76_{-0.08}^{+0.09}$ & $3.80_{-0.15}^{+0.14}$ & - & 0.000 & 1.000\\[1pt]
8 & 0 & $1623.98\pm0.24$ & $4.82\pm0.29$ & $3.57_{-0.42}^{+0.32}$ & - & 0.000 & 1.000\\[1pt]
9 & 0 & $1704.16\pm0.07$ & $4.83\pm0.05$ & $4.41_{-0.10}^{+0.11}$ & - & 0.000 & 1.000\\[1pt]
10 &  0 &  $1785.21_{-0.24}^{+0.23}$ & $5.23_{-0.03}^{+0.04}$ & $4.03_{-0.06}^{+0.05}$ & - & 0.000 & 1.000\\[1pt]
11 & 0 & $1866.91_{-0.14}^{+0.13}$ & $4.14_{-0.11}^{+0.10}$ & $3.88_{-0.20}^{+0.23}$ & - & 0.001 & 0.999\\[1pt]
12 & 0 & $1949.44_{-0.16}^{+0.18}$ & $3.26_{-0.14}^{+0.13}$ & $3.18_{-0.34}^{+0.26}$ & - & 0.000 & 1.000\\[1pt]
13 & 0 & $2032.55_{-0.30}^{+0.37}$ & $2.56_{-0.20}^{+0.18}$ & $3.71_{-0.52}^{+0.43}$ & - & 0.010 & 0.990\\[1pt]
14 & 0 & $2114.28_{-0.40}^{+0.36}$ & $2.54_{-0.16}^{+0.17}$ & $3.81_{-0.57}^{+0.51}$ & - & 0.000 & 1.000\\[1pt]
15 & 0 & $2194.96_{-0.18}^{+0.22}$ & $1.92_{-0.07}^{+0.09}$ & $3.61_{-0.31}^{+0.29}$ & - & 0.000 & 1.000\\[1pt]
16 & 0 & $2277.69_{-0.22}^{+0.21}$ & $1.70_{-0.06}^{+0.08}$ & $4.25_{-0.30}^{+0.27}$ & - & 0.525 & 0.475\\[1pt]
17 & 0 & $2360.29_{-0.52}^{+0.49}$ & $0.98_{-0.09}^{+0.10}$ & $4.47_{-0.35}^{+0.32}$ & 0.056 & 0.387 & 0.557\\[1pt]
18 & 0 & $2445.03_{-0.44}^{+0.43}$ & $1.30_{-0.09}^{+0.11}$ & $4.27_{-0.62}^{+0.53}$ & 0.000 & 0.040 & 0.960\\[1pt]
19 & 0 & $2525.94_{-0.29}^{+0.35}$ & $0.98\pm0.08$ & $4.35_{-0.25}^{+0.28}$ & 0.704 & 0.296 & 0.000\\[1pt]
 \hline
 \\[-8pt]
 0 & 1 & $1023.36_{-0.14}^{+0.13}$ & $1.99 \pm 0.07$ & $2.58_{-0.15}^{+0.14}$ & 0.056 & 0.944 & -\\[1pt]
 1 & 1 & $1101.90\pm0.09$ & $2.22\pm0.05$ & $2.43_{-0.11}^{+0.09}$ & 0.000 & 1.000 & - \\[1pt]
 2 & 1 & $1179.80\pm0.06$  & $2.64\pm0.06$ &  $1.94_{-0.12}^{+0.11}$ & - & - & -\\[1pt]
 3 & 1 & $1258.80_{-0.07}^{+0.09}$ & $3.06\pm0.06$ & $2.77_{-0.14}^{+0.15}$ & - & - & -\\[1pt]
 4 & 1 & $1339.86_{-0.06}^{+0.08}$ & $3.23_{-0.04}^{+0.06}$ & $3.36_{-0.12}^{+0.10}$ & - & - & -\\[1pt]
 5 & 1 & $1421.64\pm0.06$ & $4.24_{-0.06}^{+0.04}$ & $3.63_{-0.11}^{+0.10}$ & - & - & -\\[1pt]
 6 & 1 & $1502.23\pm0.12$  & $5.26_{-0.11}^{+0.10}$ & $3.96_{-0.32}^{+0.27}$ & - & - & -\\[1pt]
 7 & 1 & $1582.21_{-0.05}^{+0.06}$  & $5.40\pm0.05$ & $4.41_{-0.10}^{+0.09}$ & - & - & -\\[1pt]
 8 & 1 & $1662.08_{-0.07}^{+0.06}$  & $5.47_{0.06}^{+0.05}$ & $4.89\pm0.11$ & - & - & -\\[1pt]
 9 & 1 & $1742.77_{-0.06}^{+0.05}$ & $5.41\pm0.05$ & $4.16 \pm0.08$ & - & - & - \\[1pt]
10 & 1 & $1824.61_{-0.10}^{+0.09}$ & $5.11\pm0.09$ & $3.96_{-0.21}^{+0.17}$ & - & - & - \\[1pt]
11 & 1 & $1906.90\pm0.12$ & $4.36\pm0.07$ & $4.28_{-0.15}^{+0.16}$ & - & - & - \\[1pt]
12 & 1 & $1989.21_{-0.15}^{+0.14}$ & $3.94_{-0.11}^{+0.10}$ & $3.82\pm0.29$ & - & - & - \\[1pt]
13 & 1 & $2072.11_{-0.18}^{+0.17}$ & $3.44\pm0.08$ & $4.19_{-0.20}^{+0.19}$ & - & - & - \\[1pt]
14 & 1 & $2153.61_{-0.17}^{+0.19}$ & $2.75_{-0.07}^{+0.06}$ & $4.41_{-0.29}^{+0.27}$ & - & - & -\\[1pt]
15 & 1 & $2235.66_{-0.22}^{+0.21}$ & $2.38_{-0.09}^{+0.08}$ & $4.28_{-0.38}^{+0.39}$ & - & - & - \\[1pt]
16 & 1 & $2318.78_{-0.28}^{+0.26}$ & $2.00\pm0.08$ & $4.59_{-0.33}^{+0.29}$ & - & - & - \\[1pt]
17 & 1 & $2399.13_{-0.24}^{+0.26}$ & $1.65_{-0.07}^{-0.05}$ & $4.74_{-0.25}^{+0.23}$ & 0.000 & 1.000 & - \\[1pt]
18 & 1 & $2485.88_{-0.48}^{+0.47}$ & $1.37_{-0.07}^{+0.08}$ & $5.07_{-0.32}^{+0.29}$ & 0.991 & 0.009 & - \\[1pt]
 \hline
 \\[-8pt]
 0 & 2 & $979.42_{-0.48}^{+0.49}$  & $0.93\pm0.12$ & $2.13_{-0.42}^{+0.43}$ & 0.244 & 0.618 & 0.138\\[1pt]
 1 & 2 & $1055.96_{-0.67}^{+0.91}$ & $1.12\pm0.15$ & $2.10_{-0.43}^{+0.41}$ & 0.000 & 0.982 & 0.018\\[1pt]
 2 & 2 & $1137.73_{-0.29}^{+0.31}$ & $1.14\pm0.11$ & $2.21_{-0.38}^{+0.33}$ & 0.000 & 0.291 & 0.709\\[1pt]
 3 & 2 & $1216.34_{-0.33}^{+0.34}$ & $1.59_{-0.17}^{+0.18}$ & $2.40_{-0.46}^{+0.59}$ & - & 0.000 & 1.000\\[1pt]
 4 & 2 & $1294.75\pm0.29$ & $1.52_{-0.10}^{+0.11}$ & $2.34_{-0.21}^{+0.25}$ & - & 0.002 & 0.998 \\[1pt]
 5 & 2 & $1376.80_{-0.36}^{+0.32}$ & $1.77_{-0.23}^{+0.25}$ & $2.10_{-0.85}^{+0.72}$ & - & 0.000 & 1.000\\[1pt]
 6 & 2 & $1457.70_{-0.32}^{+0.31}$ & $2.07\pm0.12$ & $3.00_{-0.34}^{+0.17}$ & - & 0.000 & 1.000\\[1pt]
 7 & 2 & $1537.22_{-0.31}^{+0.40}$ & $2.41\pm0.12$ & $3.61\pm0.22$ & - & 0.000 & 1.000\\[1pt]
 8 & 2 & $1619.18_{-0.90}^{+0.89}$ & $2.72\pm0.46$ & $3.93_{-0.89}^{+0.97}$ & - & 0.000 & 1.000\\[1pt]
 9 & 2 & $1698.12\pm0.11$ & $2.97\pm0.07$ &  $3.27_{-0.15}^{+0.16}$ & - & 0.000 & 1.000\\[1pt]
10 & 2 &  $1779.42_{-0.62}^{+0.48}$ & $2.72_{-0.04}^{+0.05}$ & $3.25_{-0.10}^{+0.09}$ & - & 0.000 & 1.000\\[1pt]
11 & 2 & $1860.52_{-0.59}^{+0.44}$  & $2.03_{-0.15}^{+0.14}$ & $3.39_{-0.39}^{+0.40}$ & - & 0.001 & 0.999\\[1pt]
12 & 2 & $1944.32_{-0.28}^{+0.35}$ & $2.18_{-0.18}^{+0.20}$ & $3.11_{-0.65}^{+0.59}$ & - & 0.000 & 1.000\\[1pt]
13 & 2 &  $2027.51_{-0.57}^{+0.74}$ & $1.80_{-0.22}^{+0.28}$ & $3.64_{-0.87}^{+0.69}$ & - & 0.010 & 0.990\\[1pt]
14 & 2 & $2108.03_{-0.29}^{+0.28}$ & $1.55_{-0.20}^{+0.21}$ & $1.99_{-0.74}^{+0.63}$ & - & 0.000 & 1.000\\[1pt]
15 & 2 & $2189.96_{-0.18}^{+0.19}$ & $1.17_{-0.09}^{+0.10}$ & $1.73_{-0.35}^{+0.35}$ & - & 0.000 & 1.000\\[1pt]
16 & 2 & $2264.08_{-0.36}^{+0.40}$ & $0.79\pm0.06$ & $2.78_{-0.35}^{+0.31}$ & - & 0.525 & 0.475\\[1pt]
17 & 2 & $2353.04_{-0.57}^{+0.75}$ & $0.90_{-0.09}^{+0.09}$ & $3.19_{-0.33}^{+0.34}$ & 0.056 & 0.387 & 0.557\\[1pt]
18 & 2 & $2438.00_{-0.25}^{+0.28}$ & $0.99_{-0.10}^{+0.11}$ & $2.45_{-0.71}^{+0.69}$ & 0.000 & 0.040 & 0.960\\[1pt] 
19 & 2 & $2519.84_{-2.14}^{+1.00}$ & $0.96_{-0.09}^{+0.10}$ & $3.00_{-0.27}^{+0.36}$ & 0.704 & 0.296 & 0.000\\[1pt]
\hline                                
\end{tabular}
\label{tab:frequencies}
\end{table*}

\section{Results for peak bagging of KIC~9139163 based on A14 background}
\label{sec:results3}
In this Appendix, we report the same oscillation peaks as in Table~\ref{tab:frequencies} but with frequencies, amplitudes, and linewidths, as obtained by means of Approach 1 that is based on a different fit of the background, following the formulation given by A14, which consists of a flat noise component and a granulation Harvey-like profile. 

\begin{table*}
\caption{Median values with corresponding 68.3\,\% shortest credible intervals for the oscillation frequencies, amplitudes and linewidths of KIC~9139163, as stemmed by \diamonds\,\,for the case of the background model adopted by A14.}             
\centering                         
\begin{tabular}{l l l l l}       
\hline\hline
\\[-8pt]          
Peak \# & Angular Degree & \multicolumn{1}{c}{Frequency} & \multicolumn{1}{c}{Amplitude} & \multicolumn{1}{c}{Linewidth}\\   
 & & \multicolumn{1}{c}{($\mu$Hz)} & \multicolumn{1}{c}{(ppm)} & \multicolumn{1}{c}{($\mu$Hz)}\\
\hline
\\[-8pt]
0 & 0 &  $985.39_{-0.37}^{+0.46}$   &   $1.76_{-0.21}^{+0.22}$   &   $3.46_{-0.98}^{+1.01}$   \\[1pt]
1 & 0 &  $1065.03_{-0.14}^{+0.13}$   &   $2.70 \pm 0.11$   &   $3.51_{-0.44}^{+0.43}$   \\[1pt]
2 & 0 &  $1143.07_{-0.10}^{+0.09}$   &   $2.59_{-0.10}^{+0.08}$   &   $2.96_{-0.34}^{+0.28}$   \\[1pt]
3 & 0 &  $1221.19_{-0.18}^{+0.19}$   &   $3.26_{-0.16}^{+0.17}$   &   $4.17_{-0.54}^{+0.55}$   \\[1pt]
4 & 0 &  $1301.61_{-0.13}^{+0.14}$   &   $3.63_{-0.15}^{+0.13}$   &   $3.32_{-0.41}^{+0.38}$   \\[1pt]
5 & 0 &  $1383.06_{-0.17}^{+0.18}$   &   $4.31_{-0.18}^{+0.19}$   &   $4.45_{-0.51}^{+0.49}$   \\[1pt]
6 & 0 &  $1464.35\pm0.14$   &   $5.05\pm0.12$   &   $5.13_{-0.34}^{+0.32}$   \\[1pt]
7 & 0 &  $1544.31\pm0.16$   &   $5.79_{-0.16}^{+0.17}$   &   $5.26_{-0.32}^{+0.33}$   \\[1pt]
8 & 0 &  $1623.75_{-0.22}^{+0.20}$   &   $6.19_{-0.20}^{+0.24}$   &   $6.10\pm0.45$   \\[1pt]
9 & 0 &  $1704.31\pm0.23$   &   $5.27_{-0.29}^{+0.32}$   &   $4.62_{-0.58}^{+0.49}$   \\[1pt]
10 & 0 &  $1785.36\pm0.19$   &   $5.48_{-0.17}^{+0.20}$   &   $5.38_{-0.41}^{+0.38}$   \\[1pt]
11 & 0 &  $1867.01_{-0.16}^{+0.17}$   &   $5.26_{-0.13}^{+0.14}$   &   $5.46_{-0.33}^{+0.32}$   \\[1pt]
12 & 0 &  $1949.54_{-0.28}^{+0.29}$   &   $4.40_{-0.28}^{+0.26}$   &   $5.38_{-0.64}^{+0.59}$   \\[1pt]
13 & 0 &  $2032.80_{-0.44}^{+0.42}$   &   $4.27_{-0.22}^{+0.25}$   &   $7.08_{-0.46}^{+0.50}$   \\[1pt]
14 & 0 &  $2114.83_{-0.24}^{+0.25}$   &   $3.76\pm0.13$   &   $6.57_{-0.53}^{+0.41}$   \\[1pt]
15 & 0 &  $2196.36_{-0.81}^{+0.70}$   &   $3.28\pm0.27$   &   $7.67_{-0.83}^{+0.74}$   \\[1pt]
16 & 0 &  $2277.70_{-0.22}^{+0.25}$   &   $3.08\pm0.08$   &   $8.41_{-0.45}^{+0.50}$   \\[1pt]
17 & 0 &  $2360.73_{-0.84}^{+0.76}$   &   $1.95_{-0.21}^{+0.23}$   &   $9.03_{-0.96}^{+0.95}$   \\[1pt]
18 & 0 &  $2445.08_{-0.53}^{+0.56}$   &   $2.22_{-0.16}^{+0.17}$   &   $8.64_{-1.02}^{+1.07}$   \\[1pt]
19 & 0 &  $2526.11_{-0.60}^{+0.65}$   &   $1.34_{-0.19}^{+0.20}$   &   $9.37_{-1.01}^{+1.24}$   \\[1pt]
 \hline
 \\[-8pt]
 0 & 1 &  $1023.09\pm0.18$   &   $2.34_{-0.11}^{+0.10}$   &   $4.14_{-0.43}^{+0.44}$   \\[1pt]
1 & 1 &  $1101.80\pm0.13$   &   $2.76_{-0.08}^{+0.09}$   &   $4.15_{-0.37}^{+0.41}$   \\[1pt]
2 & 1 &  $1179.78\pm0.08$   &   $3.44_{-0.07}^{+0.08}$   &   $3.52_{-0.27}^{+0.25}$   \\[1pt]
3 & 1 &  $1258.71\pm0.10$   &   $4.10\pm0.07$   &   $4.89_{-0.22}^{+0.24}$   \\[1pt]
4 & 1 &  $1339.77_{-0.07}^{+0.08}$   &   $4.25_{-0.05}^{+0.06}$   &   $4.93_{-0.18}^{+0.19}$   \\[1pt]
5 & 1 &  $1421.65\pm0.10$   &   $5.20_{-0.08}^{+0.09}$   &   $4.87_{-0.25}^{+0.26}$   \\[1pt]
6 & 1 &  $1502.22_{-0.11}^{+0.12}$   &   $6.32_{-0.11}^{+0.10}$   &   $5.62_{-0.29}^{+0.27}$   \\[1pt]
7 & 1 &  $1582.13\pm0.06$   &   $6.45\pm0.06$   &   $5.60_{-0.12}^{+0.13}$   \\[1pt]
8 & 1 &  $1661.92_{-0.12}^{+0.13}$   &   $6.72_{-0.10}^{+0.09}$   &   $6.94_{-0.31}^{+0.28}$   \\[1pt]
9 & 1 &  $1742.66_{-0.12}^{+0.13}$   &   $6.79\pm0.10$   &   $6.37_{-0.30}^{+0.28}$   \\[1pt]
10 & 1 &  $1824.54\pm0.10$   &   $6.42\pm0.09$   &   $5.85_{-0.25}^{+0.24}$   \\[1pt]
11 & 1 &  $1906.80_{-0.13}^{+0.15}$   &   $5.75\pm0.09$   &   $7.31\pm-0.31$   \\[1pt]
12 & 1 &  $1988.93_{-0.10}^{+0.11}$   &   $5.37\pm0.06$   &   $6.51_{-0.22}^{+0.23}$   \\[1pt]
13 & 1 &  $2071.61_{-0.15}^{+0.16}$   &   $5.21_{-0.08}^{+0.07}$   &   $8.44_{-0.43}^{+0.32}$   \\[1pt]
14 & 1 &  $2153.60\pm0.13$   &   $4.52\pm0.06$   &   $8.68_{-0.24}^{+0.25}$   \\[1pt]
15 & 1 &  $2235.50_{-0.15}^{+0.14}$   &   $3.75_{-0.06}^{+0.05}$   &   $7.60_{-0.26}^{+0.21}$   \\[1pt]
16 & 1 &  $2318.00_{-0.26}^{+0.25}$   &   $3.58\pm0.07$   &   $9.48_{-0.38}^{+0.39}$   \\[1pt]
17 & 1 &  $2399.78\pm0.25$   &   $2.81\pm0.06$   &   $9.32_{-0.53}^{+0.51}$   \\[1pt]
18 & 1 &  $2486.69_{-0.40}^{+0.39}$   &   $1.97_{-0.09}^{+0.08}$   &   $7.86_{-0.84}^{+0.80}$   \\[1pt]
 \hline
 \\[-8pt]
 0 & 2 &  $979.72_{-0.78}^{+0.86}$   &   $0.96_{-0.24}^{+0.25}$   &   $2.95_{-0.94}^{+0.93}$   \\[1pt]
1 & 2 &  $1056.57_{-0.58}^{+0.63}$   &   $1.37_{-0.15}^{+0.17}$   &   $3.37_{-0.50}^{+0.52}$   \\[1pt]
2 & 2 &  $1137.57\pm0.24$   &   $1.65_{-0.11}^{+0.12}$   &   $3.25_{-0.31}^{+0.32}$   \\[1pt]
3 & 2 &  $1215.90_{-0.33}^{+0.31}$   &   $2.26_{-0.19}^{+0.20}$   &   $3.69_{-0.49}^{+0.48}$   \\[1pt]
4 & 2 &  $1294.71_{-0.42}^{+0.47}$   &   $2.42_{-0.18}^{+0.22}$   &   $4.92_{-0.70}^{+0.76}$   \\[1pt]
5 & 2 &  $1376.62_{-0.32}^{+0.30}$   &   $2.65_{-0.27}^{+0.28}$   &   $4.23\pm0.91$   \\[1pt]
6 & 2 &  $1457.08_{-0.30}^{+0.29}$   &   $2.65_{-0.19}^{+0.17}$   &   $4.22_{-0.40}^{+0.46}$   \\[1pt]
7 & 2 &  $1536.95_{-0.55}^{+0.52}$   &   $3.02_{-0.28}^{+0.23}$   &   $4.76_{-0.46}^{+0.47}$   \\[1pt]
8 & 2 &  $1616.92_{-0.77}^{+0.79}$   &   $3.01_{-0.41}^{+0.39}$   &   $5.08_{-0.84}^{+0.77}$   \\[1pt]
9 & 2 &  $1698.11_{-0.59}^{+0.48}$   &   $4.27_{-0.39}^{+0.41}$   &   $6.79_{-1.07}^{+1.17}$   \\[1pt]
10 & 2 &  $1778.87_{-0.34}^{+0.36}$   &   $3.35_{-0.28}^{+0.27}$   &   $5.13_{-0.76}^{+0.66}$   \\[1pt]
11 & 2 &  $1859.79_{-0.44}^{+0.47}$   &   $2.79_{-0.23}^{+0.22}$   &   $5.37_{-0.49}^{+0.50}$   \\[1pt]
12 & 2 &  $1943.64_{-0.50}^{+0.56}$   &   $3.16_{-0.30}^{+0.38}$   &   $5.87_{-0.56}^{+0.79}$   \\[1pt]
13 & 2 &  $2024.59_{-0.98}^{+0.79}$   &   $2.72_{-0.30}^{+0.31}$   &   $5.91_{-0.82}^{+0.70}$   \\[1pt]
14 & 2 &  $2106.98_{-0.35}^{+0.34}$   &   $2.79_{-0.15}^{+0.16}$   &   $6.37_{-0.64}^{+0.52}$   \\[1pt]
15 & 2 &  $2188.53_{-0.81}^{+1.01}$   &   $2.40_{-0.32}^{+0.34}$   &   $6.64_{-0.98}^{+0.97}$   \\[1pt]
16 & 2 &  $2264.53_{-0.42}^{+0.43}$   &   $1.68_{-0.12}^{+0.11}$   &   $6.60_{-0.55}^{+0.57}$   \\[1pt]
17 & 2 &  $2353.13_{-0.79}^{+0.88}$   &   $1.86_{-0.24}^{+0.22}$   &   $7.27_{-1.10}^{+1.11}$   \\[1pt]
18 & 2 &  $2436.78_{-0.81}^{+0.71}$   &   $1.68\pm0.20$   &   $7.51_{-1.13}^{+1.29}$   \\[1pt]
19 & 2 &  $2519.35_{-1.10}^{+1.18}$   &   $1.20_{-0.19}^{+0.18}$   &   $7.97_{-1.19}^{+1.21}$   \\[1pt]
\hline
\\[-8pt]
\end{tabular}
\label{tab:frequenciesApp}
\end{table*}

\end{document}